\documentclass[ALICE,manyauthors]{cernphprep}
\usepackage[comma,square,numbers,sort&compress]{natbib}
\usepackage{hyperref}
\usepackage{lineno}
\usepackage[T1]{fontenc}
\usepackage{xspace}
\usepackage[utf8]{inputenc}
\usepackage{booktabs}
\usepackage{multirow}
\usepackage{color}
\usepackage{euscript}
\usepackage{mathptmx}
\usepackage{silence}
\usepackage[T1]{fontenc}
\usepackage{orcidlink}

\begin{document}
%

\mathchardef\mhyphen="2D

\newcommand{\pp}           {pp\xspace}
\newcommand{\ppbar}        {\mbox{$\mathrm {p\overline{p}}$}\xspace}
\newcommand{\XeXe}         {\mbox{Xe--Xe}\xspace}
\newcommand{\PbPb}         {\mbox{Pb--Pb}\xspace}
\newcommand{\pA}           {\mbox{pA}\xspace}
\newcommand{\pPb}          {\mbox{p--Pb}\xspace}
\newcommand{\AuAu}         {\mbox{Au--Au}\xspace}
\newcommand{\dAu}          {\mbox{d--Au}\xspace}

\newcommand{\sigmapid}{$\sigma^{^{3}\mathrm{He}}_{\mathrm{d}E/\mathrm{d}x}$}
\newcommand{\nsigma} {$\left(  \mathrm{d}E/\mathrm{d}x -  \langle \mathrm{d}E/\mathrm{d}x \rangle_{^{3}\mathrm{He}} \right) /\sigma^{^{3}\mathrm{He}}_{\mathrm{d}E/\mathrm{d}x}$ }
\newcommand{\s}            {\ensuremath{\sqrt{s}}\xspace}
\newcommand{\snn}          {\ensuremath{\sqrt{s_{\mathrm{NN}}}}\xspace}
\newcommand{\pt}           {\ensuremath{p_{\rm T}}\xspace}
\newcommand{\meanpt}       {$\langle p_{\mathrm{T}}\rangle$\xspace}
\newcommand{\ycms}         {\ensuremath{y_{\rm CMS}}\xspace}
\newcommand{\ylab}         {\ensuremath{y_{\rm lab}}\xspace}
\newcommand{\etarange}[1]  {\mbox{$\left | \eta \right |~<~#1$}}
\newcommand{\yrange}[1]    {\mbox{$\left | y \right |~<~#1$}}
\newcommand{\dndy}         {\ensuremath{\mathrm{d}N_\mathrm{ch}/\mathrm{d}y}\xspace}
\newcommand{\dndeta}       {\ensuremath{\mathrm{d}N_\mathrm{ch}/\mathrm{d}\eta}\xspace}
\newcommand{\avdndeta}     {\ensuremath{\langle\dndeta\rangle}\xspace}
\newcommand{\dNdy}         {\ensuremath{\mathrm{d}N_\mathrm{ch}/\mathrm{d}y}\xspace}
\newcommand{\Npart}        {\ensuremath{N_\mathrm{part}}\xspace}
\newcommand{\Ncoll}        {\ensuremath{N_\mathrm{coll}}\xspace}
\newcommand{\dEdx}         {\ensuremath{\textrm{d}E/\textrm{d}x}\xspace}
\newcommand{\RpPb}         {\ensuremath{R_{\rm pPb}}\xspace}

\newcommand{\nineH}        {$\sqrt{s}~=~0.9$~Te\kern-.1emV\xspace}
\newcommand{\seven}        {$\sqrt{s}~=~7$~Te\kern-.1emV\xspace}
\newcommand{\twoH}         {$\sqrt{s}~=~0.2$~Te\kern-.1emV\xspace}
\newcommand{\twosevensix}  {$\sqrt{s}~=~2.76$~Te\kern-.1emV\xspace}
\newcommand{\five}         {$\sqrt{s}~=~5.02$~Te\kern-.1emV\xspace}
\newcommand{\twosevensixnn}{$\sqrt{s_{\mathrm{NN}}}~=~2.76$~Te\kern-.1emV\xspace}
\newcommand{\fivenn}       {$\sqrt{s_{\mathrm{NN}}}~=~5.02$~Te\kern-.1emV\xspace}
\newcommand{\LT}           {L{\'e}vy-Tsallis\xspace}
\newcommand{\GeVc}         {Ge\kern-.1emV/$c$\xspace}
\newcommand{\MeVc}         {Me\kern-.1emV/$c$\xspace}
\newcommand{\TeV}          {Te\kern-.1emV\xspace}
\newcommand{\GeV}          {Ge\kern-.1emV\xspace}
\newcommand{\MeV}          {Me\kern-.1emV\xspace}
\newcommand{\GeVmass}      {Ge\kern-.2emV/$c^2$\xspace}
\newcommand{\MeVmass}      {Me\kern-.2emV/$c^2$\xspace}
\newcommand{\lumi}         {\ensuremath{\mathcal{L}}\xspace}

\newcommand{\ITS}          {\rm{ITS}\xspace}
\newcommand{\TOF}          {\rm{TOF}\xspace}
\newcommand{\ZDC}          {\rm{ZDC}\xspace}
\newcommand{\ZDCs}         {\rm{ZDCs}\xspace}
\newcommand{\ZNA}          {\rm{ZNA}\xspace}
\newcommand{\ZNC}          {\rm{ZNC}\xspace}
\newcommand{\SPD}          {\rm{SPD}\xspace}
\newcommand{\SDD}          {\rm{SDD}\xspace}
\newcommand{\SSD}          {\rm{SSD}\xspace}
\newcommand{\TPC}          {\rm{TPC}\xspace}
\newcommand{\TRD}          {\rm{TRD}\xspace}
\newcommand{\VZERO}        {\rm{V0}\xspace}
\newcommand{\VZEROA}       {\rm{V0A}\xspace}
\newcommand{\VZEROC}       {\rm{V0C}\xspace}
\newcommand{\Vdecay} 	   {\ensuremath{V^{0}}\xspace}

\newcommand{\ee}           {\ensuremath{e^{+}e^{-}}} 
\newcommand{\pip}          {\ensuremath{\pi^{+}}\xspace}
\newcommand{\pim}          {\ensuremath{\pi^{-}}\xspace}
\newcommand{\kap}          {\ensuremath{\rm{K}^{+}}\xspace}
\newcommand{\kam}          {\ensuremath{\rm{K}^{-}}\xspace}
\newcommand{\pbar}         {\ensuremath{\rm\overline{p}}\xspace}
\newcommand{\kzero}        {\ensuremath{{\rm K}^{0}_{\rm{S}}}\xspace}
\newcommand{\lmb}          {\ensuremath{\Lambda}\xspace}
\newcommand{\almb}         {\ensuremath{\overline{\Lambda}}\xspace}
\newcommand{\Om}           {\ensuremath{\Omega^-}\xspace}
\newcommand{\Mo}           {\ensuremath{\overline{\Omega}^+}\xspace}
\newcommand{\X}            {\ensuremath{\Xi^-}\xspace}
\newcommand{\Ix}           {\ensuremath{\overline{\Xi}^+}\xspace}
\newcommand{\Xis}          {\ensuremath{\Xi^{\pm}}\xspace}
\newcommand{\Oms}          {\ensuremath{\Omega^{\pm}}\xspace}
\newcommand{\degree}       {\ensuremath{^{\rm o}}\xspace}

\begin{titlepage}
\PHyear{2024}       
\PHnumber{144}      
\PHdate{23 May}     

\title{Measurement of the production and elliptic flow of (anti)nuclei \\in Xe--Xe collisions at $\mathbf{\sqrt{\textit{s}_{\textup{NN}}} = 5.44}$ TeV}
\ShortTitle{(Anti)nuclei production and flow in Xe--Xe collisions at ${\sqrt{\textit{s}_{\textup{NN}}} = 5.44}$ TeV}   

\Collaboration{ALICE Collaboration\thanks{See Appendix~\ref{app:collab} for the list of collaboration members}}
\ShortAuthor{ALICE Collaboration} 

\begin{abstract}

Measurements of (anti)deuteron and (anti)$^3$He production in the rapidity range $ |y| < $ 0.5 as a function of the transverse momentum and event multiplicity in Xe--Xe collisions at a center-of-mass energy per nucleon--nucleon pair of \snn = 5.44 TeV are presented. The coalescence parameters $B_2$ and $B_3$ are measured as a function of the transverse momentum per nucleon. 
The ratios between (anti)deuteron and (anti)$^3$He yields and those of \mbox{(anti)protons} and pions are reported as a function of the mean charged-particle multiplicity density, and compared with two implementations of the statistical hadronization model and with coalescence predictions. 
The elliptic flow of (anti)deuterons is measured for the first time in Xe--Xe collisions and shows features similar to those already observed in Pb--Pb collisions, i.e., the mass ordering at low transverse momentum and the meson--baryon grouping at intermediate transverse momentum.
The production of nuclei is particularly sensitive to the chemical freeze-out temperature of the system created in the collision, which is extracted from a grand-canonical-ensemble-based thermal fit, performed for the first time including light nuclei along with light-flavor hadrons in Xe--Xe collisions. The extracted chemical freeze-out temperature \mbox{$T_{\rm chem}$ = (154.2 $\pm$ 1.1) MeV} in Xe--Xe collisions is similar to that observed in Pb--Pb collisions and close to the crossover temperature predicted by lattice quantum chromodynamics calculations.

\end{abstract}
\end{titlepage}

\setcounter{page}{2}

\section{Introduction} 
\label{sec:Introduction}

The investigation of the production mechanism of light (anti)nuclei in high-energy hadronic collisions is one of the main topics in modern nuclear physics. The ALICE Collaboration has significantly contributed to this field with systematic measurements of (anti)nucleus production in different collision systems and center-of-mass energies provided by the Large Hadron Collider (LHC)~\cite{Adam:2015pna, 3HePbPb, Acharya:2017dmc, Acharya:2017fvb, Acharya:2017bso, deuteron_pp7TeV, Acharya:2019rys, Acharya:2019ttn, Acharya:2019xmu, Acharya:2020sfy, Acharya:2020lus, deuteron_jets, nuclei_pp_13TeV, nuclei_pp_5TeV, hypertriton_pPb_ALICE, ALICE:2022boh, ALICE:2022weh, ALICE:2022jmr, ALICE:2022wpn}. These results extend previous experimental measurements at lower collision energies, from the BNL Alternating Gradient Synchrotron (AGS)~\cite{ Bennett:1998be, Ahle:1999in, Armstrong:2000gz, Armstrong:2000gd}, the CERN Super Proton Synchrotron (SPS)~\cite{ Ambrosini:1997bf}, and the BNL Relativistic Heavy Ion Collider (RHIC)~\cite{Adler:2001prl,Adler:2004uy, Arsene:2010px,Agakishiev:2011ib,Adamczyk:2016gfs,Adam:2019wnb}, to the TeV energy scale. In addition to transverse-momentum ($p_{\rm T}$) and event-multiplicity differential measurements of (anti)nucleus production yields, the ALICE Collaboration has also measured the anisotropic flow of (anti)deuterons~\cite{Acharya:2017dmc,Acharya:2020lus} and (anti)${}^{3}$He~\cite{Acharya:2019ttn}, and (anti)deuteron number fluctuations~\cite{deuteronFluctuations} in Pb--Pb collisions. 

Such a large wealth of experimental data is typically described using three categories of theoretical models: the statistical hadronization model 
(SHM)~\cite{SHM5,SHM6,SHM4,SHM2,SHM1,SHM3,Sharma:2022poi}, models that combine relativistic hydrodynamics with a subsequent hadronic afterburner~\cite{Zhao:2018lyf,Oliinychenko:2018ugs}, and baryon coalescence~\cite{Coalescence3,Coalescence1, SATO1981153, Csernai:1986qf, Coalescence2,Blum:2017qnn,Zhao:2018lyf,Mahlein:2023fmx}. In the former, light (anti)nuclei are assumed to be produced by a source in local thermal and hadrochemical equilibrium at the chemical freeze-out with a temperature of about 155 MeV~\cite{SHM1}. Light (anti)nuclei are bound states with binding energies of about 2.2 MeV for deuterons, 8.48 MeV for tritons, and 7.72 MeV for $^3$He nuclei, orders of magnitude smaller than the temperature of the environment created in the high-energy nuclear collisions. In the context of the SHM, two solutions have been proposed to explain the survival of such loosely bound states in the hot and dense hadron gas phase produced in ultrarelativistic heavy-ion collisions. The first assumes that light (anti)nuclei are created at the hadronization phase boundary as compact colorless multi-quark systems with negligible interaction cross sections with hadrons~\cite{SHM1}. The time scale for the evolution of their wave function is estimated to be longer than the lifetime of the hadron gas, which is about 10 fm/$c$. The second assumes that while the abundance of elementary hadrons, including nucleons, is determined at chemical freeze-out, the yield of light nuclei (i.e., composite objects with a binding energy much smaller than the temperature) continues to evolve in thermal equilibrium between nuclei and nucleons until the kinetic freeze-out is reached~\cite{Vovchenko:2019aoz}.

Recent model developments have tackled the unresolved question of the survival of weakly bound multi-baryon states within the hadron gas phase with intense rescattering, combining relativistic hydrodynamics with a subsequent afterburner of hadrons~\cite{Zhao:2018lyf,Oliinychenko:2018ugs}. Within these models, nucleons and light nuclei are generated at the phase transition using the Cooper-Frye formula~\cite{Cooper:1974mv}, which characterizes hadron generation based on the local energy density of the fireball. The hadron yields are set according to the values predicted by the thermal model at the chemical freeze-out temperature. The propagation of these particles through the hadronic medium is simulated using various transport codes (such as UrQMD~\cite{Bass:1998ca,Bleicher:1999xi} or SMASH~\cite{Oliinychenko:2018ugs}), considering the well-established interaction cross sections and resonant states. 

Finally, in the coalescence model, light (anti)nuclei are formed by the coalescence of nucleons that are close in phase space and with matching spin-isospin configurations, at kinetic freeze-out (occurring when the elastic interactions stop and the momentum of the particles is fixed). In the state-of-the-art implementation of the coalescence model, the formation probability is calculated by folding the phase-space distributions of (point-like) nucleons with the Wigner density distribution of the bound state. Significant progress has been made over the last decade in these phenomenological approaches driven by the growing amount of available data.   

Although both SHM and coalescence models are successful in describing different aspects of nuclear production, they are also different in some key aspects. For instance, the coalescence model is sensitive to the size of the nucleus and its wave function, and in particular to the relation between nuclear size and emission source size~\cite{SATO1981153,Coalescence2}. On the contrary, the SHM predictions depend only on the mass and on the spin degeneracy factor of the nucleus. This aspect is particularly relevant in small collision systems (e.g., pp and p--Pb collisions), since the size of the nucleus ($r_{\rm {deuteron}}$ = 1.96 fm and \mbox{$r_{\rm{ ^3He}}$ = 1.76 fm}~\cite{Ropke:2008qk}) and that of the baryon-emitting source (\mbox{$r\approx1$ fm} in pp collisions~\cite{ALICE:2018ysd} and $r\sim 1.5$ fm in p--Pb collisions~\cite{ALICE:2019hdt}) are comparable. In turn, in the SHM the evolution of the particle yields with the event multiplicity is driven solely by the baryon number conservation, assuming that the hadronization temperature is constant. For this reason, an interesting observable to test the consistency of models describing the hadron chemistry in high-energy hadronic collisions is the ratio of the integrated yields of nuclei relative to those of protons (and of pions, to test the SHM only). Such ratios increase smoothly with multiplicity, and such behavior is qualitatively reproduced by both SHM and coalescence models. In the former, the plateau of the ratios at high multiplicity is described by the grand-canonical ensemble of the statistical model~\cite{SHM1, CoalescenceTheory, Coalescence2}, where all the charges are only conserved on average, but fluctuate from one microscopic state to another. Such ensemble holds true to describe the particle yield in the case of sufficiently large reaction volumes, as in heavy-ion collisions. 
In the low-multiplicity region, corresponding to small collision systems, the exact conservation of charges from one microscopic state to another plays an important role in determining the final state particle yield. Therefore, the canonical ensemble of the SHM is used. This results in the so-called canonical suppression of the yields of particles carrying conserved charges, relative to their grand-canonical values. In the coalescence models, instead, the trend of the particle ratios with multiplicity is related to the interplay of the increasing source size and the decreasing coalescence probability with increasing multiplicity. 
Finally, in the SHM nuclei are produced at the chemical freeze-out, when the hadron species are fixed, while in the coalescence model nucleons coalesce to form nuclei between the chemical and the kinetic freeze-out. 
Many efforts have been put forward recently, in the latest developments of the SHM, to explain the survival of nuclei in a system with \mbox{$T_{\rm{chem}} \sim$ 155 MeV}, postulating the possibility of their formation at the kinetic freeze-out (which occurs later, at \mbox{$T_{\rm{kin}}\sim(100$--$120)$ MeV}~\cite{STAR:2017sal,ALICE:2019hno}), see for instance Refs.~\cite{Xu:2017akx,Xu:2018jff, Oliinychenko:2018ugs,Neidig:2021bal,Sun:2022xjr}.

The current models present contrasting scenarios when describing the post-hadronization phase of loosely bound states. Hence, it becomes important to examine the measurements of the elliptic flow of light (anti)nuclei, namely the second harmonic component ($v_2$) obtained from the Fourier decomposition of the azimuthal distribution of their momentum with respect to the reaction plane. These measurements offer valuable insights into the propagation of these particles through the hadron gas phase and the dynamics of their interactions with other particles. Specifically, according to relativistic hydrodynamics, a clear mass ordering is expected at low \pt (i.e., at fixed \pt lighter particles have larger $v_2$ compared with heavier particles), when comparing the elliptic flow of nuclei and that of other hadron species, as observed in Refs.~\cite{FlowChargedParticlesLHC,Acharya:2020lus,flowIDhadronsXeXe}. Such mass ordering arises from the interplay between the radial flow (quantifying the isotropic expansion of the system of particles created in the collision) and the anisotropic flow, and therefore provides constraints on the transport coefficients of the medium. The mass ordering of the $v_2$ can also help to shed light on the coalescence mechanism, as a scaling relation is expected to follow from a naive quark coalescence model~\cite{Molnar:2003ff} that only allows quarks with equal momentum to form a hadron~\cite{Yan:2006bx}.

In this article, the first measurements of (anti)deuteron and (anti)${}^{3}$He production in Xe--Xe collisions at \snn = 5.44 TeV and of the (anti)deuteron elliptic flow are presented. These results contribute to the global picture that emerged from previous measurements, which is characterized by a continuous evolution of the ratios of the yield of (anti)nuclei to those of pions and protons and of the coalescence parameters with the charged-particle multiplicity. The results presented in this paper cover multiplicity intervals bridging between p--Pb and Pb--Pb collisions and fill the gaps between some measurements in heavy-ion collisions. The results are discussed in the context of the statistical hadronization and baryon coalescence models. 

\section{Experimental apparatus} 
\label{sec:ExperimentalApparatus}

The exceptional capabilities of ALICE in tracking and particle identification (PID), make it well-suited for the study of light nuclei and antinuclei in various collision systems at the LHC. A detailed description of the ALICE subdetectors and their performance can be found in Refs.~\cite{ALICE_general_2, ALICEperformance} and related references.

The analysis described in this work employs four detectors: the Inner Tracking System (\ITS), the Time Projection Chamber (\TPC), the Time-Of-Flight (\TOF), and the \VZERO detector~\cite{T0_Cortese:2004aa}. These detectors (except for the \VZERO, which is installed in the forward region) are positioned in the central barrel inside a solenoidal magnet generating a magnetic field with an intensity of up to {\it B} = 0.5 T.

The \ITS~\cite{ALICE_general_2,ITS}, which covers the entire azimuthal angle and the pseudorapidity range $|\eta|$ < 0.9, serves primarily as a tracking detector, allowing for the reconstruction of charged-particle tracks in the vicinity of the collision point. This allows for the precise reconstruction of the position of the primary and secondary vertices from weak decays, starting from the reconstructed tracks. It comprises three subsystems of silicon detectors arranged cylindrically around the beam axis: the two innermost layers are Silicon Pixel Detectors (\SPD), the central two layers are Silicon Drift Detectors (\SDD), and finally the two outermost layers are made of Silicon Strip Detectors (\SSD). The \ITS plays a role in discriminating primary and secondary nuclei produced in spallation processes with the detector material and the beam vacuum tube, via the determination of the distance-of-closest approach (DCA) of the track to the primary vertex.

The primary tracking detector, namely the \TPC~\cite{TPC}, allows for charged-particle reconstruction with up to 159 three-dimensional space points and for particle identification, by measuring the specific ionization energy loss (\dEdx) in the gas. The \TPC is a cylindrical drift chamber, coaxial with the beam vacuum tube and filled with a gas mixture containing 90\% Ne and 10\% CO$_2$ at atmospheric pressure during the Xe--Xe data taking. With a radius ranging from 85 to 250 cm and a length of 500 cm in the beam direction, the \TPC volume covers the same pseudorapidity interval as the \ITS. The \TPC provides a measurement of the specific energy loss with a resolution ranging from about 5.2\% in pp collisions to about 6.5\% in central Pb--Pb collisions~\cite{ALICEperformance}, when traversed by minimum ionizing particles throughout the entire detector. Charged-particle transverse momentum is measured with a resolution ranging from about 1\% at 1 $\mathrm{GeV}/\textit{c}$ to approximately 3\% at 10 $\mathrm{GeV}/\textit{c}$ using the \TPC. 

The \TOF detector~\cite{TOF} covers the full azimuthal angle and the pseudorapidity interval $|\eta|$ < 0.9. It consists of Multi-gap Resistive Plate Chambers (MRPCs) positioned at an average distance of 3.8 m from the nominal interaction point. 
The TOF time resolution is 56 ps~\cite{Carnesecchi:2018oss}, while the event time resolution varies depending on the collision system and the track multiplicity~\cite{Adam:2016ilk}. The start time for the time-of-flight is provided by the T0 detector~\cite{T0_Cortese:2004aa} and by the TOF detector itself. The TOF-based determination of the start time is particularly useful for measurements at large multiplicities, as it is the case of Xe–Xe collisions for which a resolution better than 20 ps is achieved if more than 50 tracks are used for its determination. The T0 consists of two arrays of Cherenkov counters, T0A and T0C, located on opposite sides of the interaction point, covering the pseudorapidity regions $4.6 < \eta < 4.9$ and $-3.3 < \eta < -3.0$. A weighted average is performed when both T0 and TOF detectors have measured the start time~\cite{Adam:2016ilk}. For the particle identification, the measured time-of-flight is compared with the expected time calculated from the track momentum and length for each possible mass hypothesis. 

Finally, the \VZERO detector~\cite{T0_Cortese:2004aa} is used to define the minimum-bias (MB) trigger (requiring coincident signals in the \VZERO detectors to be synchronous with the bunch crossing time defined by the LHC clock) and to select events based on their multiplicity, which is related to the centrality of the Xe--Xe collision. It comprises two arrays of plastic scintillators (\VZEROA and \VZEROC) located at asymmetric positions, one on each side of the interaction point. They cover the pseudorapidity regions $2.8 < \eta < 5.1$ and $-3.7 < \eta < -1.7$, respectively. 
The analysis of (anti)deuterons is conducted in several centrality classes defined as percentiles of the measured amplitude distribution in the \VZEROA and \VZEROC counters~\cite{VZEROPerformance}: 0--10\%, 10--20\%, 20--40\%, 40--60\%, and 60--90\%, where 0\% corresponds to the most central collisions and 90\% to the most peripheral ones. Instead, (anti)$^3$He nuclei, due to their lower abundance, are analyzed in a single wide centrality class 0--90\%.
\section{Data analysis}
\label{sec:DataAnalysis}

\subsection{Event selection}
\label{subsec:EventSelection}
The data used for this analysis were collected in 2017 when Xenon nuclei ($^{129}_{54}\text{Xe}$) collided at a center-of-mass energy per nucleon--nucleon collision of $\sqrt{s_\text{NN}}$ = 5.44 TeV for the first time at the LHC. The recorded instantaneous luminosity was \mbox{$2\times 10^{-25}~\text{cm}^{-2}\text{s}^{-1}$}, and the detected hadronic interaction rate was around 80--150 Hz. The integrated luminosity delivered to ALICE was fractions of 1~$\mu b^{-1}$~\cite{Schaumann:2018qat}. The solenoidal magnet of the ALICE apparatus was operated in a low-field configuration of 0.2 T. 
The minimum bias event trigger was fully efficient in the centrality interval 0--90$\%$ as demonstrated by the flat centrality distribution~\cite{ALICE-PUBLIC-2018-003}. The analysis is performed in different centrality classes, selected offline, with varying widths depending on the analysis and the nucleus species. To keep the conditions of the detectors as uniform as possible and reject background collisions, the $z$-coordinate of the primary vertex along the beam axis is required to be within 10 cm from the nominal interaction point. Given the low interaction rate during data taking, the pile-up of collisions in the same bunch crossing is found to be negligible. Events with pile-up occurring during the drift time of the TPC are rejected based on the correlation between the total number of the space points (clusters) in the TPC and the number of SDD and SSD clusters, as described in Ref.~\cite{JPsiFlow}. The residual pile-up contribution, estimated using MC simulations anchored to the same data-taking period, is found to be less than 1$\%$. The relative contribution of tracks produced in out-of-bunch collisions is further reduced for (anti)deuterons because their reconstructed tracks are required to have an associated hit in the TOF detector, which is used for the (anti)deuteron identification. In the case of (anti)$^{3}$He, due to their low production rate, the probability for simultaneous productions in pile-up events is less than 6$\times 10^{-9}$. Any bias in the multiplicity of the event containing an (anti)$^{3}$He does not affect the analysis, which is performed in the full centrality interval 0--90$\%$. In total, about 1.3 million events are selected for further analysis.

\subsection{Track selection and identification of (anti)nuclei}
\label{subsec:TrackSelectionPID}

(Anti)deuteron and (anti)$^{3}$He candidates are selected from a sample of charged-particle tracks reconstructed in the ITS and TPC in the pseudorapidity range $|\eta| < 0.8$. Several track quality criteria are applied, such as having a number of TPC crossed rows larger than 70, a number of TPC clusters ($N^{\rm TPC}_{\rm cls}$) used for the \dEdx calculation larger than 50 to ensure a good \dEdx resolution, a fraction of TPC crossed rows to findable clusters larger than 80$\%$, a minimum of two reconstructed hits in the ITS with at least one located in any of the two innermost layers, and a good track fit in the TPC by requiring $\chi^{2}/N^{\rm TPC}_{\rm cls} < 4$. The contribution from secondary tracks is reduced by requiring a maximal DCA to the primary vertex in the transverse plane ($\mathrm{DCA}_{xy}$) and in the longitudinal direction ($\mathrm{DCA}_{z}$) lower than 0.1 cm.

\begin{figure}[hbt]
    \begin{center}
    \includegraphics[width = 0.49\textwidth]{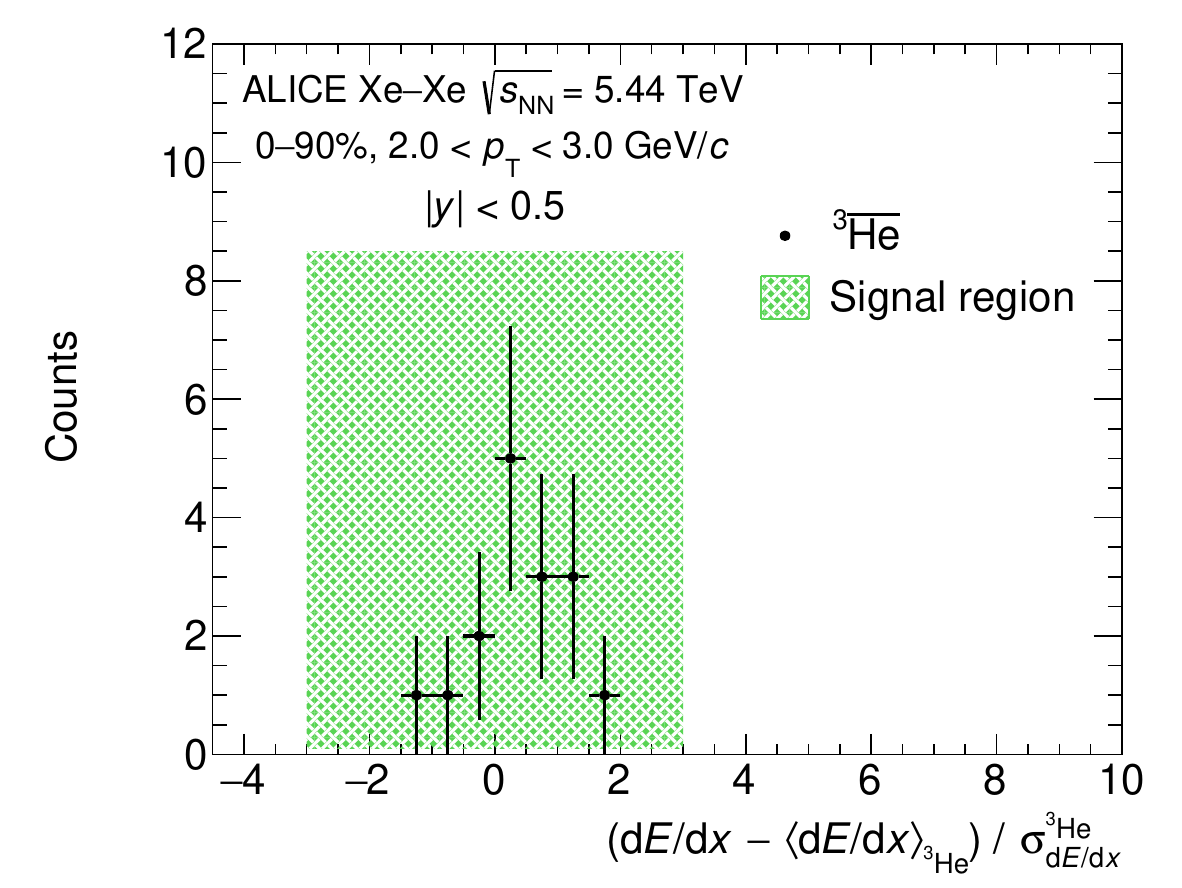}
    \includegraphics[width = 0.49\textwidth]{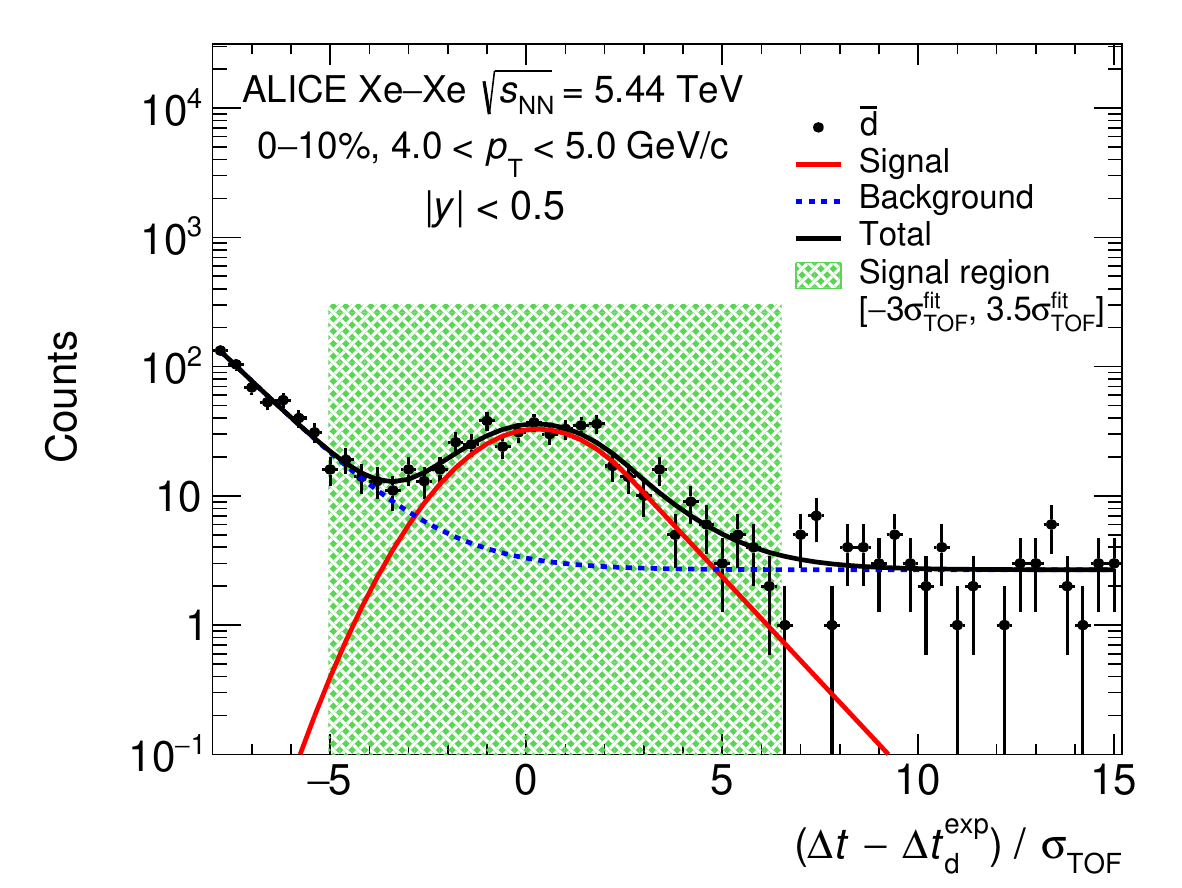}
    \end{center}
\caption{Distribution of the $\left(\mathrm{d}E/\mathrm{d}x-\langle\mathrm{d}E/\mathrm{d}x\rangle_{\rm ^{3}He}\right)/\sigma^{\rm ^{3}He}_{\mathrm{d}E/\mathrm{d}x}$ method used to extract the signal of $^3\overline{\rm{He}}$ (left), and distribution of $(\Delta t - \Delta t_{\rm d}^{\rm exp})/\sigma_{\rm TOF}$ used to extract the signal of $\overline{\rm{d}}$ (right). The regions highlighted in green represent the regions of the signal integration, different for the two signal extraction methods. }
\label{fig:signalExtraction}        
\end{figure}

The (anti)$^{3}$He identification is based on the energy loss per unit of track length measured by the TPC~\cite{Adam:2015pna, 3HePbPb, Acharya:2017fvb, Acharya:2019rys, Acharya:2019xmu, nuclei_pp_13TeV, nuclei_pp_5TeV, ALICE:2022boh, ALICE:2022weh}. The (anti)$^{3}$He candidates are selected by requiring that their measured \dEdx is within 3$\sigma^{\rm ^{3}He}_{\mathrm{d}E/\mathrm{d}x}$ from the expected average value, calculated using the Bethe–Bloch parametrization~\cite{BetheBloch}, where $\sigma^{\rm ^{3}He}_{\mathrm{d}E/\mathrm{d}x}$ is the \dEdx resolution for (anti)$^{3}$He. The contamination from other particle species is found to be negligible over the full transverse-momentum range covered by this measurement, which is $1<p_{\rm T}<7$ GeV/$c$. In the case of (anti)deuterons, both TPC and TOF information are used to extract signal candidates. A sample of deuteron candidates is selected applying a 3$\sigma$ selection on the difference between the measured $\mathrm{d}E/\mathrm{d}x$ in the TPC and the expected average value for deuterons. Then the (anti)deuteron signal is extracted using the TOF information of preselected deuteron candidates~\cite{Adam:2015pna, Acharya:2017dmc, Acharya:2017fvb, deuteron_pp7TeV, Acharya:2019rys, Acharya:2020sfy, deuteron_jets, nuclei_pp_13TeV, nuclei_pp_5TeV, ALICE:2022boh, ALICE:2022weh, ALICE:2022jmr}, from a fit to the $(\Delta t-\Delta t_{\rm exp})/\sigma_{\rm TOF}$ distribution, where $\Delta t$ is the time-of-flight measured by the TOF detector, $\Delta t_{\rm exp}$ is its expected value for deuterons, and $\sigma_{\rm TOF}$ is the time-of-flight resolution of TOF. To model the signal the fit is executed using a Gaussian function with an exponential tail, in the range $-7 < (\Delta t - \Delta t_{\rm exp})/\sigma_{\rm TOF} <$ 15. To model the background the fit uses an exponential, for $p_{\rm T}<4$ GeV/$c$, or the sum of an exponential and a linear function, for $p_{\rm T}>4$ GeV/$c$. The (anti)deuteron yield is calculated by integrating the signal counts (counting the entries in the histogram after subtracting the background fit function) in the range [$-3\sigma^{\rm fit}_{\mathrm{TOF}}$, $+3.5\sigma^{\rm fit}_{\mathrm{TOF}}$], being $\sigma^{\rm fit}_{\mathrm{TOF}}$ the TOF resolution extracted from the fit. The reason for an asymmetric interval is the presence of an exponential tail towards higher values of $(\Delta t-\Delta t_{\rm exp})/\sigma_{\rm TOF}$, that reflects the TOF detector time response~\cite{TOF}. 
As an example, the signal extraction procedure, for both (anti)$^{3}$He and (anti)deuterons, is shown in Fig.~\ref{fig:signalExtraction}.

\subsection{Corrections based on simulations}
\label{subsec:Corrections}

The product of the geometrical acceptance and the reconstruction and selection efficiency (Acceptance $\times$ Efficiency) of (anti)nuclei and the contamination from secondary nuclei from spallation are calculated using Monte Carlo (MC) simulations. The HIJING~\cite{HIJING} event generator is used to simulate Xe--Xe collision events. Nuclei and antinuclei are embedded into the simulated events with a uniform distribution in transverse momentum and rapidity, within \mbox{0 < \pt < 10 $\mathrm{GeV}/\textit{c}$} and --1 < $y$ < 1, respectively. The interactions of (anti)nuclei with the detector materials are simulated using GEANT4~\cite{GEANT4} as transport code. Considering that the $p_{\rm T}$ distributions of nuclei and their relative abundances in the simulation are different from data, centrality and $p_{\rm T}$-dependent corrections are applied in the MC simulations using weights. These are defined as the ratio of measured and generated \pt spectra in the simulation. 
A Blast-Wave (BW) parametrization~\cite{BlastWave1} is used to describe the \pt shape in real data in the full range of transverse momentum. The parameters of the Blast-Wave functions are obtained in the first iteration using the efficiency calculated based on the generated \pt distributions. The obtained BW parametrizations in different centrality intervals are then used to reweight the input shape in the MC and these updated efficiencies are used to correct the raw spectra. This iterative procedure converges already after two iterations. 

The Acceptance $\times$ Efficiency is calculated, in each centrality and \pt interval, as the ratio of reconstructed and generated primary (anti)nuclei in the simulations. The generated (anti)nuclei are counted in $|y|<0.5$, in order to normalize the corrected yield to one unit of rapidity. The same track selection and particle identification criteria as in data are applied to the reconstructed sample in the MC. Particles from pile-up events are removed from the generated sample and kept in the reconstructed sample to properly incorporate the pile-up correction into the efficiency. The (anti)deuteron efficiency varies from about 20\% and 30$\%$ for antimatter and matter, respectively for \mbox{\pt$<1.2$ GeV/$c$} to about 45\% for antideuterons and 50$\%$ deuterons, for \mbox{\pt$>1.2$ GeV/$c$}, with a small centrality dependence, while that of (anti)$^{3}$He ranges between about 60$\%$ for \mbox{\pt$<2 $ GeV/$c$} to about 80$\%$ at larger \pt. The $^{3}\overline{\rm He}$ efficiency is a few percent smaller than that of $^{3}$He because of antimatter annihilation in the detector material, but still the difference between the $^{3}$He and $^{3}\overline{\rm He}$ efficiencies is smaller compared with the statistical precision of data. 

The fraction of primary deuterons is extracted in different centrality and \pt intervals from fits to the $\mathrm{DCA}_{xy}$ distributions with templates. Antideuterons from data are used as templates for primary deuterons, since measured antideuterons are all primary as there are no secondary antideuterons produced in the interactions in the material. The $\mathrm{DCA}_{xy}$ templates for secondary deuterons from spallation are taken from MC simulations. The $\mathrm{DCA}_{xy}$ distributions in data are filled using a high-purity sample of deuterons selected by requiring that the measured \dEdx and time-of-flight are within 3 standard deviations from their expected values. The fraction of primary deuterons is calculated by integrating the primary-deuteron template from the fit in the interval \mbox{$\mathrm{DCA}_{xy} < 0.1$} cm and dividing it by the number of deuterons in the same $\mathrm{DCA}_{xy}$ range. The fractions of primary deuterons extracted from the $\mathrm{DCA}_{xy}$ fits depend on centrality, with values ranging from 50$\%$ in central to almost 90$\%$ in peripheral collisions at low \pt. These fractions converge to 100$\%$ for \pt $>$ 1.6 GeV/$c$. The fraction of secondary $^{3}$He from spallation could not be extracted due to the very limited number of counts. For this reason, the  $^{3}$He spectrum is measured only for \pt $>2$ GeV/$c$ where the contribution from spallation is negligible.

\subsection{Elliptic-flow measurement}
\label{subsec:FlowMeasurement}

The azimuthal distribution of (anti)deuterons produced in the Xe--Xe collisions with respect to the \textit{n}-th order flow symmetry plane $\Psi_{n}$~\cite{Voloshin:2008dg,Ollitrault:2009ie,Alver:2010gr,Qiu:2011iv} can be expressed as a Fourier series 
 \begin{equation}
 E \frac{\mathrm{ d^3} N}{\mathrm{d} p^3} = \frac{1}{2\pi} \frac{\mathrm {d^2} N}{p_\mathrm{T} \mathrm{d} p_{\rm{T}} \mathrm{d} y } \left( 1 + \sum_{n=1}^{\infty} 2 v_n \cos \left( n \left( \varphi -  \Psi_{n} \right) \right) \right),
\end{equation}
where $E$ is the energy of the particle, $p$ the momentum, $\varphi$ the azimuthal angle, $y$ the rapidity, and  
\begin{equation}
v_n = \langle \cos \left( n(\varphi - \Psi_{n}) \right)\rangle.
\end{equation}
The second coefficient of the Fourier series ($v_2$) is called elliptic flow and is related to the initial geometrical spatial asymmetry of the matter created in heavy-ion collisions. 
The elliptic-flow coefficients are measured using the Scalar Product (SP) method~\cite{Adler:2002pu,Voloshin:2008dg}, as done in Refs.~\cite{Acharya:2017dmc,Acharya:2020lus}. This is a two-particle correlation technique based on the scalar product of the unit flow vector of the particle of interest, $k$, and the $Q$-vector. The unit flow vector is denoted by \textbf{u}$_{n,k} = \mathrm{exp}(in\varphi_k)$, where $\varphi_k$ is the azimuthal angle of the particle $k$. The $Q$-vector is computed from the azimuthal distribution of a set of charged particles (so called reference flow particles) as 
\begin{equation}
    \textbf{\textit{Q}}_n = \sum w_i e^{in\varphi_i} ,
\end{equation}
where $w_i$ is a weight applied to correct for non-uniform acceptance and efficiency of the detector (as done in Refs.~\cite{FlowChargedParticlesLHC, Acharya:2020lus, protonV2}), $n$ is the order of the harmonic, and $\varphi_i$ is the azimuthal angle for the $i$th reference flow particle.
The $v_n$ flow coefficients are calculated as
\begin{equation}
    v_n \{\mathrm{SP}\} = \frac{\left\langle \langle \textbf{u}_{n,k}{\textbf{\textit{Q}}^*_n} \rangle \right\rangle}{\sqrt{\frac{\langle \textbf{\textit{Q}}_n \textbf{\textit{Q}}^{A*}_n \rangle \langle \textbf{\textit{Q}}_n \textbf{\textit{Q}}^{B*}_n \rangle} {\langle \textbf{\textit{Q}}^A_n \textbf{\textit{Q}}^{B*}_n \rangle}}} .
\end{equation}
Single brackets $\langle ... \rangle$ denote an average over all events, while double brackets $\langle \langle ... \rangle \rangle$ indicate an average over all particles in all events, and ${}^*$ denotes the complex conjugate. The denominator is a correction factor that is introduced to take into account the resolution of the $\textbf{\textit{Q}}_n$ vector. In this analysis, the $\textbf{\textit{Q}}_n$ vector is calculated from the azimuthal distribution of the energy deposition measured in the V0A, while the $\textbf{\textit{Q}}_n^A$ and $\textbf{\textit{Q}}_n^B$ vectors (used to estimate the resolution of $\textbf{\textit{Q}}_n$) are determined from the azimuthal distribution of the energy deposited in the V0C and that of tracks reconstructed in the TPC, respectively. Using these detectors, a pseudorapidity gap $|\Delta \eta| > 2$ between the particle of interest and the reference flow particles is introduced. Such a pseudorapidity gap reduces non-flow effects, which are correlations not arising from the collective expansion of the system (e.g., resonance decays and jets). 

\begin{figure}[hbt]
    \begin{center}
    \includegraphics[width = 0.49\textwidth]{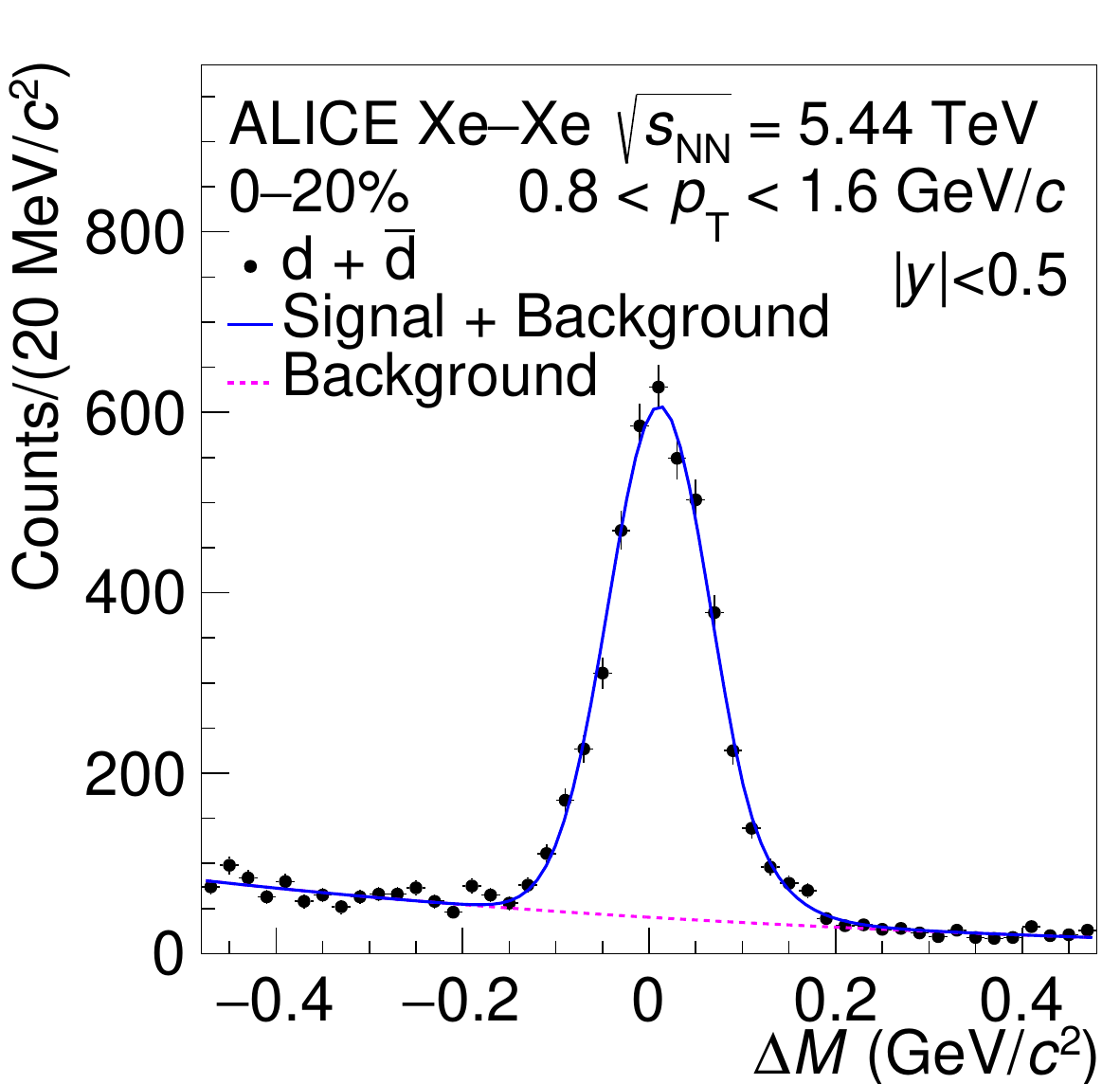}
    \includegraphics[width = 0.49\textwidth]{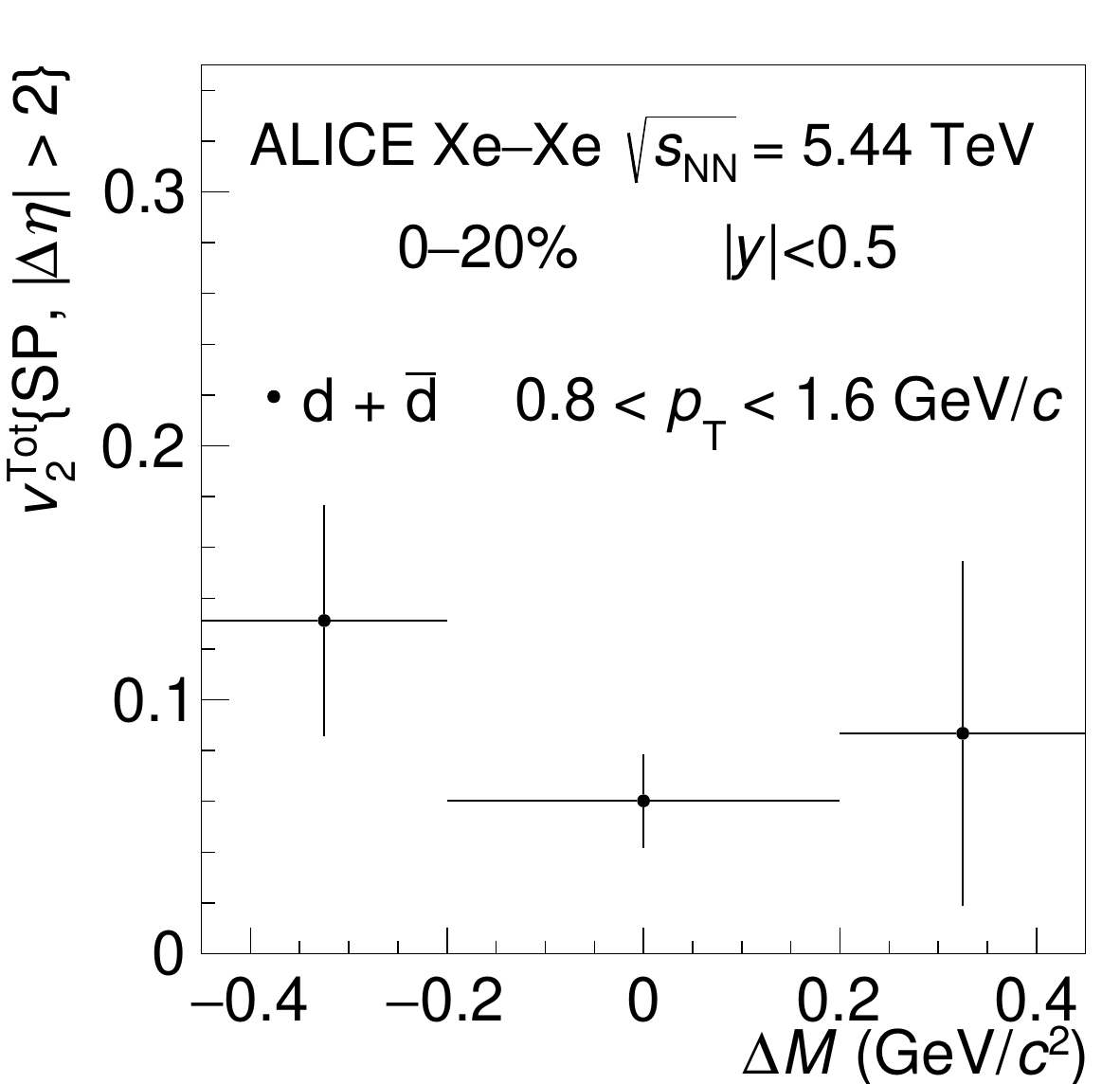}
    \end{center}
\caption{The $\Delta M$ distribution (left) and the extracted $v_2$ (right) of deuteron candidates as a function of $\Delta M$ for $0.8~<$~\pt~$<1.6$~\GeVc in the centrality interval 0--20\% in Xe--Xe collisions. The statistical uncertainties on the data points are represented by bars. The measurement is carried out at midrapidity, i.e., $\left|y\right|<0.5$. }
\label{fig:v2ExtractionMethod}        
\end{figure}

The $v_{2}$ coefficients of (anti)deuterons are calculated in three different ranges of $\Delta M = M-M_{\rm d}$, where $M$ is the particle mass obtained from the measured momentum and time-of-flight of each deuteron candidate track, and $M_{\rm d}$ is the deuteron mass taken from CODATA~\cite{CODATA}. The $v_{2}^{\rm Tot}$($\Delta M$) contains contributions from the signal ($v_2^{\mathrm{sig}}$) and from the background ($v_2^{\mathrm{bkg}}$)
\begin{equation} 
v_2^{\rm Tot} (\Delta M) = v_2^{\mathrm{sig}}  \frac{N^{\mathrm{sig}}}{N^{\mathrm{tot}}}(\Delta M) +  v_2^{\mathrm{bkg}} \frac{N^{\mathrm{bkg}}}{N^{\mathrm{tot}}}(\Delta M),
\label{eq:v2tot}
\end{equation}
where $N^{\rm{sig}}$ is the number of \mbox{deuterons}, $N^{\rm{bkg}}$ the number of background particles and $N^{\rm{tot}}$ is their sum. 
The $v_{2}^{\rm sig}$ is extracted from the value of $v_{2}^{\rm Tot}$ measured in the central interval of the $\Delta M$ distribution, in which $v_2^{\mathrm{bkg}}$ is evaluated using the average between the left and right $\Delta M$ intervals and propagating the statistical uncertainties accordingly. 
The $N^{\rm{sig}}$ and $N^{\rm{bkg}}$ are obtained from the fit to the $\Delta M$ distribution using a Gaussian with an exponential tail for the signal and an exponential for the background. 
The signal extraction procedure is illustrated in Fig.~\ref{fig:v2ExtractionMethod} for $0.8<p_{\rm T} <1.6$ GeV/$c$ in the centrality interval \mbox{0--20$\%$}. To avoid contamination from secondary deuterons produced by spallation in the detector material, for the flow measurement only antideuterons are used for $p_{\rm T} <1$ GeV/$c$. 

\section{Systematic uncertainties}
\label{sec:SystematicUncertainties}

The dominant sources of systematic uncertainties on the (anti)nucleus \pt spectra and elliptic flow are related to the event and track selections, ITS--TPC matching, particle identification, signal extraction, absorption of (anti)nuclei in the detector material, and material budget. These are summarized in Table~\ref{Table:syst_uncertainty_summary}, reporting the uncertainties in the lowermost and uppermost \pt intervals, as the systematic uncertainties evolve monotonically with \pt. The final systematic uncertainty assigned to the data points is obtained by summing in quadrature all individual contributions. The methods used to estimate the systematic uncertainties from these sources are illustrated in the following.

\begin{table}
\centering
\caption{Summary of the different contributions to the systematic uncertainties on the (anti)nucleus \pt-differential yields and (anti)deuteron elliptic flow, in the lowermost and uppermost \pt intervals. All values are in percentage. }
\begin{tabular}{ccccccc}
\toprule
 & \multicolumn{2}{c}{\rm{d ($\overline{\rm d}$)     yield}} & \multicolumn{2}{c}{\rm{d + $\overline{\rm d}$   elliptic flow}} & \multicolumn{2}{c}{\rm{ $^3$He ($^3\overline{\mathrm{He}}$) yield}} \\ 
\midrule
  \pt range (GeV/$c$)& \rm{0.8 -- 1.2} & \rm{4.0 -- 5.0} & \rm{0.8 -- 1.6} & \rm{ 4.0 -- 5.0}  & \rm{1.0 -- 2.0} & \rm{5.0 -- 7.0} \\
\midrule
  \rm{Source of uncertainty} & & & & & &\\
\midrule
Event selection & 1 (1) & 1 (1)  &  1 & 1  & 1 (1) & 1 (1) \\
Tracking and PID  	& 1 (2) & 1 (2) & 2  &  2 		& 1 (4) & 5 (6) \\
ITS--TPC matching & 2 (2) & 2 (2)  &  2 & 2 & 2 (2) & 2 (2) \\
Signal extraction & 1 (1) & 5 (5) & 2  &  7 & 	-- & -- \\
Absorption & 1 (6) & 0.6 (3) &  6   & 3 &  0.3 (2) & 0.3 (1)     \\
Material Budget	& 1 (1) & 1 (1)  &  1 & 1   & 1 (1) & 1 (1) \\
\bottomrule
\end{tabular}
\label{Table:syst_uncertainty_summary}
\end{table}

\subsection{Event selection uncertainty}
\label{subsec:EventSelectionUnc}

The systematic uncertainty from the event selection is estimated by varying the event selection criteria, i.e., the requirement on the $z$-coordinate of the primary vertex and the pile-up rejection. This contribution is found to be about 1$\%$ for both \pt spectra and elliptic-flow analyses.

\subsection{Track selection, PID and ITS--TPC matching efficiency uncertainty}
\label{subsec:TrackSelectionPIDUnc}

The systematic uncertainties related to track selection and particle identification are estimated by repeating the full analysis chain using 50 different settings with different selection criteria. These are chosen by sampling the analysis parameters randomly from uniform probability distributions. The standard deviation of the distribution of fully-corrected yields is taken as systematic uncertainty in each \pt and centrality interval. This method, based on the frequentist approach to explore the probability density function of measurements, takes into account possible correlations between the different analysis parameters. The contribution from ITS--TPC matching efficiency is estimated from charged-particle tracks by comparing the probabilities of prolonging a track from the TPC to the ITS in data and MC~\cite{Acharya:2020sfy}. This uncertainty is approximately 2$\%$ and is added in quadrature to the previous contributions.

\subsection{Signal extraction uncertainty}
\label{subsec:SignalExtractionUnc}

The signal extraction uncertainty is assigned only to (anti)deuteron \pt spectra and elliptic flow, while for $^3$He it is found to be negligible. Such contribution is estimated by repeating the fit to the TOF signal using different intervals. The default fit range is $-7 < (\Delta t-\Delta t_{\rm exp})/\sigma_{\rm TOF} < 15$, while for systematic uncertainties the limits are varied with a uniform random sampling in $[-10,-7]$ for the lower limit and in $[10,15]$ for the upper limit.
The standard deviation of the extracted yields is taken as a systematic uncertainty in each \pt and centrality interval.
Additionally, only for flow measurement, the $v_2^{\mathrm{bkg}}$
was evaluated by using only the left or the right $\Delta M$ interval for $v_2^{\mathrm{bkg}}$ in Eq.~\ref{eq:v2tot}.

\subsection{Absorption and material budget uncertainties}
\label{subsec:AbsorptionMaterialBudgetUnc}

The contribution due to the uncertainty on the measured inelastic cross section of (anti)nuclei with the detector material is estimated using MC productions with increased and decreased cross section according to the measured uncertainties~\cite{absorption1,absorption2,absorption3,absorption4,antideuteronInelCS,antiHe3Absorption}, as done in Ref.~\cite{3HePbPb}. The ALICE detector material is known with a precision of 4.5$\%$~\cite{ALICEperformance}. The systematic uncertainty on the (anti)nucleus yield related to the uncertainty in the material budget is taken from the measurement of (anti)nucleus production in Pb--Pb collisions at \snn = 5.02 TeV~\cite{ALICE:2022boh}.

\section{Results} 
\label{sec:Results}

\subsection{Transverse-momentum spectra}

\begin{figure}[!hbt]
\centering
\includegraphics[width=0.49\textwidth]{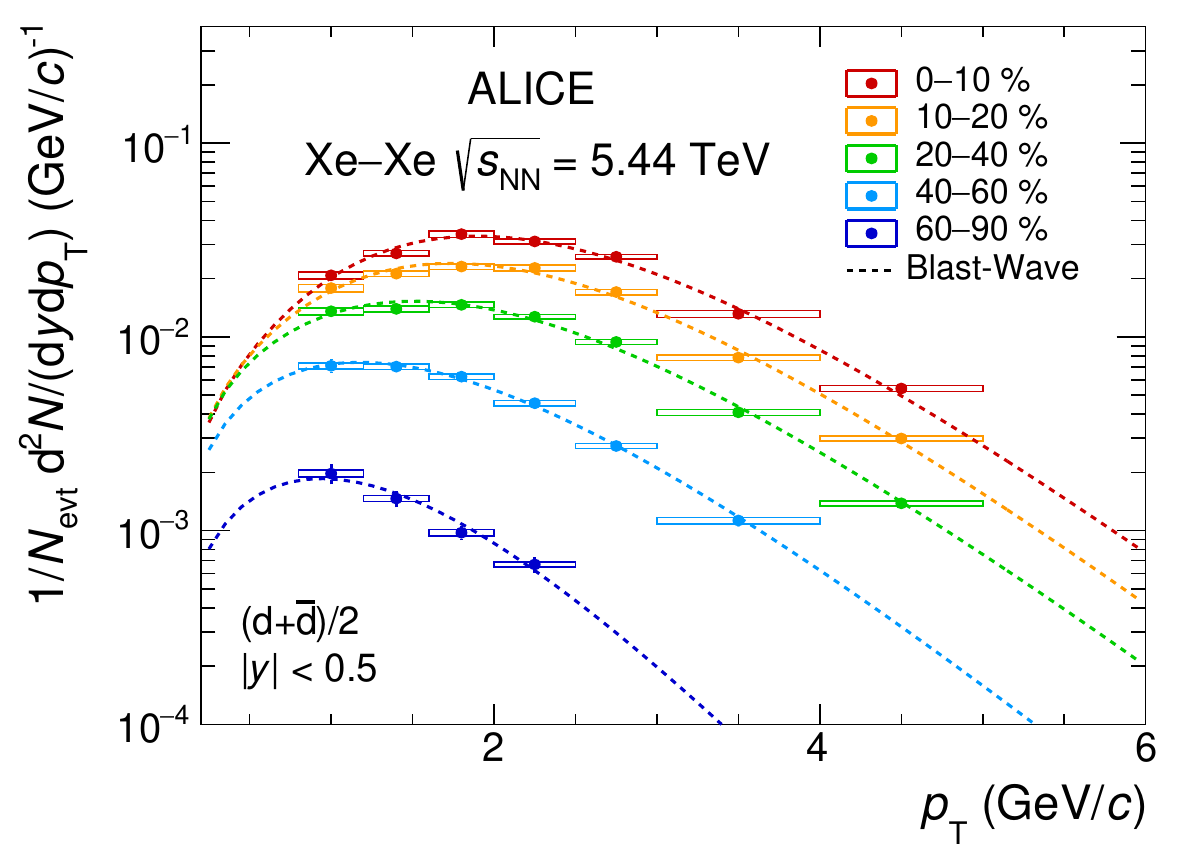}
\includegraphics[width=0.49\textwidth]{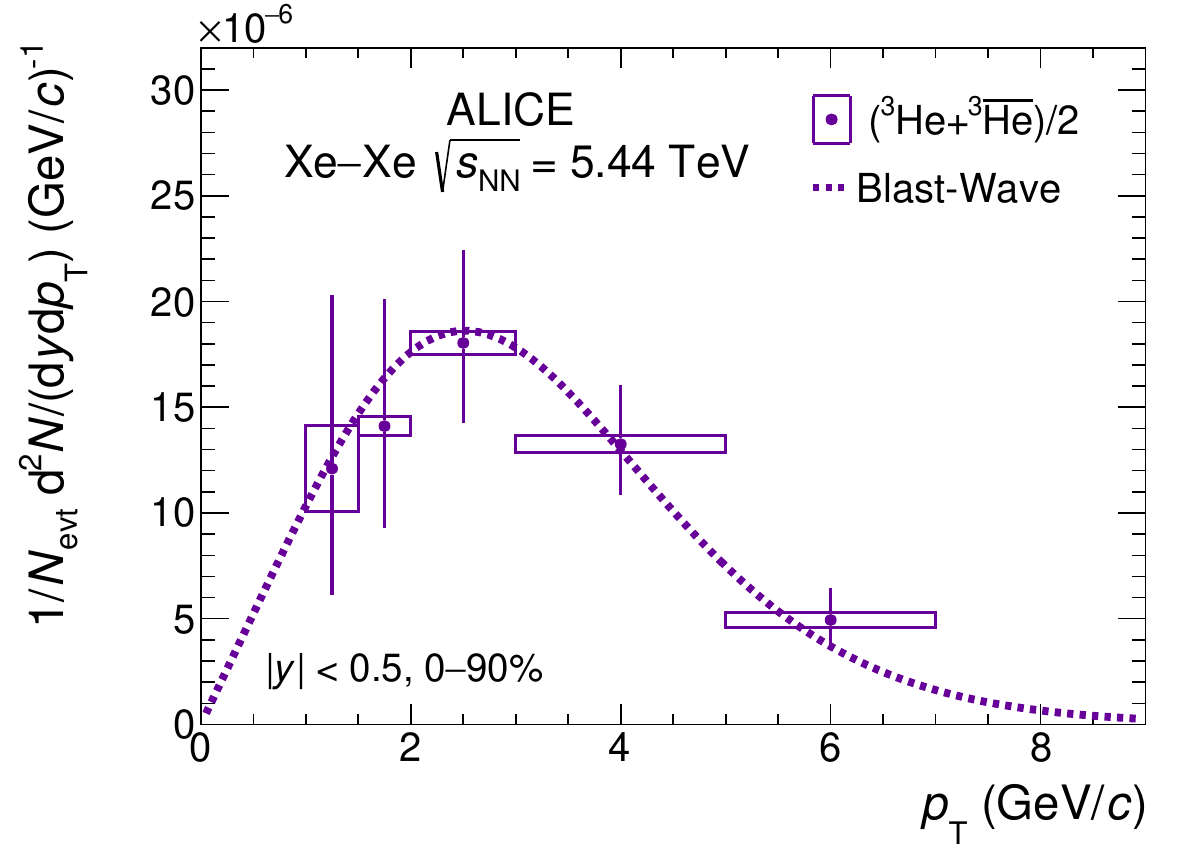}
\caption{Transverse-momentum differential yields of the average of deuterons and antideuterons in different centrality classes (left panel), and of the average of $^3\mathrm{He}$ and $^3\overline{\mathrm{He}}$ (right panel) in the centrality class (0--90\%). Dotted lines show the Blast-Wave individual fit to the \pt spectra. Statistical and systematic uncertainties are shown as vertical bars and boxes, respectively.}
\label{fig:deuteron_pTspectra}
\end{figure}

The transverse-momentum spectra of d and $\overline{\mathrm{d}}$, as well as those of $^3$He and $^3\overline{\mathrm{He}}$ are found to be consistent within the uncertainties in all centrality classes, as expected in the case of vanishing baryochemical potential at midrapidity at LHC energies. Hence, the average \pt-differential yields of deuterons and antideuterons, and of $^3$He and $^3\overline{\mathrm{He}}$, are calculated, as presented in Fig.~\ref{fig:deuteron_pTspectra}. 
The slope of the transverse-momentum distributions flattens when moving from peripheral to central collisions, i.e., with increasing event multiplicity, as it was already observed in p--Pb collisions at \snn = 5.02 TeV~\cite{Acharya:2019rys} and at \snn = 8.16 TeV~\cite{ALICE:2022weh}, and in Pb--Pb collisions at \snn = 2.76 TeV~\cite{3HePbPb,Acharya:2017dmc} and at \snn = 5.02 TeV~\cite{ALICE:2022boh}. 

To extrapolate the yields in the unmeasured \pt intervals, a Blast-Wave function~\cite{BlastWave1} is fitted to the \pt distributions of (anti)d and (anti)$^3$He. For the case of the $^3$He, the parameters of the Blast-wave fit function except for the normalization are constrained to those of the deuteron~\cite{ALICE:2023qyl}. The resulting \pt-integrated yields (d$N$/d$y$) of (anti)deuterons and (anti)$^{3}$He are calculated summing the two contributions obtained by integrating the data points in the measured \pt region and by integrating the fit function in the unmeasured one. The difference between the integral of the fit function and the integral of the data points is accounted for in the systematic uncertainty of the final yields. The resulting d$N$/d$y$ are summarized in Table~\ref{tab:YieldDeuteron}. The fraction of yield obtained from the extrapolation of the fit function varies between 8\% and 35\% from central to peripheral collisions for (anti)d, and $\approx9\%$ for (anti)$^3$He. 

The systematic uncertainties on the integrated yields were obtained extrapolating the uncertainties associated with the \pt spectra, treating differently the \pt-correlated and -uncorrelated systematic uncertainties. 
The uncertainties on material budget, event selection, signal extraction, ITS--TPC matching efficiency, and absorption are found to be highly correlated in \pt and are considered as fully \pt correlated, whereas the uncertainties due to track selection and PID are found to be mostly uncorrelated with transverse momentum. 
For the computation of the systematic uncertainties on the integrated yields three contributions have been considered, as done in previous similar analyses~\cite{Acharya:2019rys, ALICE:2022weh}: (i) the \pt-uncorrelated contribution was evaluated with the Gaussian sampling method; (ii) the correlated part of the systematic uncertainties of the \pt spectra was used to shift accordingly the data points up and down with respect to the mean value, allowing the refit of the \pt spectra; (iii) the uncertainty on the extrapolation was evaluated fitting the spectra with several alternative functions (i.e., Boltzmann~\cite{BHALLA1981446}, $m_{\mathrm T}$-exponential~\cite{Adler:2003cb}, and Lévy-Tsallis~\cite{Tsallis:1987eu}) and taking into account the spread with respect to the Blast-Wave results. In case (i), the systematic uncertainty was taken as the RMS of the distribution of the varied integrated yield values, whereas in cases (ii) and (iii) half of the difference between the maximum and the minimum yields was considered as uncertainty. The final uncertainty was obtained as the sum in quadrature of the different contributions. 

\begin{table}[!hbt]
\centering
\caption{Integrated yields (d$N$/d$y$) of the average of deuterons and antideuterons and of the average of $^{3}$He and $^{3}\overline{\rm He}$, for each centrality class measured at midrapidity ($|y|<0.5$). The values of the average charged-particle multiplicity density ($\langle$d$N_{\rm ch}$/d$\eta$$\rangle \big|_{|\eta|<0.5}$) are taken from Ref.~\cite{ALICE:2018cpu}. The first uncertainty on d$N$/d$y$ is statistical and the second is systematic. }
\begin{tabular}{cccc}
\toprule
Centrality class & $\langle$d$N_{\rm ch}$/d$\eta$$\rangle \big|_{|\eta|<0.5}$ & d$N$/d$y$ [(d+$\overline{\mathrm d}$)/2]  & d$N$/d$y$ [($^{3}$He+$^{3}\overline{\mathrm{He}}$)/2]\\
\midrule
0--10$\%$     & 1053 $\pm$ 25  &  (8.72 $\pm$ 0.17 $\pm$ 0.52) $\times$ $10^{-2}$ & \\
10--20$\%$   & 706 $\pm$ 17 &  (6.19 $\pm$ 0.12 $\pm$ 0.37) $\times$ $10^{-2}$ & \\
20--40$\%$   & 397 $\pm$ 9 &  (3.86 $\pm$ 0.08 $\pm$ 0.23) $\times$ $10^{-2}$ & \\
40--60$\%$ & 158 $\pm$ 4  &  (1.64 $\pm$ 0.03 $\pm$ 0.10) $\times$ $10^{-2}$ & \\
60--90$\%$ & 40 $\pm$ 1  &  (3.25 $\pm$ 0.06 $\pm$ 0.19) $\times$ $10^{-3}$ & \\
0--90$\%$    & 333 $\pm$ 5 &  (2.99 $\pm$ 0.06 $\pm$ 0.19) $\times$ $10^{-2}$ & (7.45 $\pm$ 0.94  $\pm$ 0.67 )$\times  10^{-5}$\\
\bottomrule
\end{tabular}
\label{tab:YieldDeuteron}
\end{table}

\subsection{Ratio to proton and pion yields}
\begin{figure}[!hbtp]
\centering
\begin{minipage}[b]{0.85\textwidth}
	\includegraphics[width=\textwidth]{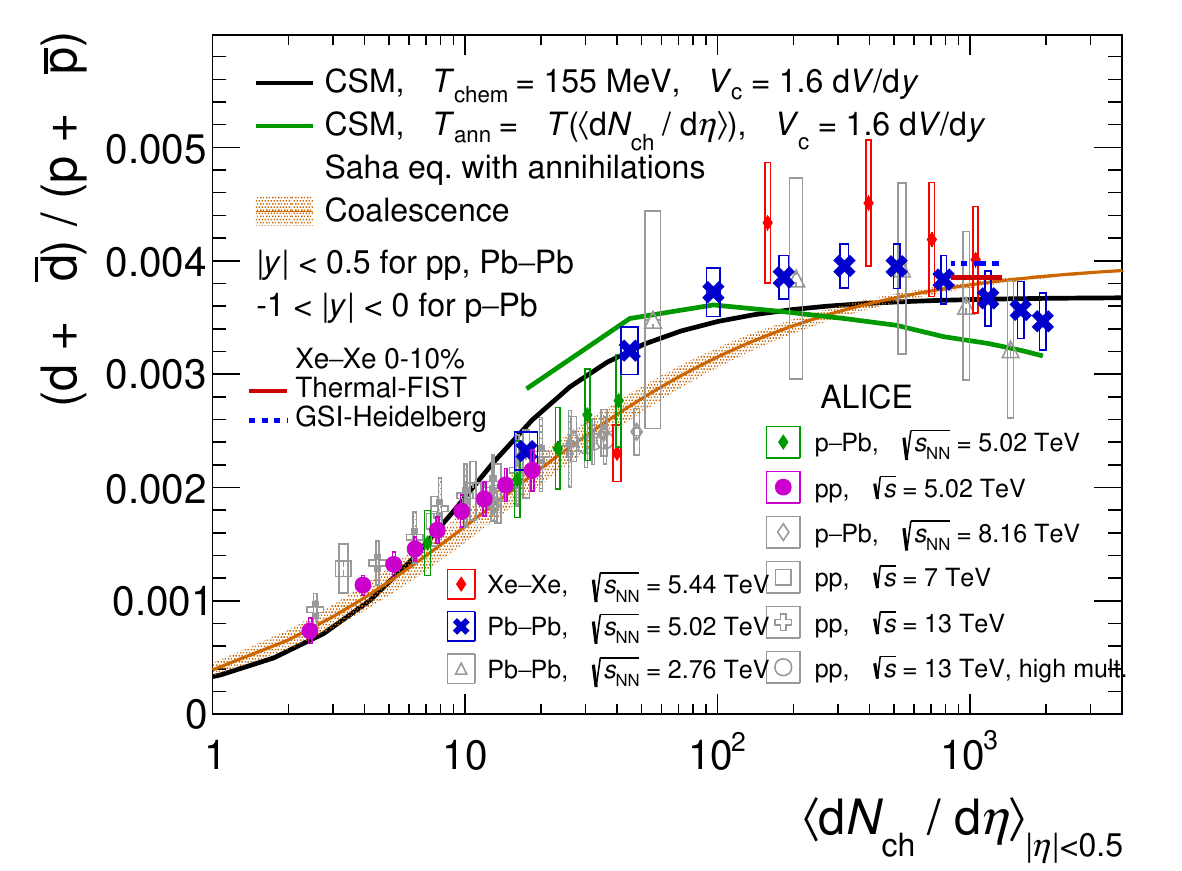}
\end{minipage}
  \hfill
\begin{minipage}[b]{0.85\textwidth}
    \includegraphics[width=\textwidth]{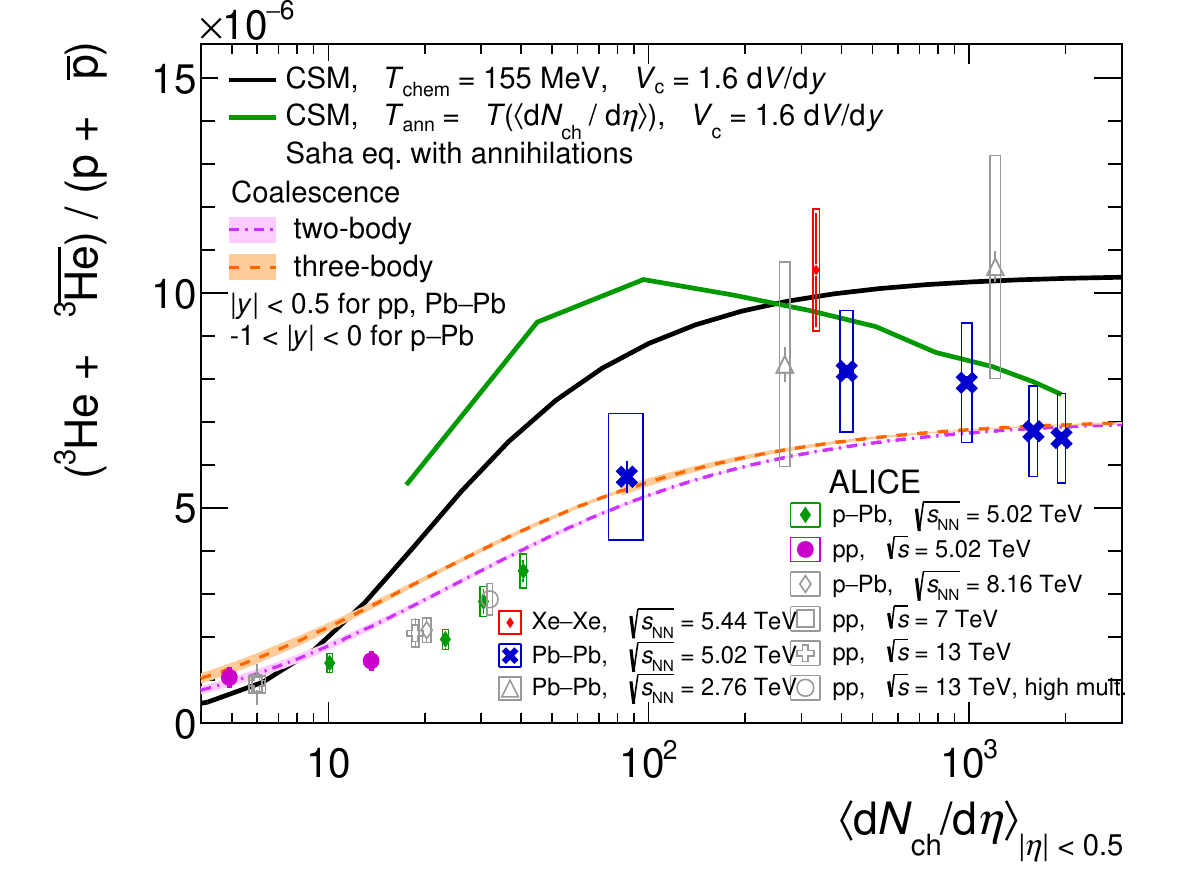}
\end{minipage}   
\caption{Deuteron-to-proton (top panel) and $^3$He-to-proton (bottom panel) yield ratios as a function of the average charged-particle multiplicity density. All yields are measured at midrapidity, i.e., $\left|y\right|<$ 0.5 for pp, Xe--Xe and Pb--Pb collisions, and $-1<\left|y\right|<0$ for p--Pb collisions. Statistical uncertainties are shown as vertical lines, whereas systematic ones are shown as boxes. The coalescence predictions are shown as bands, whose width represents the model uncertainty. The black and green lines correspond to the expectations of two different hypotheses of the CSM. }
\label{fig:NucleiOverP}
\end{figure}

The consistency of models describing the hadron chemistry in high-energy hadronic collisions can be tested by comparing the model predictions with the measured yields for different hadron species or, alternatively, with yield ratios of different hadrons. The ratios of the measured yields of nuclei and those of protons are sensitive to the production mechanism as they remove the dependence on the volume of the system while keeping the dependence on the freeze-out temperature. 
In Fig.~\ref{fig:NucleiOverP} the deuteron-to-proton (top panel) and $^3$He-to-proton (bottom panel) yield ratios measured in all available collision systems at the LHC as a function of the average charged-particle pseudorapidity density $\langle$d$N_{\mathrm{ch}}/$d$\eta\rangle$~\cite{deuteron_pp7TeV, Acharya:2019rys, 3HePbPb, Acharya:2017fvb, nuclei_pp_13TeV, ALICE:2022weh} are compared with the predictions from the statistical hadronization model with canonical ensemble (CSM) and the coalescence one. The results from all available collision systems at \snn$=$ 5.02 and 5.44 TeV are highlighted in color and show a common and smooth increasing trend with increasing $\langle$d$N_{\mathrm{ch}}/$d$\eta\rangle$, in both d/p and $^3$He/p yield ratio cases. 
The results at \snn$=$ 5.02 and 5.44 TeV follow the same trend as the results at different collision energies (shown in gray in Fig.~\ref{fig:NucleiOverP}). 

For the coalescence predictions, the probability of forming an (anti)nucleus is given by the overlap of the phase-space distributions of the constituent nucleons in the emission source with the Wigner density of the bound state. The latter is calculated approximating the (anti)nucleus internal wave function with a Gaussian function~\cite{coalescenceSmallSystems}. From the studies shown in Ref.~\cite{Mahlein:2023fmx}, up to a factor 50\% difference in the coalescence predictions for the momentum distribution of deuterons can be expected using a Gaussian wave function instead of a more realistic one, such as Argonne $v_{18}$. 
For $^3$He, the coalescence process can happen in two ways, either as a two-step process in which the $^{3}$He is formed by the coalescence of a deuteron and a proton (two-body coalescence), or as a three-body coalescence process in which three nucleons combine to form an $^{3}$He nucleus~\cite{coalescenceSmallSystems}. 
In the case of the $^3$He/p, the measured ratio in Xe--Xe collisions is higher than the prediction of the coalescence models by about $2\sigma$. Notably, the calculations with two- and three-body coalescence yield similar results for the $^3$He/p ratios at the multiplicity of Xe--Xe collisions. The difference between the two predictions is larger at lower multiplicity, i.e., in the range covered by pp and p--Pb collisions. In this multiplicity region, the two-body coalescence overshoots the data by $\approx3\sigma$, with the exception of the 20--40\% centrality class of p--Pb collisions at \snn = 5.02 TeV (corresponding to $\langle$d$N_{\mathrm{ch}}/$d$\eta\rangle \approx$ 23), that is 5$\sigma$ away from the model. The three-body coalescence curve is systematically overshooting the data in the low and intermediate multiplicity region, and the agreement with the measurements worsens up to 7$\sigma$ for the 20--40\% centrality class of p--Pb collisions at \mbox{\snn = 5.02 TeV}. None of the coalescence curves predicts the decreasing trend of data at very high multiplicity ($\langle$d$N_{\mathrm{ch}}/$d$\eta\rangle> 300$), for neither the deuteron-to-proton nor the $^3$He-to-proton yield ratios. 

In the CSM, exact conservation of baryon number (B), electric charge (Q), and strangeness (S) are required in a specified volume known as correlation volume $V_{\mathrm{c}}$. Two versions of the CSM, both using $V_{\mathrm{c}}=$ 1.6 d$V$/d$y$, are shown in comparison with data: (i) the first one assumes the system to be in full chemical equilibrium and the chemical freeze-out temperature to be independent of multiplicity \mbox{($T_{\rm{chem}}$ = 155 MeV)}~\cite{Vovchenko:2018fiy}, (ii) the second uses the Saha equation approach~\cite{Vovchenko:2019aoz} to compute light-nucleus abundances in the hadronic phase, and considers the annihilation freeze-out temperature to be dependent on $\langle$d$N_{\mathrm{ch}}/$d$\eta\rangle$, evolving from 160 MeV for multiplicities of about 20 to \mbox{132 MeV} for multiplicity of about 2000~\cite{Vovchenko:2022xil}. For these models, the extension of the correlation volume was investigated through the event-by-event antideuteron number fluctuation measurement~\cite{deuteronFluctuations}. In such study it was found out that the correlation volume needed to describe the net-deuteron number fluctuations in central Pb--Pb collisions is $V_{\mathrm{c}}$ = (1.6 $\pm$ 0.3) d$V$/d$y$, well smaller than the one needed to describe the net-proton number fluctuations in the same collision system, $V_{\mathrm{c}}$ = (3--5) d$V$/d$y$~\cite{ALICE:2019nbs,Vovchenko:2020kwg,ALICE:2022xpf,vanillaCSM}. 
In the case of the deuteron-to-proton yield ratios, the expectations of CSM (i) show good agreement with data in the full multiplicity range, with some tension in the intermediate multiplicity, where the most peripheral Xe--Xe data point is lower than the model by $\approx3.5\sigma$. In the multiplicity region covered by heavy-ion collisions ($\langle$d$N_{\mathrm{ch}}/$d$\eta\rangle>10^2$), the predictions from CSM (i) show a plateau that matches the predictions of the grand-canonical statistical model~\cite{SHM1, CoalescenceTheory, Coalescence2}, while data hint towards a decreasing trend for the highest multiplicities, as observed in the case of the p/$\pi$ ratio~\cite{ALICE:2021lsv}. Such a trend is qualitatively reproduced by CSM (ii), which assumes that such suppression is entirely due to baryon annihilation in the hadronic phase~\cite{Vovchenko:2022xil}. In summary, both the coalescence and the CSM (i) predictions describe the absolute values of the yield ratios within $\approx25\%$, and both reproduce the overall smoothly increasing trend of the d/p ratio with increasing multiplicities, up to $\langle$d$N_{\mathrm{ch}}/$d$\eta\rangle \approx $ 300.
The CSM (ii) model reproduces the decreasing trend of data at large $\langle$d$N_{\mathrm{ch}}/$d$\eta\rangle$ values ($>$300), but is further from data as compared with the other two models in the remaining multiplicity range ($\langle$d$N_{\mathrm{ch}}/$d$\eta\rangle<$300).

In the case of the $^3$He/p yield ratios, at high multiplicities ($10^2<\langle$d$N_{\mathrm{ch}}/$d$\eta\rangle<10^3$), the result from Xe--Xe collisions is in agreement within 1$\sigma$ with both the CSM predictions. In the low multiplicity range ($\langle$d$N_{\mathrm{ch}}/$d$\eta\rangle<$ 10), the CSM (i) hypothesis is in agreement with the data within $\approx1.5\sigma$, while in the intermediate multiplicity range it is excluded by up to $13\sigma$. The CSM (ii) captures the decreasing trend of the data at high multiplicity, with an agreement within 1$\sigma$ for $\langle$d$N_{\mathrm{ch}}/$d$\eta\rangle>$ 100, although it is further from the data as compared to CSM (i) and coalescence for $\langle$d$N_{\mathrm{ch}}/$d$\eta\rangle<$ 100. Overall, the coalescence models reproduce the yield ratios within $\approx60\%$ and the general increasing trend of the data points, although significant discrepancies remain. 

\begin{figure}[!hbt]
\centering
\begin{minipage}[b]{0.85\textwidth}
	\includegraphics[width=\textwidth]{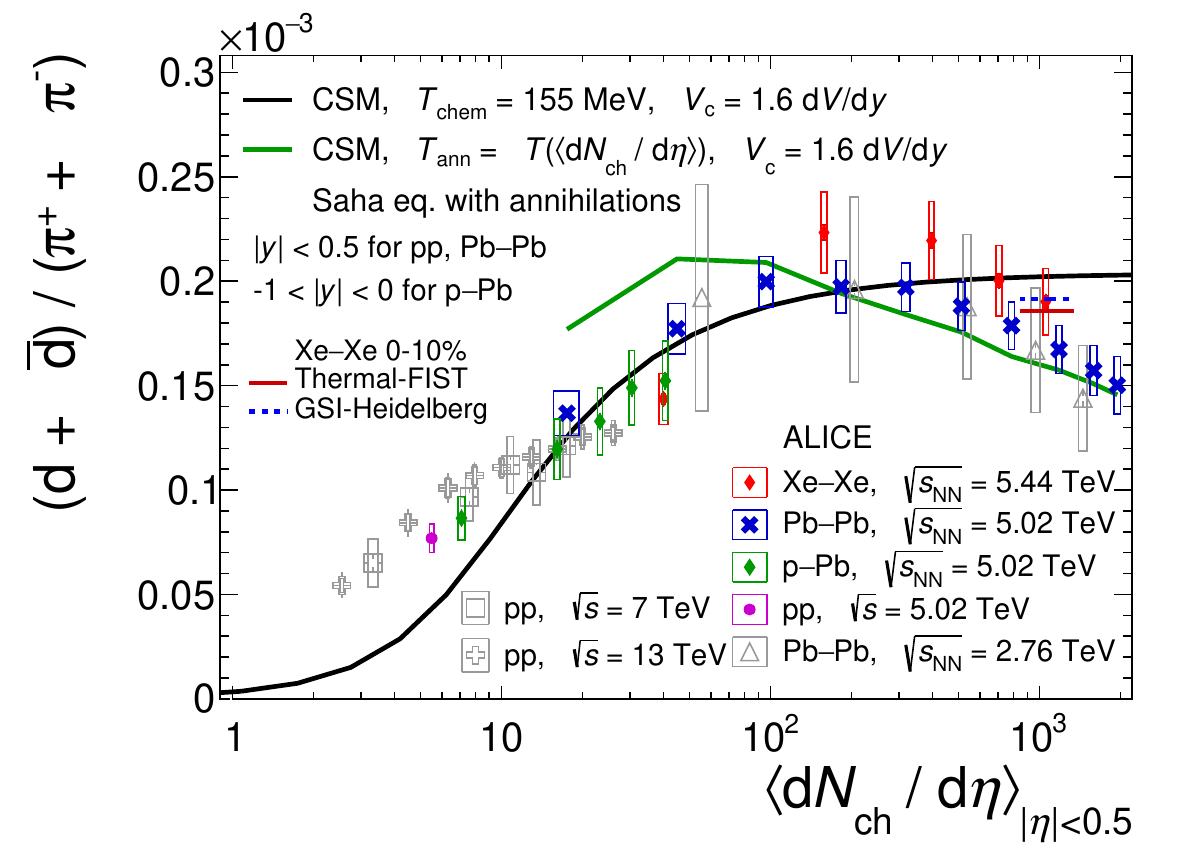}
\end{minipage}
  \hfill
\begin{minipage}[b]{0.85\textwidth}
    \includegraphics[width=\textwidth]{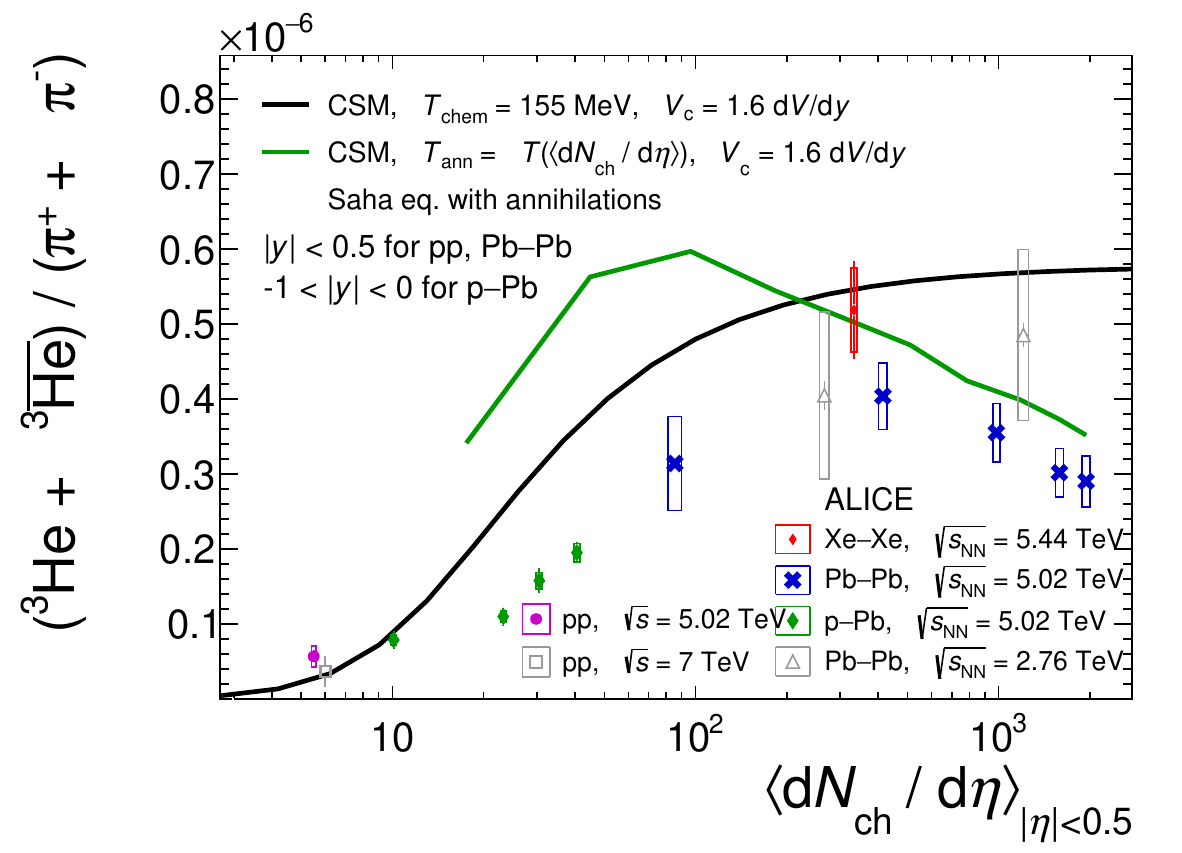}
\end{minipage}   
\caption{Deuteron-to-pion (top panel), and $^3$He-to-pion (bottom panel) yield ratios as a function of the charged-particle multiplicity density. All yields are measured at midrapidity, i.e., $\left|y\right|<$ 0.5 for pp, Xe--Xe and Pb--Pb collisions, and $-1<\left|y\right|<0$ for p--Pb collisions. Lines represent statistical uncertainties, whereas boxes show systematic ones. The solid lines correspond to the expectations of two implementations of the CSM, see text for details.}
\label{fig:NucleiOverPions}
\end{figure}

The decreasing trend of the particle ratios, for increasing multiplicity and $\langle$d$N_{\mathrm{ch}}/$d$\eta\rangle>$ 100, is even more pronounced when looking at the nucleus-to-pion yield ratios, as shown in Fig.~\ref{fig:NucleiOverPions}. Also in this case, the colored markers show the results from all available collision systems at \snn$=$ 5.02 and 5.44 TeV, while the results at different energies are shown in gray. All results follow a common trend with increasing $\langle$d$N_{\mathrm{ch}}/$d$\eta\rangle$. The data are compared with the predictions from the CSM models (i) and (ii). 
For deuterons, the measured ratios are described only qualitatively by the CSM (i) at low multiplicities ($\langle$d$N_{\mathrm{ch}}/$d$\eta\rangle<$ 10), while the model (i) predictions are consistent with the data at intermediate and high multiplicities, within $\approx1\sigma$ and 3$\sigma$, respectively, being the most different for the most central Pb--Pb collisions at \mbox{\snn = 5.02 TeV}. The predictions from CSM (ii) capture the trend of the data at high multiplicity and agree with the Pb--Pb experimental data within 1$\sigma$ and with the Xe--Xe results within 1.6$\sigma$, for $\langle$d$N_{\mathrm{ch}}/$d$\eta\rangle>$ 100, while overshooting the data points for $\langle$d$N_{\mathrm{ch}}/$d$\eta\rangle<$ 100.
For nuclei with $A$ = 3, the ratio to pion yields shows discrepancies with CSM (i) of up to $\approx17\sigma$ in the intermediate multiplicity range and up to $\approx8\sigma$ with the most central Pb--Pb data at \snn = 5.02 TeV. The CSM (ii) predictions agree within 2$\sigma$ with the data for $\langle$d$N_{\mathrm{ch}}/$d$\eta\rangle>$ 100, but fail in reproducing the data at $\langle$d$N_{\mathrm{ch}}/$d$\eta\rangle<$ 100. 

The decrease of the yield ratio at high average charged-particle multiplicity values is larger in Fig.~\ref{fig:NucleiOverPions} with respect to the corresponding ratios in Fig.~\ref{fig:NucleiOverP}. This is due to the decreasing trend of both the deuteron-to-pion and proton-to-pion yield ratios, leading to the partial cancellation of this trend in the deuteron-to-proton yield ratio. Among all presented models only the CSM (ii) captures the decreasing trend of the nucleus-to-proton and nucleus-to-pion yield ratios as well as their absolute magnitude.  

Finally, combining the integrated yields of $^3$He and deuterons with the yield of protons measured in Ref.~\cite{ALICE:2021lsv}, one can obtain the ratio $(N_{^3\rm{He}}\times N_{\rm{p}})/N_{\rm{d}}^2 = 0.589 \pm 0.134$ for the multiplicity integrated class 0--90\% (corresponding to $\langle$d$N_{\rm ch}$/d$\eta$$\rangle =333\pm5$), where the uncertainty is obtained by propagating the sum in quadrature of the statistical and systematic uncertainties on the single yields. This ratio in Xe--Xe collisions is consistent within 1$\sigma$ with the ratio of $(N_{\rm{t}}\times N_{\rm{p}})/N_{\rm{d}}^2$ measured by ALICE in Pb--Pb collisions at \mbox{\snn$=$ 5.02 TeV and 2.76 TeV}, corresponding to a multiplicity range spanning from about 85 to about 1800. This ratio is also compatible within 1$\sigma$ with the ratio measured by STAR in Au--Au collisions at similar multiplicity (of about 300), and within 2$\sigma$ with the rest of the ratios measured at lower energies (\snn = 7.7--200 GeV) shown in Ref.~\cite{STAR:2022hbp}. Notably, the ratio in Xe--Xe collisions makes use of the yield of $^3$He instead of that of triton, as triton was not measured in such a collision system. However, the yields of triton and $^3$He have been measured in Pb--Pb collisions at \mbox{\snn$= 5.02$ TeV} and found to be consistent within uncertainties~\cite{ALICE:2022boh}. Hence, such a ratio is assumed to be qualitatively valid for comparing the results from different energies and collision systems. Such a ratio is a powerful tool to test the production models, as argued in Ref.~\cite{STAR:2022hbp}, since its trend as a function of multiplicity and its overall value are expected to be different in SHM and coalescence predictions.
    
\subsection{Coalescence parameters}
In the coalescence models the key observable is the coalescence parameter $B_A$, that quantifies the probability for $A$ nucleons to bind together forming a nucleus of mass number $A$. This parameter can be obtained experimentally through the ratio of the invariant yield of the nucleus with mass number $A$ and that of protons. The former is expressed as a function of the \pt of the nucleus, denoted as $p_{\mathrm T}^{A}$, while the latter is evaluated at the transverse momentum denoted as $p_{\mathrm T}^{\rm p}$, with $p_{\mathrm T}^{\mathrm p}$ = $p_{\mathrm T}^{A}$/$A$. The coalescence parameter is obtained as follows

\begin{equation}
B_{A} = { \biggl( \dfrac{1}{2 \pi p^{A}_{\mathrm T}} \biggl( \dfrac{ \rm{d}^2\it{N}} {\rm{d}\it{y}\rm{d}\it{p}_{\mathrm T}} \biggr)_{A}  \biggr)}  \bigg/{  \biggl( \dfrac{1}{2 \pi p^{\rm p}_{\mathrm T}} \biggl( \dfrac{\rm{d}^2\it{N}}{\rm{d}\it{y}\rm{d}\it{p}_{\mathrm T}}  \biggr)_{\rm p}  \biggr)^A}  \, ,
\label{eq:BA}
\end{equation}

assuming that protons and neutrons have the same \pt distributions, as they belong to the same isospin doublet. 
The coalescence parameter is related to the production probability of the nucleus via this process and can be theoretically evaluated from the overlap of the nucleus wave function and the phase space distribution of the constituents via the Wigner function formalism~\cite{Coalescence2}. This formalism allows one to account for the source size and the quantum-mechanical properties of the nucleons in the calculation of $B_A$, using a realistic wave function for the nucleus. 

\begin{figure}[!hbt]
\centering
\begin{minipage}[b]{0.49\textwidth}
    \includegraphics[width=\textwidth]{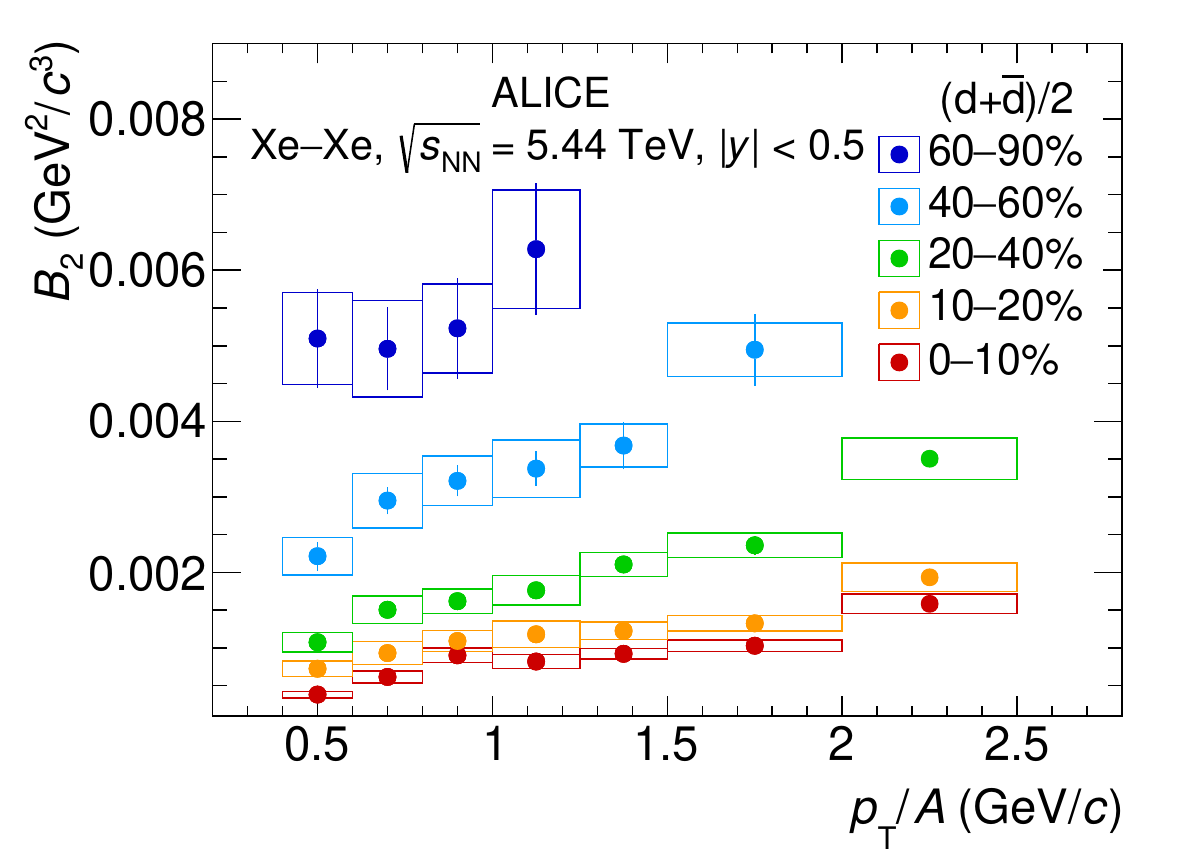}
\end{minipage}   
\hfill
\begin{minipage}[b]{0.49\textwidth}
	\includegraphics[width=\textwidth]{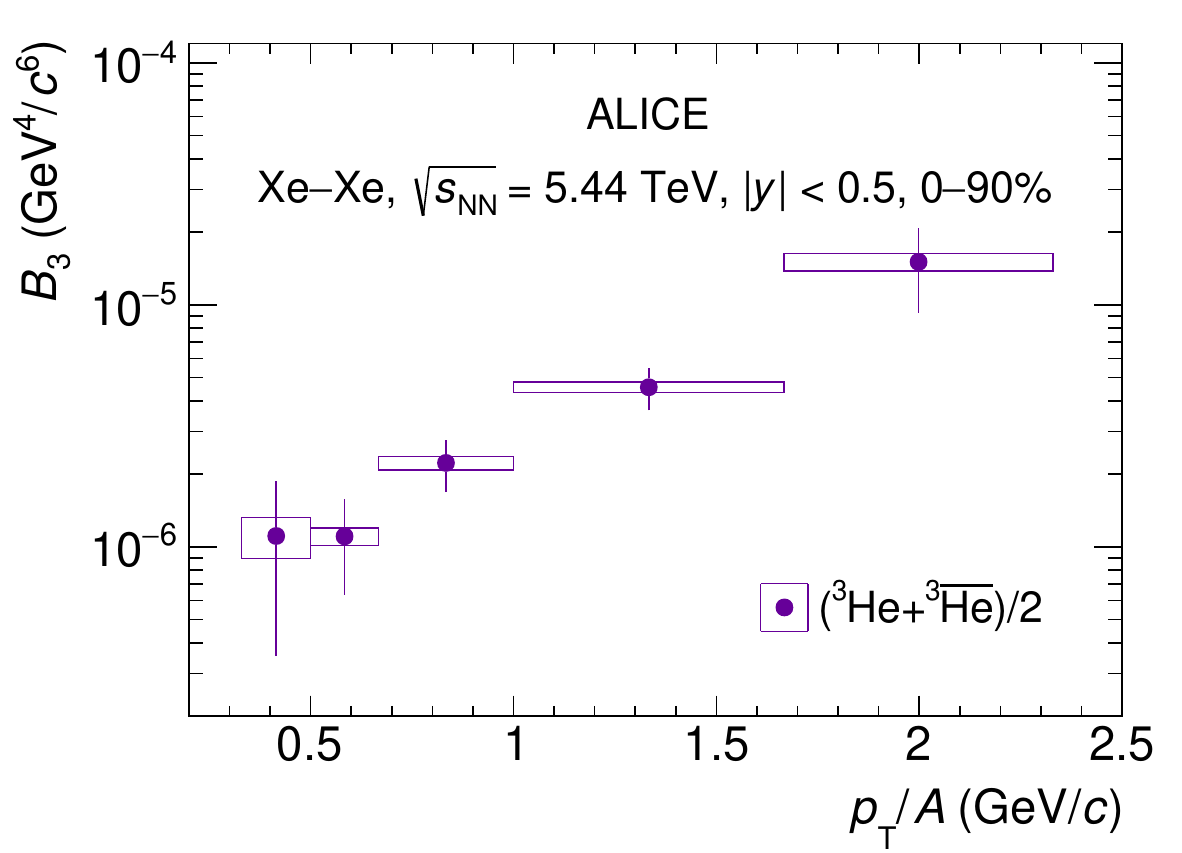}
\end{minipage}
\caption{Coalescence parameters $B_2$ (left panel) and $B_3$ (right panel) as a function of $p_{\mathrm T}$/$A$, measured for deuterons and $^3$He, respectively. Statistical uncertainties are represented as vertical lines whereas boxes represent the systematic ones. }
\label{fig:CoalescenceParameters}
\end{figure}

In Fig.~\ref{fig:CoalescenceParameters} the coalescence parameters $B_2$ and $B_3$ are shown as a function of the transverse momentum per nucleon (\pt/$A$), also for different centrality classes in the case of $B_2$. The transverse-momentum spectra of protons used for the calculation of $B_A$ are taken from Ref.~\cite{ALICE:2021lsv}. 
The coalescence parameters increase with increasing \pt/$A$ in all centrality classes. This trend was already observed in previous measurements from ALICE, in both heavy-ion and pp collisions~\cite{ALICE:2022boh, 3HePbPb, deuteron_pp7TeV, Acharya:2019rys, Acharya:2019xmu, Acharya:2020sfy, nuclei_pp_13TeV, nuclei_pp_5TeV, ALICE:2022weh}. Such behavior is not expected by simple coalescence models that neglect the spatial extension of the baryon-emitting source. However, it is well reproduced by state-of-the-art coalescence models that use the Wigner function formalism and properly take into account the source size~\cite{Mahlein:2023fmx}. Hence, the rising trend of the coalescence parameters with \pt/$A$ reflects the decreasing trend of the source size with $p_{\rm{T}}$~\cite{ALICE:2020ibs}: smaller source sizes (at higher \pt) correspond to larger coalescence parameters.

\subsection{Elliptic flow of deuterons}

\begin{figure}[hbt]
    \begin{center}
    \includegraphics[width = 0.9\textwidth]{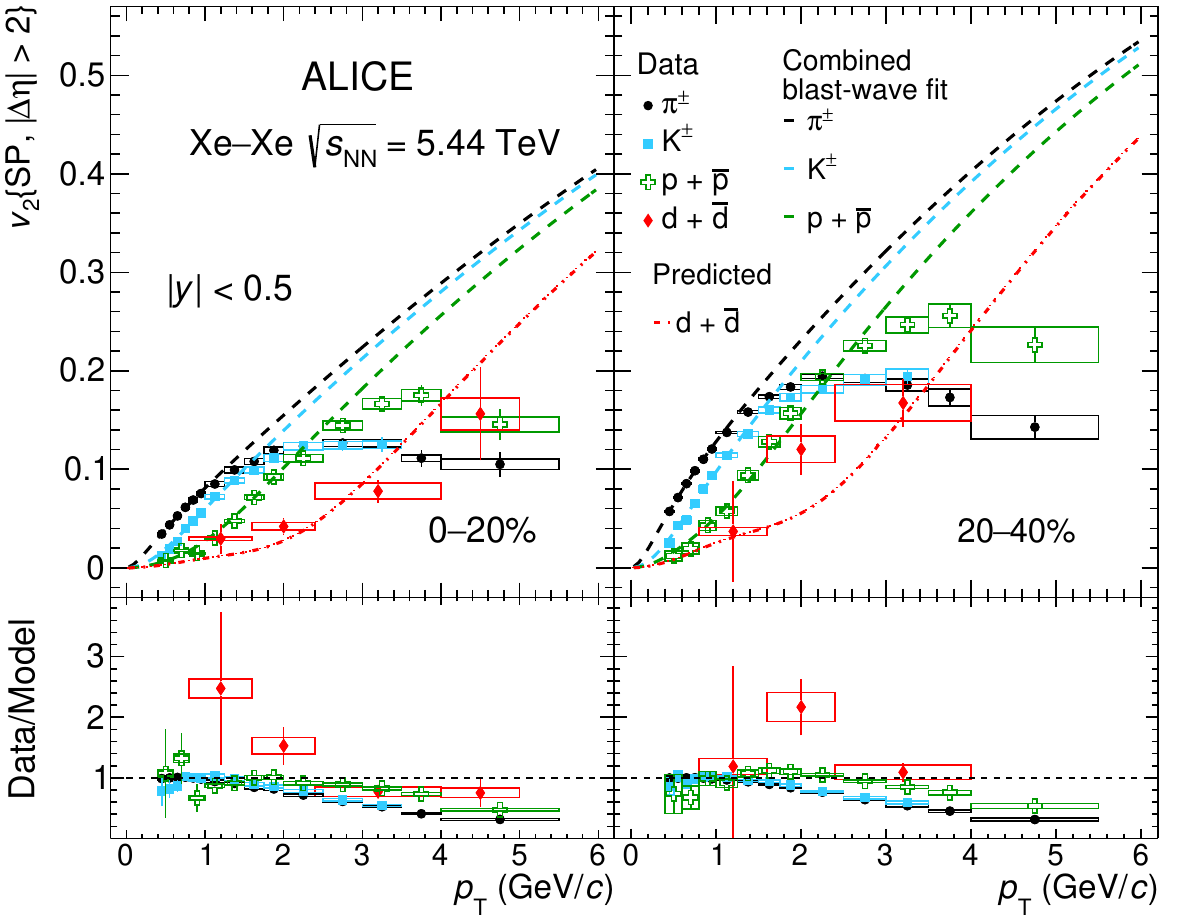}
    \end{center}
\caption{(Anti)deuteron $v_2$ measured at $|y| < 0.5$ in the centrality classes 0--20$\%$ (left) and 20--40$\%$ (right) in comparison with the Blast-Wave predictions obtained from simultaneous fits of the \pt spectra and $v_2$ of pions, kaons, and protons~\cite{flowIDhadronsXeXe}. Statistical and systematic uncertainties are represented by bars and boxes, respectively. }
\label{fig:dataVsBW}        
\end{figure}

The deuteron elliptic-flow coefficients measured in the centrality classes 0--20$\%$ and 20--40$\%$ are shown in Fig.~\ref{fig:dataVsBW} as a function of \pt in comparison with the $v_2$ of pions, kaons, and protons taken from Ref.~\cite{flowIDhadronsXeXe}. These are measured in narrower centrality intervals as compared with the measurements presented in this paper. The values corresponding to the same centrality classes are obtained as weighted averages using the \pt spectra taken from Ref.~\cite{ALICE:2021lsv} as weights. A clear mass ordering is observed at low \pt, as expected from relativistic hydrodynamics~\cite{relativisticHydro}, with a slower rise with \pt of the (anti)deuteron $v_2$ as compared with that of lighter hadrons. The mass ordering of $v_2$ at low \pt develops mostly by final-state rescattering in the hadronic phase~\cite{Hirano:2007ei}, reflecting the redistribution of the momentum anisotropy generated in the early QGP states among the different hadron species, driven by the radial acceleration and cooling of matter during the hadronic rescattering phase.
Assuming that (anti)nuclei are produced as compact multi-quark systems with negligible final-state interactions with hadrons, their flow would be fully developed in the partonic phase. On the other hand, if (anti)nuclei are produced by coalescence, their flow could also have a contribution from the hadronic rescattering phase, which is inherited from that of the coalescing nucleons. 

In Fig.~\ref{fig:dataVsBW}, the deuteron $v_{2}$ measurements are compared with the expectations from the Blast-Wave and simple coalescence models using the same procedure described in Refs.~\cite{Acharya:2017dmc,Acharya:2020lus,Acharya:2019ttn}. The Blast-Wave predictions are obtained from a simultaneous fit of the $v_{2}$ and the \pt spectra of pions, kaons, and protons measured in Xe--Xe collisions at \snn = 5.44 TeV~\cite{ALICE:2021lsv,flowIDhadronsXeXe} in the transverse-momentum ranges $(0.5 < p^{\pi}_{\mathrm{T}} < 1)\ \mathrm{GeV}/\textit{c}$, $(0.7 < p^{\mathrm{K}}_{\mathrm{T}} < 2)\ \mathrm{GeV}/\textit{c}$, and  $(0.7 < p^{\mathrm{p}}_{\mathrm{T}} < 2.5)\ \mathrm{GeV}/\textit{c}$, respectively, and in the same centrality classes. The four parameters of the Blast-Wave fits represent the kinetic freeze-out temperature ($T_{\mathrm{kin}}$), the mean transverse expansion rapidity ($\rho_{0}$), the amplitude of its azimuthal variation ($\rho_{\mathrm{a}}$), and the variation in the azimuthal density of the source ($s_{2}$), as described in Ref.~\cite{BlastWave2}. 

The values of the Blast-Wave parameters extracted from the fits are reported in \autoref{Table:BlastWaveParameters} for each centrality interval. The elliptic flow of (anti)deuterons is calculated using the parameters obtained from the simultaneous fit and the deuteron mass, i.e., assuming the same kinetic freeze-out conditions.

\begin{table}[!hbt]
\begin{center}
\centering
\renewcommand{\arraystretch}{1.1}
\caption{Blast-Wave parameters extracted from the simultaneous fits of the \pt spectra and $v_{2}$ of pions, kaons, and protons. See text for details. }
\begin{tabular}{cccc}
\hline
Fit parameters &  Centrality 0--20$\%$  &  Centrality  20--40$\%$  &  \\
\hline
     $T_{\mathrm{kin}}$ (MeV)             & $112  \pm 2$  & $111  \pm 2$    \\
     $\rho_{0}$           & $1.29 \pm 0.05$  & $1.31 \pm 0.05$   \\
     $\rho_{\mathrm{a}}$ $\times10^{-2}$  & $5.87 \pm 0.02$  & $8.85 \pm 0.02$  \\
     $s_{2}$  $\times10^{-2}$             & $4.94 \pm 0.02$  & $8.86 \pm 0.02$   \\
\hline
\end{tabular}
\label{Table:BlastWaveParameters}
\end{center}
\end{table} 

The simple coalescence model used in this context is based on the assumption that the invariant yield of (anti)deuterons with transverse momentum \pt is proportional to the product of the invariant yields of its constituent nucleons with transverse momentum $p_{\mathrm{T}}/2$ and on isospin symmetry, due to which the proton and neutron $v_{2}$ are identical. 
Considering only elliptical anisotropies of the constituent nucleons, i.e., neglecting higher-order harmonics, the (anti)deuteron $v_2$ from coalescence is obtained from that of (anti)protons using the scaling law~\cite{Molnar:2003ff}

\begin{equation}
v_{2,\mathrm{d}}(p_{\rm T}) = \frac{2 v_{2,\mathrm{p}} (p_{\rm {T}}/2)}{1+2{v^2_{2,\rm{p}}} (p_{\mathrm {T}}/2)}.
\label{eq:v2coal}
\end{equation} 

\begin{figure}[hbt]
    \begin{center}
    \includegraphics[width = 0.8\textwidth]{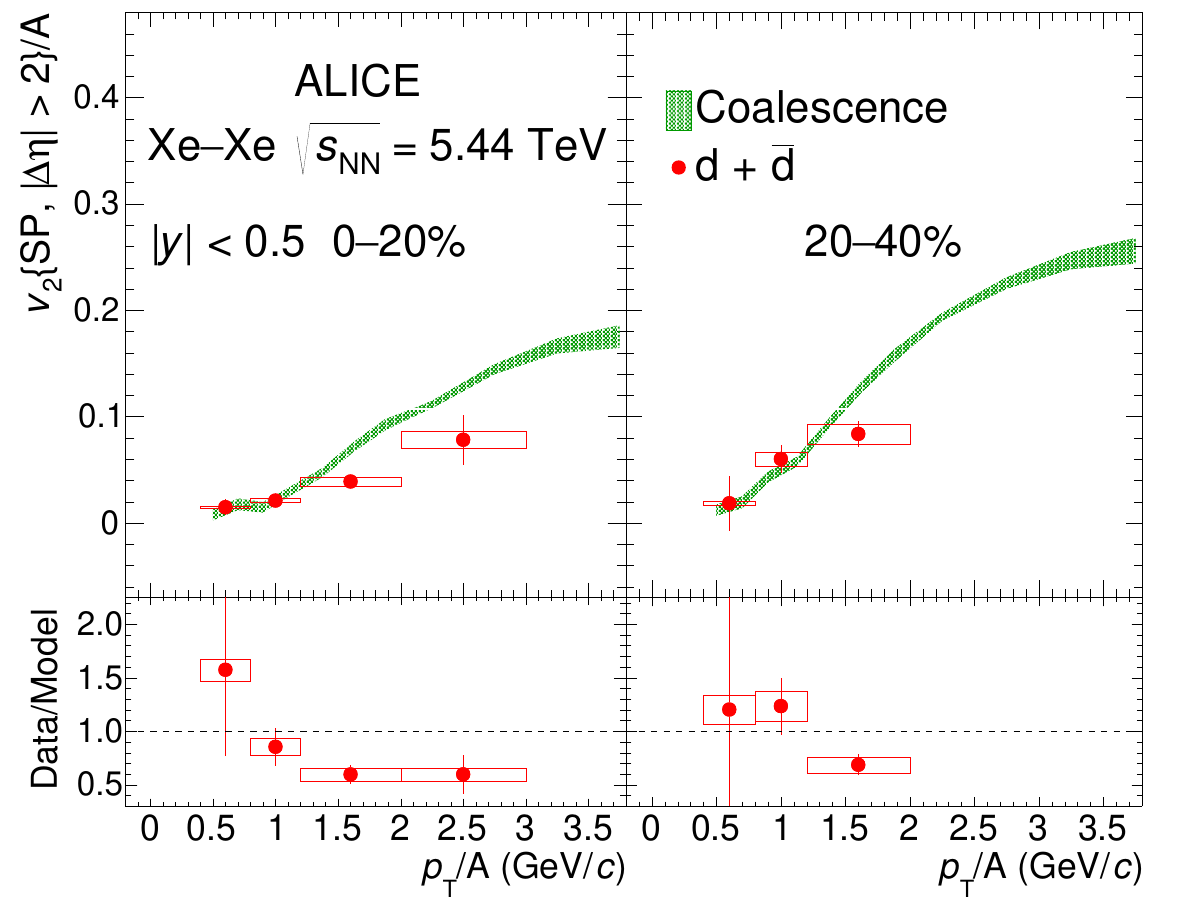}
    \end{center}
\caption{(Anti)deuteron $v_2$ measured in the centrality classes 0--20$\%$ (left) and 20--40$\%$ (right) in comparison with the coalescence predictions based on the mass-number scaling. Statistical and systematic uncertainties are shown as bars and boxes, respectively. The flow of deuterons is measured at midrapidity, i.e., $\left|y\right|<$ 0.5.}
\label{fig:dataVsCoalescence}        
\end{figure}

The comparison between the (anti)deuteron $v_2$ measurements and the simple coalescence predictions is shown in Fig.~\ref{fig:dataVsCoalescence}. As observed already in elliptic-flow measurements in Pb--Pb collisions for both (anti)deuterons and (anti)$^{3}$He, the data are closer to the Blast-Wave predictions in more central collisions and to the coalescence model in more peripheral collisions. These two simple models represent approximate limits for (anti)nucleus flow which describe the measurements in complementary centrality intervals. 

As noted in Ref.~\cite{flowIDhadronsXeXe}, the flow coefficients at low \pt are expected to be smaller in Pb--Pb collisions than in Xe--Xe ones, due to the larger radial flow in the Pb--Pb case. Such an effect is expected to be larger in central collisions and for heavier particles, such as light nuclei. The results of the $v_2$ of deuteron in Xe--Xe collisions, however, seem to be above the corresponding measurements in Pb--Pb collisions, as already observed in the case of the flow of lighter identified particles~\cite{flowIDhadronsXeXe}. However, the present results have large statistical uncertainties, which do not allow for quantitative comparisons with the Pb--Pb results. In Ref.~\cite{flowIDhadronsXeXe}, more quantitative comparisons between the flow coefficients of light-flavor hadrons in Pb--Pb collisions and those in Xe--Xe collisions are discussed and interpreted as due to the interplay of the nuclear deformation of the Xe nucleus with respect to the Pb one, and of initial-state fluctuations proportional to the square root of the mass number of the colliding nucleus. 

\subsection{Thermal fit}
\begin{figure}[!ht]
\centering
\begin{minipage}[b]{0.7\textwidth}
    \includegraphics[width=\textwidth]{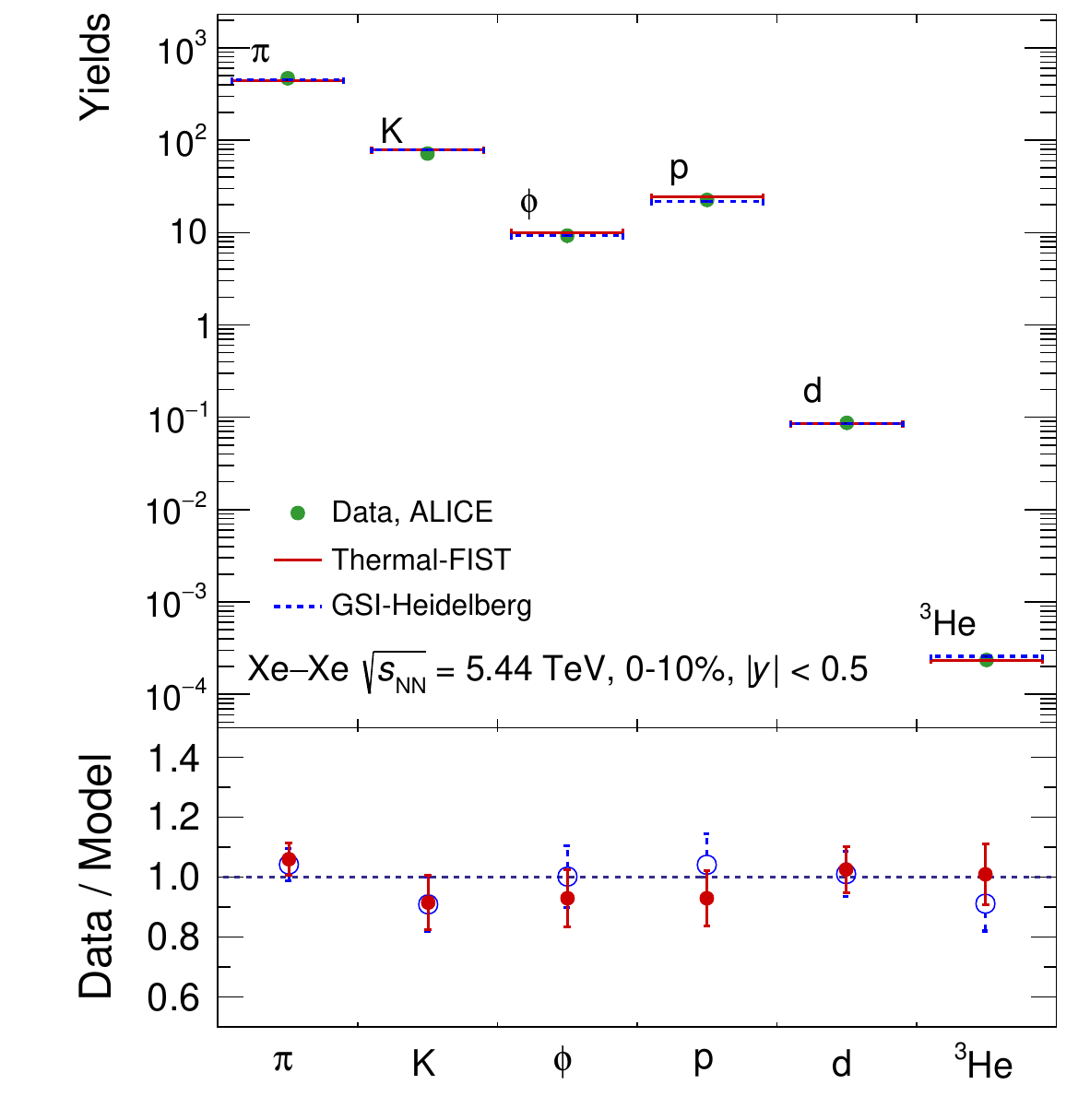}
\end{minipage}   
\caption{Thermal fit to the measured hadron yields in Xe--Xe collisions at \snn = 5.44 TeV, performed with two different implementations of the statistical hadronization model, namely the Thermal-FIST~\cite{Vovchenko:2019pjl} and the GSI-Heidelberg model~\cite{Andronic:2008gu}. The yields shown for each given species are the average of the particle and antiparticle yields. In the bottom panel, the ratios of data to the model calculations are shown.}
\label{fig:ThermalFit}
\end{figure}

The \pt-integrated yields of d and $^3$He for central (0--10\%) Xe--Xe collisions are shown in Fig.~\ref{fig:ThermalFit} together with those of $\pi$, K, p, and $\phi$.
At the LHC, the production of most light-flavor hadrons and light (anti)(hyper)nuclei in nucleus–nucleus collisions can be described in the framework of the statistical hadronization approach via thermal models based on the grand canonical ensemble with a single chemical freeze-out temperature ($T_{\rm chem}$), and a given volume $V$ of the fireball.
Thermal model fits to the ALICE data for the yields of $\pi$, K, $\phi$, p~\cite{ALICE:2021lsv}, d, and $^3$He measured in central (0--10\%) Xe--Xe collisions at \mbox{$\sqrt{s_{\rm NN}}=$ 5.44 TeV} are performed using the open source Thermal-FIST package~\cite{Vovchenko:2019pjl}. The fit converges at \mbox{$T_{\rm chem}$ = (154.2 $\pm$ 1.1) MeV}, $V =$ (3626 $\pm$ 298) fm$^3$, and \mbox{$\chi^2/$NDF = 0.83}. 
All other parameters are fixed using the parameterizations reported in Ref.~\cite{vanillaCSM}. 
The calculations are carried out using the energy-dependent Breit--Wigner treatment of resonance widths with constant branching ratios~\cite{Vovchenko:2018fmh}. Enabling the partial chemical equilibrium~\cite{Vovchenko:2022xil}, without baryon annihilation, does not significantly change the results of the calculated yields of light nuclei, once the $T_{\rm chem}$ and $T_{\rm kin}$ are fixed to 155 MeV and 100 MeV, respectively. 
The results for the yields obtained from the thermal fit are shown as solid lines in Fig.~\ref{fig:ThermalFit}. 
Similar fits to the measured yields have been carried out using the GSI-Heidelberg thermal model~\cite{Andronic:2008gu}. For such fit, the temperature of the system and the baryochemical potential are fixed to the results obtained in the most central Pb--Pb collisions at \mbox{\snn = 2.76 TeV} with the S-matrix correction for the protons ($T_{\rm chem}$ = 156.6 MeV, $\mu_{B}$ = 0.7 MeV), while the volume $V$ is a free parameter~\cite{Andronic:2018qqt}.
The fit to the Xe--Xe data converges at \mbox{$V =$ (2996 $\pm$ 102) fm$^3$} ($|y|<$0.5), and \mbox{$\chi^2/$NDF = 0.55}. 
\\The main difference between the two models is in the treatment of the interactions: while the GSI-Heidelberg approach implements the S-matrix formalism~\cite{Dashen:1969ep,Venugopalan:1992hy,Weinhold:1997ig,Huovinen:2016xxq,Lo:2017sde} to include non-resonant pion--nucleon interactions at LHC energies, Thermal-FIST uses the energy-dependent Breit--Wigner resonance widths~\cite{Breit:1936zzb}. Since the pion is the most abundant particle at freeze-out and the corresponding cross sections for hadronic interactions are large, the handling of $\pi$N interactions is a key aspect when analyzing observables involving nucleons in ultra-relativistic nucleus--nucleus collisions. The different treatment of interactions is reflected in the different results of both the volume of the system, which is driven by the abundance of pions, and its temperature. The interplay between these two parameters is anticorrelated. Indeed, fixing in the Thermal-FIST model the $T_{\rm chem}$ to a higher value, e.g., the value used for the GSI-Heidelberg fit (\mbox{$T_{\rm chem}=$ 156.6 MeV}), the resulting volume becomes smaller (\mbox{$V =$ (3095 $\pm$ 101) fm$^3$}), and comparable with that obtained with the GSI-Heidelberg approach.
\\The resulting yields for the GSI-Heidelberg model are shown as dotted lines in Fig.~\ref{fig:ThermalFit}, being well compatible with the Thermal-FIST results. Light hadron yields are taken from Ref.~\cite{ALICE:2021lsv}. The yields of all given species are the average of the particle and antiparticle yields. The yield of $^3$He in the 0--10\% centrality class is obtained using the same scaling used for alpha nuclei in Ref.~\cite{Acharya:2017bso}. 
The results of the models prove that the production of light nuclei in Xe--Xe collisions is sensitive to the temperature of the system at chemical freeze-out. 
Similar fits to the most central \mbox{Pb--Pb} collisions at \mbox{\snn $=$ 2.76 TeV} in one unit of rapidity at midrapidity, including the S-matrix, gave as results $T_{\rm chem}$ = 156.6 MeV, $\mu_{B}$ = 0.7 MeV, and \mbox{$V =$ 4175 fm$^3$}~\cite{Andronic:2018qqt}. The $T_{\rm{chem}}$ obtained from the fit of the Thermal-FIST model to the measured light-flavor hadron yields in central Xe--Xe collisions is similar to the $T_{\rm{chem}}$ obtained in central Pb--Pb collisions using S-matrix based thermal model and consistent with the pseudocritical temperature of the chiral crossover transition predicted by lattice QCD calculations, \mbox{$T =$ (156.5 $\pm$ 1.5) MeV}~\cite{HotQCD:2018pds}. 
Moreover, the final state light nuclei yields can also be modified by the baryon annihilation~\cite{Vovchenko:2022xil}, which is not included in this model comparison. However, the effect of baryon annihilation is expected to be more sensitive in central Pb--Pb collisions compared with central Xe--Xe collisions, due to the higher charged-particle multiplicity in Pb--Pb events compared with Xe--Xe~\cite{Vovchenko:2022xil}. This can be seen in Fig.~\ref{fig:NucleiOverP} and~\ref{fig:NucleiOverPions}, in which the two model calculations agree with the measurements, considering the current precision of the data, within 2$\sigma$ for the ratio of deuteron yields and 1$\sigma$ for those of $^3$He.

\section{Summary} 
\label{sec:Conclusions}

Measurements of (anti)deuteron and (anti)$^{3}$He production and of the elliptic-flow of (anti)deuteron in Xe--Xe collisions at \snn = 5.44 TeV are presented. These results contribute to the understanding of the light (anti)nucleus production mechanism by complementing the existing picture, which includes measurements in different collision systems and at different center-of-mass energies per nucleon--nucleon pair. 
The hadrochemistry of light nuclei, i.e., the relative abundance of particles, in small collision systems (pp and p--Pb) was found to be independent of collision energies and systems and mostly driven by $\langle$d$N_{\mathrm{ch}}/$d$\eta\rangle$. This is further investigated for the first time in heavy-ion collisions at LHC energies by comparing the measurements in Xe--Xe collisions, and the measurements are found to be consistent across the two different collision systems at similar charged particle multiplicity densities.
The canonical statistical model and the coalescence predictions are tested by comparing the multiplicity dependence of (anti)deuteron and (anti)$^{3}$He yields, relative to proton and pion yields. The Xe--Xe results fit the behavior of the previous ALICE measurements, which are qualitatively described by the CSM and, in the case of the ratio of deuterons to protons, also by coalescence, across all multiplicities. A version of the CSM with $V_{\mathrm{c}}$= 1.6 d$V$/d$y$, a multiplicity-dependent annihilation freeze-out temperature and the Saha equation approach~\cite{Vovchenko:2019aoz} to compute light-nucleus abundances in the hadronic phase, captures the decreasing trend of data at high multiplicity, corresponding to central Pb--Pb and Xe--Xe collisions, in all reported particle ratios. This suggests that such suppression is due to baryon annihilations in the hadronic phase. However, the same implementation of the CSM model fails to simultaneously reproduce the trend of the particle ratios at intermediate multiplicities. This shows that at the moment there is no common implementation and available parametrization of the CSM able to reproduce all particle ratios from small collision systems to central heavy-ion collisions.  
The yields of light-flavor hadrons produced in Xe--Xe collisions are described, in the framework of the statistical hadronization model, with different implementations of the thermal fit, namely the Thermal-FIST package~\cite{Vovchenko:2019pjl} and the GSI-Heidelberg model~\cite{Andronic:2008gu}. In the former case, the fit converged at \mbox{$T_{\rm chem}$ = (154.2 $\pm$ 1.1) MeV}, $V =$ (3626 $\pm$ 298) fm$^3$, while in the latter case the temperature and baryochemical potential are fixed to $T_{\rm chem}$= 156.6 MeV, $\mu_{B}$= 0.7 MeV and the fit converges at a volume \mbox{$V =$ (2996 $\pm$ 102) fm$^3$}. 
The results from the two models are consistent with each other within 2$\sigma$, with the results obtained in Pb--Pb collisions \mbox{($T_{\rm chem}$ = 156.6 MeV)}~\cite{Andronic:2018qqt}, and with the pseudo-critical temperature for the transition from QGP to hadron gas predicted by lattice QCD calculations \mbox{($T =$ (156.5 $\pm$ 1.5) MeV)}~\cite{HotQCD:2018pds}. 
Finally, the results of the elliptic flow of (anti)deuterons are compared with the predictions of a simple coalescence model and to the expectations of a Blast-Wave fit of $v_2$ and $p_{\rm T}$ of light-flavor hadrons measured in the same collision system and at the same energy. Such comparison shows that the data are closer to the Blast-Wave predictions in more central collisions and to the coalescence model in more peripheral collisions, as already observed in the results of (anti)deuterons and (anti)$^{3}$He flow measured in Pb--Pb collisions.

\newenvironment{acknowledgement}{\relax}{\relax}
\begin{acknowledgement}
\section*{Acknowledgements}

The ALICE Collaboration would like to thank all its engineers and technicians for their invaluable contributions to the construction of the experiment and the CERN accelerator teams for the outstanding performance of the LHC complex.
The ALICE Collaboration gratefully acknowledges the resources and support provided by all Grid centres and the Worldwide LHC Computing Grid (WLCG) collaboration.
The ALICE Collaboration acknowledges the following funding agencies for their support in building and running the ALICE detector:
A. I. Alikhanyan National Science Laboratory (Yerevan Physics Institute) Foundation (ANSL), State Committee of Science and World Federation of Scientists (WFS), Armenia;
Austrian Academy of Sciences, Austrian Science Fund (FWF): [M 2467-N36] and Nationalstiftung f\"{u}r Forschung, Technologie und Entwicklung, Austria;
Ministry of Communications and High Technologies, National Nuclear Research Center, Azerbaijan;
Conselho Nacional de Desenvolvimento Cient\'{\i}fico e Tecnol\'{o}gico (CNPq), Financiadora de Estudos e Projetos (Finep), Funda\c{c}\~{a}o de Amparo \`{a} Pesquisa do Estado de S\~{a}o Paulo (FAPESP) and Universidade Federal do Rio Grande do Sul (UFRGS), Brazil;
Bulgarian Ministry of Education and Science, within the National Roadmap for Research Infrastructures 2020-2027 (object CERN), Bulgaria;
Ministry of Education of China (MOEC) , Ministry of Science \& Technology of China (MSTC) and National Natural Science Foundation of China (NSFC), China;
Ministry of Science and Education and Croatian Science Foundation, Croatia;
Centro de Aplicaciones Tecnol\'{o}gicas y Desarrollo Nuclear (CEADEN), Cubaenerg\'{\i}a, Cuba;
Ministry of Education, Youth and Sports of the Czech Republic, Czech Republic;
The Danish Council for Independent Research | Natural Sciences, the VILLUM FONDEN and Danish National Research Foundation (DNRF), Denmark;
Helsinki Institute of Physics (HIP), Finland;
Commissariat \`{a} l'Energie Atomique (CEA) and Institut National de Physique Nucl\'{e}aire et de Physique des Particules (IN2P3) and Centre National de la Recherche Scientifique (CNRS), France;
Bundesministerium f\"{u}r Bildung und Forschung (BMBF) and GSI Helmholtzzentrum f\"{u}r Schwerionenforschung GmbH, Germany;
General Secretariat for Research and Technology, Ministry of Education, Research and Religions, Greece;
National Research, Development and Innovation Office, Hungary;
Department of Atomic Energy Government of India (DAE), Department of Science and Technology, Government of India (DST), University Grants Commission, Government of India (UGC) and Council of Scientific and Industrial Research (CSIR), India;
National Research and Innovation Agency - BRIN, Indonesia;
Istituto Nazionale di Fisica Nucleare (INFN), Italy;
Japanese Ministry of Education, Culture, Sports, Science and Technology (MEXT) and Japan Society for the Promotion of Science (JSPS) KAKENHI, Japan;
Consejo Nacional de Ciencia (CONACYT) y Tecnolog\'{i}a, through Fondo de Cooperaci\'{o}n Internacional en Ciencia y Tecnolog\'{i}a (FONCICYT) and Direcci\'{o}n General de Asuntos del Personal Academico (DGAPA), Mexico;
Nederlandse Organisatie voor Wetenschappelijk Onderzoek (NWO), Netherlands;
The Research Council of Norway, Norway;
Pontificia Universidad Cat\'{o}lica del Per\'{u}, Peru;
Ministry of Science and Higher Education, National Science Centre and WUT ID-UB, Poland;
Korea Institute of Science and Technology Information and National Research Foundation of Korea (NRF), Republic of Korea;
Ministry of Education and Scientific Research, Institute of Atomic Physics, Ministry of Research and Innovation and Institute of Atomic Physics and Universitatea Nationala de Stiinta si Tehnologie Politehnica Bucuresti, Romania;
Ministry of Education, Science, Research and Sport of the Slovak Republic, Slovakia;
National Research Foundation of South Africa, South Africa;
Swedish Research Council (VR) and Knut \& Alice Wallenberg Foundation (KAW), Sweden;
European Organization for Nuclear Research, Switzerland;
Suranaree University of Technology (SUT), National Science and Technology Development Agency (NSTDA) and National Science, Research and Innovation Fund (NSRF via PMU-B B05F650021), Thailand;
Turkish Energy, Nuclear and Mineral Research Agency (TENMAK), Turkey;
National Academy of  Sciences of Ukraine, Ukraine;
Science and Technology Facilities Council (STFC), United Kingdom;
National Science Foundation of the United States of America (NSF) and United States Department of Energy, Office of Nuclear Physics (DOE NP), United States of America.
In addition, individual groups or members have received support from:
Czech Science Foundation (grant no. 23-07499S), Czech Republic;
European Research Council (grant no. 950692), European Union;
ICSC - Centro Nazionale di Ricerca in High Performance Computing, Big Data and Quantum Computing, European Union - NextGenerationEU;
Academy of Finland (Center of Excellence in Quark Matter) (grant nos. 346327, 346328), Finland.

\end{acknowledgement}

\bibliographystyle{utphys}   
\bibliography{bibliography}

\providecommand{\href}[2]{#2}\begingroup\raggedright\begin{thebibliography}{100}

\bibitem{Adam:2015pna}
{\bfseries ALICE} Collaboration, J.~Adam {\em et~al.}, ``{Precision measurement
  of the mass difference between light nuclei and anti-nuclei}'',
  \href{http://dx.doi.org/10.1038/nphys3432}{{\em Nature Phys.} {\bfseries 11}
  (2015) 811--814}, \href{http://arxiv.org/abs/1508.03986}{{\ttfamily
  arXiv:1508.03986 [nucl-ex]}}.

\bibitem{3HePbPb}
{\bfseries ALICE} Collaboration, J.~Adam {\em et~al.}, ``{Production of light
  nuclei and anti-nuclei in pp and Pb--Pb collisions at energies available at
  the CERN Large Hadron Collider}'',
  \href{http://dx.doi.org/10.1103/PhysRevC.93.024917}{{\em Phys. Rev.}
  {\bfseries C93} (2016) 024917},
\href{http://arxiv.org/abs/1506.08951}{{\ttfamily arXiv:1506.08951 [nucl-ex]}}.

\bibitem{Acharya:2017dmc}
{\bfseries ALICE} Collaboration, S.~Acharya {\em et~al.}, ``{Measurement of
  deuteron spectra and elliptic flow in Pb--Pb collisions at $\sqrt{s_{\mathrm
  {NN}}}$ = 2.76 TeV at the LHC}'',
  \href{http://dx.doi.org/10.1140/epjc/s10052-017-5222-x}{{\em Eur. Phys. J.}
  {\bfseries C77} (2017) 658},
\href{http://arxiv.org/abs/1707.07304}{{\ttfamily arXiv:1707.07304 [nucl-ex]}}.

\bibitem{Acharya:2017fvb}
{\bfseries ALICE} Collaboration, S.~Acharya {\em et~al.}, ``{Production of
  deuterons, tritons, $^{3}$He nuclei and their antinuclei in pp collisions at
  $\mathbf{\sqrt{{\textit s}}}$ = 0.9, 2.76 and 7 TeV}'',
  \href{http://dx.doi.org/10.1103/PhysRevC.97.024615}{{\em Phys. Rev.}
  {\bfseries C97} (2018) 024615},
\href{http://arxiv.org/abs/1709.08522}{{\ttfamily arXiv:1709.08522 [nucl-ex]}}.

\bibitem{Acharya:2017bso}
{\bfseries ALICE} Collaboration, S.~Acharya {\em et~al.}, ``{Production of
  $^{4}$He and $^{4}\overline{\textrm{He}}$ in Pb--Pb collisions at
  \mbox{$\sqrt{s_{\mathrm{NN}}}$ = 2.76 TeV} at the LHC}'',
  \href{http://dx.doi.org/10.1016/j.nuclphysa.2017.12.004}{{\em Nucl. Phys. A}
  {\bfseries 971} (2018) 1--20},
  \href{http://arxiv.org/abs/1710.07531}{{\ttfamily arXiv:1710.07531
  [nucl-ex]}}.

\bibitem{deuteron_pp7TeV}
{\bfseries ALICE} Collaboration, S.~Acharya {\em et~al.}, ``{Multiplicity
  dependence of (anti-)deuteron production in pp collisions at $\sqrt{s}$ = 7
  TeV}'', \href{http://dx.doi.org/10.1016/j.physletb.2019.05.028}{{\em Phys.
  Lett. B} {\bfseries 794} (2019) 50--63},
  \href{http://arxiv.org/abs/1902.09290}{{\ttfamily arXiv:1902.09290
  [nucl-ex]}}.

\bibitem{Acharya:2019rys}
{\bfseries ALICE} Collaboration, S.~Acharya {\em et~al.}, ``{Multiplicity
  dependence of light (anti-)nuclei production in p--Pb collisions at
  $\sqrt{s_{\rm{NN}}}$ = 5.02 TeV}'',
  \href{http://dx.doi.org/10.1016/j.physletb.2019.135043}{{\em Phys. Lett. B}
  {\bfseries 800} (2020) 135043},
  \href{http://arxiv.org/abs/1906.03136}{{\ttfamily arXiv:1906.03136
  [nucl-ex]}}.

\bibitem{Acharya:2019ttn}
{\bfseries ALICE} Collaboration, S.~Acharya {\em et~al.}, ``{Measurement of the
  (anti-)$^{3}$He elliptic flow in Pb--Pb collisions at $\sqrt{s_{\rm{NN}}}$ =
  5.02 TeV}'', \href{http://dx.doi.org/10.1016/j.physletb.2020.135414}{{\em
  Phys. Lett. B} {\bfseries 805} (2020) 135414},
\href{http://arxiv.org/abs/1910.09718}{{\ttfamily arXiv:1910.09718 [nucl-ex]}}.

\bibitem{Acharya:2019xmu}
{\bfseries ALICE} Collaboration, S.~Acharya {\em et~al.}, ``{Production of
  (anti-)$^3$He and (anti-)$^3$H in p--Pb collisions at $\sqrt{s_{\rm{NN}}}$ =
  5.02 TeV}'', \href{http://dx.doi.org/10.1103/PhysRevC.101.044906}{{\em Phys.
  Rev.} {\bfseries C101} (2020) 044906},
\href{http://arxiv.org/abs/1910.14401}{{\ttfamily arXiv:1910.14401 [nucl-ex]}}.

\bibitem{Acharya:2020sfy}
{\bfseries ALICE} Collaboration, S.~Acharya {\em et~al.}, ``{(Anti-)deuteron
  production in pp collisions at $\sqrt{s}=13 \ \mathrm{TeV}$}'',
  \href{http://dx.doi.org/10.1140/epjc/s10052-020-8256-4}{{\em Eur. Phys. J.}
  {\bfseries C80} (2020) 889},
\href{http://arxiv.org/abs/2003.03184}{{\ttfamily arXiv:2003.03184 [nucl-ex]}}.

\bibitem{Acharya:2020lus}
{\bfseries ALICE} Collaboration, S.~Acharya {\em et~al.}, ``{Elliptic and
  triangular flow of (anti)deuterons in Pb--Pb collisions at
  $\sqrt{s_{\mathrm{NN}}}$ = 5.02 TeV}'',
  \href{http://dx.doi.org/10.1103/PhysRevC.102.055203}{{\em Phys. Rev. C}
  {\bfseries 102} (2020) 055203},
  \href{http://arxiv.org/abs/2005.14639}{{\ttfamily arXiv:2005.14639
  [nucl-ex]}}.

\bibitem{deuteron_jets}
{\bfseries ALICE} Collaboration, S.~Acharya {\em et~al.}, ``{Jet-associated
  deuteron production in pp collisions at $\sqrt{s}$ = 13 TeV}'',
  \href{http://dx.doi.org/10.1016/j.physletb.2021.136440}{{\em Phys. Lett. B}
  {\bfseries 819} (2021) 136440},
  \href{http://arxiv.org/abs/2011.05898}{{\ttfamily arXiv:2011.05898
  [nucl-ex]}}.

\bibitem{nuclei_pp_13TeV}
{\bfseries ALICE} Collaboration, S.~Acharya {\em et~al.}, ``{Production of
  light (anti)nuclei in pp collisions at $ \sqrt{s} $ = 13 TeV}'',
  \href{http://dx.doi.org/10.1007/JHEP01(2022)106}{{\em JHEP} {\bfseries 01}
  (2022) 106}, \href{http://arxiv.org/abs/2109.13026}{{\ttfamily
  arXiv:2109.13026 [nucl-ex]}}.

\bibitem{nuclei_pp_5TeV}
{\bfseries ALICE} Collaboration, S.~Acharya {\em et~al.}, ``{Production of
  light (anti)nuclei in pp collisions at $\sqrt{s} = 5.02$~TeV}'',
  \href{http://dx.doi.org/10.1140/epjc/s10052-022-10241-z}{{\em Eur. Phys. J.
  C} {\bfseries 82} (2022) 289},
  \href{http://arxiv.org/abs/2112.00610}{{\ttfamily arXiv:2112.00610
  [nucl-ex]}}.

\bibitem{hypertriton_pPb_ALICE}
{\bfseries ALICE} Collaboration, S.~Acharya {\em et~al.}, ``{Hypertriton
  Production in p--Pb Collisions at $\sqrt {s_{\rm NN}}=5.02$\,\,TeV}'',
  \href{http://dx.doi.org/10.1103/PhysRevLett.128.252003}{{\em Phys. Rev.
  Lett.} {\bfseries 128} (2022) 252003},
  \href{http://arxiv.org/abs/2107.10627}{{\ttfamily arXiv:2107.10627
  [nucl-ex]}}.

\bibitem{ALICE:2022boh}
{\bfseries ALICE} Collaboration, S.~Acharya {\em et~al.}, ``{Light (anti)nuclei
  production in Pb--Pb collisions at $\sqrt{s_{\mathrm{NN}}}=5.02$ TeV}'',
  \href{http://dx.doi.org/10.1103/PhysRevC.107.064904}{{\em Phys. Rev. C}
  {\bfseries 107} (2023) 064904},
  \href{http://arxiv.org/abs/2211.14015}{{\ttfamily arXiv:2211.14015
  [nucl-ex]}}.

\bibitem{ALICE:2022weh}
{\bfseries ALICE} Collaboration, S.~Acharya {\em et~al.}, ``{Measurement of the
  production of (anti)nuclei in p\textendash{}Pb collisions at \mbox{\snn =
  8.16 TeV}}'', \href{http://dx.doi.org/10.1016/j.physletb.2023.137795}{{\em
  Phys. Lett. B} {\bfseries 846} (2023) 137795},
  \href{http://arxiv.org/abs/2212.04777}{{\ttfamily arXiv:2212.04777
  [nucl-ex]}}.

\bibitem{ALICE:2022jmr}
{\bfseries ALICE} Collaboration, S.~Acharya {\em et~al.}, ``{Enhanced Deuteron
  Coalescence Probability in Jets}'',
  \href{http://dx.doi.org/10.1103/PhysRevLett.131.042301}{{\em Phys. Rev.
  Lett.} {\bfseries 131} (2023) 042301},
  \href{http://arxiv.org/abs/2211.15204}{{\ttfamily arXiv:2211.15204
  [nucl-ex]}}.

\bibitem{ALICE:2022wpn}
{\bfseries ALICE} Collaboration, S.~Acharya {\em et~al.}, ``{The ALICE
  experiment: a journey through QCD}'',
  \href{http://dx.doi.org/10.1140/epjc/s10052-024-12935-y}{{\em Eur. Phys. J.
  C} {\bfseries 84} (2024) 813},
  \href{http://arxiv.org/abs/2211.04384}{{\ttfamily arXiv:2211.04384
  [nucl-ex]}}.

\bibitem{Bennett:1998be}
{\bfseries E878} Collaboration, M.~J. Bennett {\em et~al.}, ``{Light nuclei
  production in relativistic Au + nucleus collisions}'',
  \href{http://dx.doi.org/10.1103/PhysRevC.58.1155}{{\em Phys. Rev. C}
  {\bfseries 58} (1998) 1155--1164}.

\bibitem{Ahle:1999in}
{\bfseries E802} Collaboration, L.~Ahle {\em et~al.}, ``{Proton and deuteron
  production in Au + Au reactions at 11.6 A GeV/$c$}'',
  \href{http://dx.doi.org/10.1103/PhysRevC.60.064901}{{\em Phys. Rev. C}
  {\bfseries 60} (1999) 064901}.

\bibitem{Armstrong:2000gz}
{\bfseries E864} Collaboration, T.~A. Armstrong {\em et~al.}, ``{Measurements
  of light nuclei production in 11.5 A GeV/$c$ Au + Pb heavy ion collisions}'',
  \href{http://dx.doi.org/10.1103/PhysRevC.61.064908}{{\em Phys. Rev. C}
  {\bfseries 61} (2000) 064908},
  \href{http://arxiv.org/abs/nucl-ex/0003009}{{\ttfamily
  arXiv:nucl-ex/0003009}}.

\bibitem{Armstrong:2000gd}
{\bfseries E864} Collaboration, T.~A. Armstrong {\em et~al.}, ``{Anti-deuteron
  yield at the AGS and coalescence implications}'',
  \href{http://dx.doi.org/10.1103/PhysRevLett.85.2685}{{\em Phys. Rev. Lett.}
  {\bfseries 85} (2000) 2685--2688},
  \href{http://arxiv.org/abs/nucl-ex/0005001}{{\ttfamily
  arXiv:nucl-ex/0005001}}.

\bibitem{Ambrosini:1997bf}
{\bfseries NA52 (NEWMASS)} Collaboration, G.~Ambrosini {\em et~al.}, ``{Baryon
  and anti-baryon production in lead-lead collisions at 158 A GeV/$c$}'',
  \href{http://dx.doi.org/10.1016/S0370-2693(97)01383-X}{{\em Phys. Lett. B}
  {\bfseries 417} (1998) 202--210}.

\bibitem{Adler:2001prl}
{\bfseries STAR} Collaboration, C.~Adler {\em et~al.}, ``{Anti-deuteron and
  anti-He-3 production in \snn = 130 GeV Au+Au collisions}'',
  \href{http://dx.doi.org/10.1103/PhysRevLett.87.262301}{{\em Phys. Rev. Lett.}
  {\bfseries 87} (2001) 262301},
  \href{http://arxiv.org/abs/nucl-ex/0108022}{{\ttfamily
  arXiv:nucl-ex/0108022}}. [Erratum: $Phys.Rev.Lett.$ \textbf{87} (2001)
  279902].

\bibitem{Adler:2004uy}
{\bfseries PHENIX} Collaboration, S.~S. Adler {\em et~al.}, ``{Deuteron and
  antideuteron production in Au + Au collisions at 200 GeV}'',
  \href{http://dx.doi.org/10.1103/PhysRevLett.94.122302}{{\em Phys. Rev. Lett.}
  {\bfseries 94} (2005) 122302},
  \href{http://arxiv.org/abs/nucl-ex/0406004}{{\ttfamily
  arXiv:nucl-ex/0406004}}.

\bibitem{Arsene:2010px}
{\bfseries BRAHMS} Collaboration, I.~Arsene {\em et~al.}, ``{Rapidity
  dependence of deuteron production in Au+Au collisions at $\sqrt{s_{\rm NN}}$
  = 200 GeV}'', \href{http://dx.doi.org/10.1103/PhysRevC.83.044906}{{\em Phys.
  Rev. C} {\bfseries 83} (2011) 044906},
  \href{http://arxiv.org/abs/1005.5427}{{\ttfamily arXiv:1005.5427 [nucl-ex]}}.

\bibitem{Agakishiev:2011ib}
{\bfseries STAR} Collaboration, H.~Agakishiev {\em et~al.}, ``{Observation of
  the antimatter helium-4 nucleus}'',
  \href{http://dx.doi.org/10.1038/nature10079}{{\em Nature} {\bfseries 473}
  (2011) 353}, \href{http://arxiv.org/abs/1103.3312}{{\ttfamily arXiv:1103.3312
  [nucl-ex]}}. [Erratum: $Nature$ \textbf{475} (2011) 412].

\bibitem{Adamczyk:2016gfs}
{\bfseries STAR} Collaboration, L.~Adamczyk {\em et~al.}, ``{Measurement of
  elliptic flow of light nuclei at $\sqrt{s_{\rm NN}}=$ 200, 62.4, 39, 27,
  19.6, 11.5, and 7.7 GeV at the BNL Relativistic Heavy Ion Collider}'',
  \href{http://dx.doi.org/10.1103/PhysRevC.94.034908}{{\em Phys. Rev. C}
  {\bfseries 94} (2016) 034908},
  \href{http://arxiv.org/abs/1601.07052}{{\ttfamily arXiv:1601.07052
  [nucl-ex]}}.

\bibitem{Adam:2019wnb}
{\bfseries STAR} Collaboration, J.~Adam {\em et~al.}, ``{Beam energy dependence
  of (anti-)deuteron production in Au + Au collisions at the BNL Relativistic
  Heavy Ion Collider}'',
  \href{http://dx.doi.org/10.1103/PhysRevC.99.064905}{{\em Phys. Rev. C}
  {\bfseries 99} (2019) 064905},
  \href{http://arxiv.org/abs/1903.11778}{{\ttfamily arXiv:1903.11778
  [nucl-ex]}}.

\bibitem{deuteronFluctuations}
{\bfseries ALICE} Collaboration, S.~Acharya {\em et~al.}, ``{First Measurement
  of Antideuteron Number Fluctuations at Energies Available at the Large Hadron
  Collider}'', \href{http://dx.doi.org/10.1103/PhysRevLett.131.041901}{{\em
  Phys. Rev. Lett.} {\bfseries 131} (2023) 041901},
  \href{http://arxiv.org/abs/2204.10166}{{\ttfamily arXiv:2204.10166
  [nucl-ex]}}.

\bibitem{SHM5}
A.~Andronic, P.~Braun-Munzinger, J.~Stachel, and H.~St$\ddot{\mathrm{o}}$cker,
  ``{Production of light nuclei, hypernuclei and their antiparticles in
  relativistic nuclear collisions}'',
  \href{http://dx.doi.org/10.1016/j.physletb.2011.01.053}{{\em Phys. Lett.}
  {\bfseries B697} (2011) 203--207},
\href{http://arxiv.org/abs/1010.2995}{{\ttfamily arXiv:1010.2995 [nucl-th]}}.

\bibitem{SHM6}
J.~Cleymans, S.~Kabana, I.~Kraus, H.~Oeschler, K.~Redlich, and N.~Sharma,
  ``{Antimatter production in proton-proton and heavy-ion collisions at
  ultrarelativistic energies}'',
  \href{http://dx.doi.org/10.1103/PhysRevC.84.054916}{{\em Phys. Rev.}
  {\bfseries C84} (2011) 054916},
\href{http://arxiv.org/abs/1105.3719}{{\ttfamily arXiv:1105.3719 [hep-ph]}}.

\bibitem{SHM4}
F.~Becattini, E.~Grossi, M.~Bleicher, J.~Steinheimer, and R.~Stock,
  ``{Centrality dependence of hadronization and chemical freeze-out conditions
  in heavy ion collisions at $\sqrt{ s_{\rm NN}}$ = 2.76 TeV}'',
  \href{http://dx.doi.org/10.1103/PhysRevC.90.054907}{{\em Phys. Rev. C}
  {\bfseries 90} (2014) 054907},
\href{http://arxiv.org/abs/1405.0710}{{\ttfamily arXiv:1405.0710 [nucl-th]}}.

\bibitem{SHM2}
V.~Vovchenko and H.~St$\ddot{\mathrm{o}}$cker, ``{Examination of the
  sensitivity of the thermal fits to heavy-ion hadron yield data to the
  modeling of the eigenvolume interactions}'',
  \href{http://dx.doi.org/10.1103/PhysRevC.95.044904}{{\em Phys. Rev.}
  {\bfseries C95} (2017) 044904},
\href{http://arxiv.org/abs/1606.06218}{{\ttfamily arXiv:1606.06218 [hep-ph]}}.

\bibitem{SHM1}
A.~Andronic, P.~Braun-Munzinger, K.~Redlich, and J.~Stachel, ``{Decoding the
  phase structure of QCD via particle production at high energy}'',
  \href{http://dx.doi.org/10.1038/s41586-018-0491-6}{{\em Nature} {\bfseries
  561} (2018) 321--330},
\href{http://arxiv.org/abs/1710.09425}{{\ttfamily arXiv:1710.09425 [nucl-th]}}.

\bibitem{SHM3}
N.~Sharma, J.~Cleymans, B.~Hippolyte, and M.~Paradza, ``{A Comparison of pp,
  p--Pb, Pb--Pb Collisions in the Thermal Model: Multiplicity Dependence of
  Thermal Parameters}'',
  \href{http://dx.doi.org/10.1103/PhysRevC.99.044914}{{\em Phys. Rev.}
  {\bfseries C99} (2019) 044914},
\href{http://arxiv.org/abs/1811.00399}{{\ttfamily arXiv:1811.00399 [hep-ph]}}.

\bibitem{Sharma:2022poi}
N.~Sharma, L.~Kumar, P.~M. Lo, and K.~Redlich, ``{Light-nuclei production in pp
  and pA collisions in the baryon canonical ensemble approach}'',
  \href{http://dx.doi.org/10.1103/PhysRevC.107.054903}{{\em Phys. Rev. C}
  {\bfseries 107} (2023) 054903},
  \href{http://arxiv.org/abs/2210.15617}{{\ttfamily arXiv:2210.15617
  [nucl-th]}}.

\bibitem{Zhao:2018lyf}
W.~Zhao, L.~Zhu, H.~Zheng, C.~M. Ko, and H.~Song, ``{Spectra and flow of light
  nuclei in relativistic heavy ion collisions at energies available at the BNL
  Relativistic Heavy Ion Collider and at the CERN Large Hadron Collider}'',
  \href{http://dx.doi.org/10.1103/PhysRevC.98.054905}{{\em Phys. Rev. C}
  {\bfseries 98} (2018) 054905},
  \href{http://arxiv.org/abs/1807.02813}{{\ttfamily arXiv:1807.02813
  [nucl-th]}}.

\bibitem{Oliinychenko:2018ugs}
{\bfseries SMASH} Collaboration, D.~Oliinychenko, L.-G. Pang, H.~Elfner, and
  V.~Koch, ``{Microscopic study of deuteron production in Pb--Pb collisions at
  \snn = 2.76 TeV via hydrodynamics and a hadronic afterburner}'',
  \href{http://dx.doi.org/10.1103/PhysRevC.99.044907}{{\em Phys. Rev. C}
  {\bfseries 99} (2019) 044907},
  \href{http://arxiv.org/abs/1809.03071}{{\ttfamily arXiv:1809.03071
  [hep-ph]}}.

\bibitem{Coalescence3}
S.~T. Butler and C.~A. Pearson, ``{Deuterons from High-Energy Proton
  Bombardment of Matter}'',
\href{http://dx.doi.org/10.1103/PhysRev.129.836}{{\em Phys. Rev.} {\bfseries
  129} (1963) 836--842}.

\bibitem{Coalescence1}
J.~I. Kapusta, ``Mechanisms for deuteron production in relativistic nuclear
  collisions'', \href{http://dx.doi.org/10.1103/PhysRevC.21.1301}{{\em Phys.
  Rev.} {\bfseries C21} (1980) 1301--1310}.

\bibitem{SATO1981153}
H.~Sato and K.~Yazaki, ``On the coalescence model for high energy nuclear
  reactions'',
  \href{http://dx.doi.org/https://doi.org/10.1016/0370-2693(81)90976-X}{{\em
  Phys. Lett. B} {\bfseries 98} (1981) 153 -- 157}.

\bibitem{Csernai:1986qf}
L.~Csernai and J.~I. Kapusta, ``{Entropy and Cluster Production in Nuclear
  Collisions}'', \href{http://dx.doi.org/10.1016/0370-1573(86)90031-1}{{\em
  Phys. Rept.} {\bfseries 131} (1986) 223--318}.

\bibitem{Coalescence2}
R.~Scheibl and U.~W. Heinz, ``{Coalescence and flow in ultrarelativistic heavy
  ion collisions}'', \href{http://dx.doi.org/10.1103/PhysRevC.59.1585}{{\em
  Phys. Rev.} {\bfseries C59} (1999) 1585--1602},
\href{http://arxiv.org/abs/nucl-th/9809092}{{\ttfamily arXiv:nucl-th/9809092
  [nucl-th]}}.

\bibitem{Blum:2017qnn}
K.~Blum, K.~C.~Y. Ng, R.~Sato, and M.~Takimoto, ``{Cosmic rays, antihelium, and
  an old navy spotlight}'',
  \href{http://dx.doi.org/10.1103/PhysRevD.96.103021}{{\em Phys. Rev. D}
  {\bfseries 96} (2017) 103021},
  \href{http://arxiv.org/abs/1704.05431}{{\ttfamily arXiv:1704.05431
  [astro-ph.HE]}}.

\bibitem{Mahlein:2023fmx}
M.~Mahlein, L.~Barioglio, F.~Bellini, L.~Fabbietti, C.~Pinto, B.~Singh, and
  S.~Tripathy, ``{A realistic coalescence model for deuteron production}'',
  \href{http://dx.doi.org/10.1140/epjc/s10052-023-11972-3}{{\em Eur. Phys. J.
  C} {\bfseries 83} (2023) 804},
  \href{http://arxiv.org/abs/2302.12696}{{\ttfamily arXiv:2302.12696
  [hep-ex]}}.

\bibitem{Vovchenko:2019aoz}
V.~Vovchenko, K.~Gallmeister, J.~Schaffner-Bielich, and C.~Greiner,
  ``{Nucleosynthesis in heavy-ion collisions at the LHC via the Saha
  equation}'', \href{http://dx.doi.org/10.1016/j.physletb.2019.135131}{{\em
  Phys. Lett. B} {\bfseries 800} (2020) 135131},
  \href{http://arxiv.org/abs/1903.10024}{{\ttfamily arXiv:1903.10024
  [hep-ph]}}.

\bibitem{Cooper:1974mv}
F.~Cooper and G.~Frye, ``{Comment on the Single Particle Distribution in the
  Hydrodynamic and Statistical Thermodynamic Models of Multiparticle
  Production}'', \href{http://dx.doi.org/10.1103/PhysRevD.10.186}{{\em Phys.
  Rev. D} {\bfseries 10} (1974) 186}.

\bibitem{Bass:1998ca}
S.~A. Bass {\em et~al.}, ``{Microscopic models for ultrarelativistic heavy ion
  collisions}'', \href{http://dx.doi.org/10.1016/S0146-6410(98)00058-1}{{\em
  Prog. Part. Nucl. Phys.} {\bfseries 41} (1998) 255--369},
  \href{http://arxiv.org/abs/nucl-th/9803035}{{\ttfamily
  arXiv:nucl-th/9803035}}.

\bibitem{Bleicher:1999xi}
M.~Bleicher {\em et~al.}, ``{Relativistic hadron hadron collisions in the
  ultrarelativistic quantum molecular dynamics model}'',
  \href{http://dx.doi.org/10.1088/0954-3899/25/9/308}{{\em J. Phys. G}
  {\bfseries 25} (1999) 1859--1896},
  \href{http://arxiv.org/abs/hep-ph/9909407}{{\ttfamily arXiv:hep-ph/9909407}}.

\bibitem{Ropke:2008qk}
G.~Ropke, ``{Light nuclei quasiparticle energy shift in hot and dense nuclear
  matter}'', \href{http://dx.doi.org/10.1103/PhysRevC.79.014002}{{\em Phys.
  Rev. C} {\bfseries 79} (2009) 014002},
  \href{http://arxiv.org/abs/0810.4645}{{\ttfamily arXiv:0810.4645 [nucl-th]}}.

\bibitem{ALICE:2018ysd}
{\bfseries ALICE} Collaboration, S.~Acharya {\em et~al.}, ``{p-p, p-$\Lambda$
  and $\Lambda$-$\Lambda$ correlations studied via femtoscopy in pp reactions
  at $\sqrt{s}$ = 7 TeV}'',
  \href{http://dx.doi.org/10.1103/PhysRevC.99.024001}{{\em Phys. Rev. C}
  {\bfseries 99} (2019) 024001},
  \href{http://arxiv.org/abs/1805.12455}{{\ttfamily arXiv:1805.12455
  [nucl-ex]}}.

\bibitem{ALICE:2019hdt}
{\bfseries ALICE} Collaboration, S.~Acharya {\em et~al.}, ``{First Observation
  of an Attractive Interaction between a Proton and a Cascade Baryon}'',
  \href{http://dx.doi.org/10.1103/PhysRevLett.123.112002}{{\em Phys. Rev.
  Lett.} {\bfseries 123} (2019) 112002},
  \href{http://arxiv.org/abs/1904.12198}{{\ttfamily arXiv:1904.12198
  [nucl-ex]}}.

\bibitem{CoalescenceTheory}
F.~Bellini and A.~P. Kalweit, ``{Testing production scenarios for
  (anti-)(hyper-)nuclei and exotica at energies available at the CERN Large
  Hadron Collider}'', \href{http://dx.doi.org/10.1103/PhysRevC.99.054905}{{\em
  Phys. Rev. C} {\bfseries 99} (2019) 054905},
  \href{http://arxiv.org/abs/1807.05894}{{\ttfamily arXiv:1807.05894
  [hep-ph]}}.

\bibitem{STAR:2017sal}
{\bfseries STAR} Collaboration, L.~Adamczyk {\em et~al.}, ``{Bulk Properties of
  the Medium Produced in Relativistic Heavy-Ion Collisions from the Beam Energy
  Scan Program}'', \href{http://dx.doi.org/10.1103/PhysRevC.96.044904}{{\em
  Phys. Rev. C} {\bfseries 96} (2017) 044904},
  \href{http://arxiv.org/abs/1701.07065}{{\ttfamily arXiv:1701.07065
  [nucl-ex]}}.

\bibitem{ALICE:2019hno}
{\bfseries ALICE} Collaboration, S.~Acharya {\em et~al.}, ``{Production of
  charged pions, kaons, and (anti-)protons in Pb--Pb and inelastic pp
  collisions at $\sqrt {s_{\rm{NN}}}$ = 5.02 TeV}'',
  \href{http://dx.doi.org/10.1103/PhysRevC.101.044907}{{\em Phys. Rev. C}
  {\bfseries 101} (2020) 044907},
  \href{http://arxiv.org/abs/1910.07678}{{\ttfamily arXiv:1910.07678
  [nucl-ex]}}.

\bibitem{Xu:2017akx}
J.~Xu and C.~M. Ko, ``{Chemical freeze-out in relativistic heavy-ion
  collisions}'', \href{http://dx.doi.org/10.1016/j.physletb.2017.06.061}{{\em
  Phys. Lett. B} {\bfseries 772} (2017) 290--293},
  \href{http://arxiv.org/abs/1704.04934}{{\ttfamily arXiv:1704.04934
  [nucl-th]}}.

\bibitem{Xu:2018jff}
X.~Xu and R.~Rapp, ``{Production of light nuclei at thermal freezeout in
  ultrarelativistic heavy-ion collisions}'',
  \href{http://dx.doi.org/10.1140/epja/i2019-12757-7}{{\em Eur. Phys. J. A}
  {\bfseries 55} (2019) 68}, \href{http://arxiv.org/abs/1809.04024}{{\ttfamily
  arXiv:1809.04024 [nucl-th]}}.

\bibitem{Neidig:2021bal}
T.~Neidig, K.~Gallmeister, C.~Greiner, M.~Bleicher, and V.~Vovchenko,
  ``{Towards solving the puzzle of high temperature light (anti)-nuclei
  production in ultra-relativistic heavy ion collisions}'',
  \href{http://dx.doi.org/10.1016/j.physletb.2022.136891}{{\em Phys. Lett. B}
  {\bfseries 827} (2022) 136891},
  \href{http://arxiv.org/abs/2108.13151}{{\ttfamily arXiv:2108.13151
  [hep-ph]}}.

\bibitem{Sun:2022xjr}
K.-J. Sun, R.~Wang, C.~M. Ko, Y.-G. Ma, and C.~Shen, ``{Unveiling the dynamics
  of little-bang nucleosynthesis}'',
  \href{http://dx.doi.org/10.1038/s41467-024-45474-x}{{\em Nature Commun.}
  {\bfseries 15} (2024) 1074},
  \href{http://arxiv.org/abs/2207.12532}{{\ttfamily arXiv:2207.12532
  [nucl-th]}}.

\bibitem{FlowChargedParticlesLHC}
{\bfseries ALICE} Collaboration, B.~Abelev {\em et~al.}, ``{Elliptic flow of
  identified hadrons in Pb--Pb collisions at $ \sqrt{s_{\mathrm{NN}}}=2.76 $
  TeV}'', \href{http://dx.doi.org/10.1007/JHEP06(2015)190}{{\em JHEP}
  {\bfseries 06} (2015) 190},
\href{http://arxiv.org/abs/1405.4632}{{\ttfamily arXiv:1405.4632 [nucl-ex]}}.

\bibitem{flowIDhadronsXeXe}
{\bfseries ALICE} Collaboration, S.~Acharya {\em et~al.}, ``{Anisotropic flow
  of identified hadrons in Xe--Xe collisions at $ \sqrt{s_{\mathrm{NN}}} $ =
  5.44 TeV}'', \href{http://dx.doi.org/10.1007/JHEP10(2021)152}{{\em JHEP}
  {\bfseries 10} (2021) 152}, \href{http://arxiv.org/abs/2107.10592}{{\ttfamily
  arXiv:2107.10592 [nucl-ex]}}.

\bibitem{Molnar:2003ff}
D.~Molnar and S.~A. Voloshin, ``{Elliptic flow at large transverse momenta from
  quark coalescence}'',
  \href{http://dx.doi.org/10.1103/PhysRevLett.91.092301}{{\em Phys. Rev. Lett.}
  {\bfseries 91} (2003) 092301},
  \href{http://arxiv.org/abs/nucl-th/0302014}{{\ttfamily
  arXiv:nucl-th/0302014}}.

\bibitem{Yan:2006bx}
T.~Z. Yan {\em et~al.}, ``{Scaling of anisotropic flow and momentum-space
  densities for light particles in intermediate energy heavy ion collisions}'',
  \href{http://dx.doi.org/10.1016/j.physletb.2006.05.018}{{\em Phys. Lett. B}
  {\bfseries 638} (2006) 50--54},
  \href{http://arxiv.org/abs/nucl-th/0605022}{{\ttfamily
  arXiv:nucl-th/0605022}}.

\bibitem{ALICE_general_2}
{\bfseries ALICE} Collaboration, K.~Aamodt {\em et~al.}, ``{The ALICE
  experiment at the CERN LHC}'',
\href{http://dx.doi.org/10.1088/1748-0221/3/08/S08002}{{\em JINST} {\bfseries
  3} (2008) S08002}.

\bibitem{ALICEperformance}
{\bfseries ALICE} Collaboration, B.~Abelev {\em et~al.}, ``{Performance of the
  ALICE Experiment at the CERN LHC}'',
  \href{http://dx.doi.org/10.1142/S0217751X14300440}{{\em Int. J. Mod. Phys.}
  {\bfseries A29} (2014) 1430044},
\href{http://arxiv.org/abs/1402.4476}{{\ttfamily arXiv:1402.4476 [nucl-ex]}}.

\bibitem{T0_Cortese:2004aa}
{\bfseries ALICE} Collaboration, P.~Cortese {\em et~al.}, ``{ALICE forward
  detectors: FMD, T0 and V0: Technical Design Report}'', 2004.
\newblock \url{https://cds.cern.ch/record/781854}. Submitted on 10 Sep 2004.

\bibitem{ITS}
{\bfseries ALICE} Collaboration, K.~Aamodt {\em et~al.}, ``{Alignment of the
  ALICE Inner Tracking System with cosmic-ray tracks}'',
  \href{http://dx.doi.org/10.1088/1748-0221/5/03/P03003}{{\em JINST} {\bfseries
  5} (2010) P03003},
\href{http://arxiv.org/abs/1001.0502}{{\ttfamily arXiv:1001.0502
  [physics.ins-det]}}.

\bibitem{TPC}
{\bfseries ALICE} Collaboration, J.~Alme {\em et~al.}, ``{The ALICE TPC, a
  large 3-dimensional tracking device with fast readout for ultra-high
  multiplicity events}'',
  \href{http://dx.doi.org/10.1016/j.nima.2010.04.042}{{\em Nucl. Instrum.
  Meth.} {\bfseries A622} (2010) 316--367},
\href{http://arxiv.org/abs/1001.1950}{{\ttfamily arXiv:1001.1950
  [physics.ins-det]}}.

\bibitem{TOF}
{\bfseries ALICE} Collaboration, A.~Akindinov {\em et~al.}, ``{Performance of
  the ALICE Time-Of-Flight detector at the LHC}'',
\href{http://dx.doi.org/10.1140/epjp/i2013-13044-x}{{\em Eur. Phys. J. Plus}
  {\bfseries 128} (2013) 44}.

\bibitem{Carnesecchi:2018oss}
{\bfseries ALICE} Collaboration, F.~Carnesecchi, ``{Performance of the ALICE
  Time-Of-Flight detector at the LHC}'',
  \href{http://dx.doi.org/10.1088/1748-0221/14/06/C06023}{{\em JINST}
  {\bfseries 14} (2019) C06023},
  \href{http://arxiv.org/abs/1806.03825}{{\ttfamily arXiv:1806.03825
  [physics.ins-det]}}.

\bibitem{Adam:2016ilk}
{\bfseries ALICE} Collaboration, J.~Adam {\em et~al.}, ``{Determination of the
  event collision time with the ALICE detector at the LHC}'',
  \href{http://dx.doi.org/10.1140/epjp/i2017-11279-1}{{\em Eur. Phys. J. Plus}
  {\bfseries 132} (2017) 99}, \href{http://arxiv.org/abs/1610.03055}{{\ttfamily
  arXiv:1610.03055 [physics.ins-det]}}.

\bibitem{VZEROPerformance}
{\bfseries ALICE} Collaboration, E.~Abbas {\em et~al.}, ``{Performance of the
  ALICE VZERO system}'',
  \href{http://dx.doi.org/10.1088/1748-0221/8/10/P10016}{{\em JINST} {\bfseries
  8} (2013) P10016},
\href{http://arxiv.org/abs/1306.3130}{{\ttfamily arXiv:1306.3130 [nucl-ex]}}.

\bibitem{Schaumann:2018qat}
M.~Schaumann {\em et~al.},
  \href{http://dx.doi.org/doi:10.18429/JACoW-IPAC2018-MOPMF039}{``{F}irst
  {X}enon{-X}enon {C}ollisions in the {LHC}'',} in {\em Proc. 9th International
  Particle Accelerator Conference (IPAC'18), Vancouver, BC, Canada, April
  29-May 4, 2018}, no.~9 in International Particle Accelerator Conference,
  pp.~180--183.
\newblock JACoW Publishing, Geneva, Switzerland, June, 2018.

\bibitem{ALICE-PUBLIC-2018-003}
{\bfseries ALICE} Collaboration, ``{Centrality determination using the Glauber
  model in Xe--Xe collisions at $\sqrt{s_{\rm NN}} = 5.44$ TeV}'',.
  \url{https://cds.cern.ch/record/2315401}.

\bibitem{JPsiFlow}
{\bfseries ALICE} Collaboration, S.~Acharya {\em et~al.}, ``$\mathrm{J}/\psi$
  elliptic and triangular flow in $\mathrm{Pb--Pb}$ collisions at
  $\sqrt{s_{\mathrm{NN}}} = 5.02 \ \mathrm{TeV}$'',
  \href{http://dx.doi.org/10.1007/JHEP10(2020)141}{{\em JHEP} {\bfseries 10}
  (2020) 141}, \href{http://arxiv.org/abs/2005.14518}{{\ttfamily
  arXiv:2005.14518 [nucl-ex]}}.

\bibitem{BetheBloch}
L.~C. Yuan and C.-S. Wu, eds., {\em {Fundamental Principles and Methods of
  Particle Detection}}, vol.~5 of {\em Methods in Experimental Physics},
  \href{http://dx.doi.org/https://doi.org/10.1016/S0076-695X(08)60442-1}{pp.~1--288}.
\newblock Academic Press, 1961.

\bibitem{HIJING}
X.-N. Wang and M.~Gyulassy, ``{HIJING: A Monte Carlo model for multiple jet
  production in $\mathrm{pp}$, $\mathrm{pA}$, and $\mathrm{AA}$ collisions}'',
  \href{http://dx.doi.org/10.1103/PhysRevD.44.3501}{{\em Phys. Rev.} {\bfseries
  D44} (1991) 3501--3516}.

\bibitem{GEANT4}
S.~Agostinelli {\em et~al.}, ``$\mathrm{GEANT4}$ -- a simulation toolkit'',
  \href{http://dx.doi.org/https://doi.org/10.1016/S0168-9002(03)01368-8}{{\em
  Nucl. Instrum. Meth. A} {\bfseries 506} (2003) 250--303}.

\bibitem{BlastWave1}
E.~Schnedermann, J.~Sollfrank, and U.~W. Heinz, ``{Thermal phenomenology of
  hadrons from 200 A GeV S+S collisions}'',
  \href{http://dx.doi.org/10.1103/PhysRevC.48.2462}{{\em Phys. Rev.} {\bfseries
  C48} (1993) 2462--2475},
\href{http://arxiv.org/abs/nucl-th/9307020}{{\ttfamily arXiv:nucl-th/9307020
  [nucl-th]}}.

\bibitem{Voloshin:2008dg}
S.~A. Voloshin, A.~M. Poskanzer, and R.~Snellings, ``{Collective phenomena in
  non-central nuclear collisions}'',
  \href{http://dx.doi.org/10.1007/978-3-642-01539-7_10}{{\em Landolt-Bornstein}
  {\bfseries 23} (2010) 293--333},
\href{http://arxiv.org/abs/0809.2949}{{\ttfamily arXiv:0809.2949 [nucl-ex]}}.

\bibitem{Ollitrault:2009ie}
J.-Y. Ollitrault, A.~M. Poskanzer, and S.~A. Voloshin, ``{Effect of flow
  fluctuations and nonflow on elliptic flow methods}'',
  \href{http://dx.doi.org/10.1103/PhysRevC.80.014904}{{\em Phys. Rev.}
  {\bfseries C80} (2009) 014904},
\href{http://arxiv.org/abs/0904.2315}{{\ttfamily arXiv:0904.2315 [nucl-ex]}}.

\bibitem{Alver:2010gr}
B.~Alver and G.~Roland, ``{Collision geometry fluctuations and triangular flow
  in heavy-ion collisions}'',
  \href{http://dx.doi.org/10.1103/PhysRevC.82.039903,
  10.1103/PhysRevC.81.054905}{{\em Phys. Rev.} {\bfseries C81} (2010) 054905},
  \href{http://arxiv.org/abs/1003.0194}{{\ttfamily arXiv:1003.0194 [nucl-th]}}.
[Erratum: $Phys. Rev.$ \textbf{C82} (2010) 039903].

\bibitem{Qiu:2011iv}
Z.~Qiu and U.~W. Heinz, ``{Event-by-event shape and flow fluctuations of
  relativistic heavy-ion collision fireballs}'',
  \href{http://dx.doi.org/10.1103/PhysRevC.84.024911}{{\em Phys. Rev.}
  {\bfseries C84} (2011) 024911},
\href{http://arxiv.org/abs/1104.0650}{{\ttfamily arXiv:1104.0650 [nucl-th]}}.

\bibitem{Adler:2002pu}
{\bfseries STAR} Collaboration, C.~Adler {\em et~al.}, ``{Elliptic flow from
  two and four particle correlations in Au+Au collisions at \snn = 130 GeV}'',
  \href{http://dx.doi.org/10.1103/PhysRevC.66.034904}{{\em Phys. Rev.}
  {\bfseries C66} (2002) 034904},
\href{http://arxiv.org/abs/nucl-ex/0206001}{{\ttfamily arXiv:nucl-ex/0206001
  [nucl-ex]}}.

\bibitem{protonV2}
{\bfseries ALICE} Collaboration, S.~Acharya {\em et~al.}, ``{Anisotropic flow
  of identified particles in Pb--Pb collisions at $\sqrt{s_{\mathrm{NN}}}=5.02$
  TeV}'', \href{http://dx.doi.org/10.1007/JHEP09(2018)006}{{\em JHEP}
  {\bfseries 09} (2018) 006},
\href{http://arxiv.org/abs/1805.04390}{{\ttfamily arXiv:1805.04390 [nucl-ex]}}.

\bibitem{CODATA}
P.~J. Mohr, D.~B. Newell, and B.~N. Taylor, ``{CODATA Recommended Values of the
  Fundamental Physical Constants: 2014}'',
  \href{http://dx.doi.org/10.1103/RevModPhys.88.035009}{{\em Rev. Mod. Phys.}
  {\bfseries 88} (2016) 035009},
  \href{http://arxiv.org/abs/1507.07956}{{\ttfamily arXiv:1507.07956
  [physics.atom-ph]}}.

\bibitem{absorption1}
J.~Jaros {\em et~al.}, ``{Nucleus-nucleus total cross sections for light nuclei
  at 1.55 and 2.89 GeV/$c$ per nucleon}'',
  \href{http://dx.doi.org/10.1103/PhysRevC.18.2273}{{\em Phys. Rev. C}
  {\bfseries 18} (1978) 2273--2292}.

\bibitem{absorption2}
A.~Auce {\em et~al.}, ``{Reaction cross sections for 38, 65, and 97 MeV
  deuterons on targets from $^{9}\mathrm{Be}$ to $^{208}\mathrm{Pb}$}'',
  \href{http://dx.doi.org/10.1103/PhysRevC.53.2919}{{\em Phys. Rev. C}
  {\bfseries 53} (1996) 2919--2925}.

\bibitem{absorption3}
F.~Binon {\em et~al.}, ``{Absorption cross-sections of 25 GeV/$c$ antideuterons
  in Li, C, Al, Cu and Pb}'',
  \href{http://dx.doi.org/https://doi.org/10.1016/0370-2693(70)90112-7}{{\em
  Phys. Lett. B} {\bfseries 31} (1970) 230 -- 232}.

\bibitem{absorption4}
S.~Denisov {\em et~al.}, ``{Measurements of antideuteron absorption and
  stripping cross sections at the momentum 13.3 GeV/$c$}'',
  \href{http://dx.doi.org/https://doi.org/10.1016/0550-3213(71)90229-X}{{\em
  Nucl. Phys. B} {\bfseries 31} (1971) 253 -- 260}.

\bibitem{antideuteronInelCS}
{\bfseries ALICE} Collaboration, S.~Acharya {\em et~al.}, ``{Measurement of the
  low-energy antideuteron inelastic cross section}'',
  \href{http://dx.doi.org/10.1103/PhysRevLett.125.162001}{{\em Phys. Rev.
  Lett.} {\bfseries 125} (2020) 162001},
  \href{http://arxiv.org/abs/2005.11122}{{\ttfamily arXiv:2005.11122
  [nucl-ex]}}.

\bibitem{antiHe3Absorption}
{\bfseries ALICE} Collaboration, S.~Acharya {\em et~al.}, ``{Measurement of
  anti-$^{3}$He nuclei absorption in matter and impact on their propagation in
  the Galaxy}'', \href{http://dx.doi.org/10.1038/s41567-022-01804-8}{{\em
  Nature Phys.} {\bfseries 19} (2023) 61--71},
  \href{http://arxiv.org/abs/2202.01549}{{\ttfamily arXiv:2202.01549
  [nucl-ex]}}.

\bibitem{ALICE:2023qyl}
{\bfseries ALICE} Collaboration, S.~Acharya {\em et~al.}, ``{Measurement of
  (anti)alpha production in central Pb-Pb collisions at $\sqrt{s_{\rm NN}}$ =
  5.02 TeV}'', \href{http://dx.doi.org/10.1016/j.physletb.2024.138943}{{\em
  Phys. Lett. B} {\bfseries 858} (2024) 138943},
  \href{http://arxiv.org/abs/2311.11758}{{\ttfamily arXiv:2311.11758
  [nucl-ex]}}.

\bibitem{BHALLA1981446}
{\bfseries JAIPUR-JAMMU-LUND} Collaboration, K.~B. Bhalla {\em et~al.},
  ``{Relativistic $\alpha$ Particles Emitted in Fe Emulsion Interactions at
  1.7A GeV}'', \href{http://dx.doi.org/10.1016/0375-9474(81)90659-X}{{\em Nucl.
  Phys. A} {\bfseries 367} (1981) 446--458}.

\bibitem{Adler:2003cb}
{\bfseries PHENIX} Collaboration, S.~S. Adler {\em et~al.}, ``{Identified
  charged particle spectra and yields in Au+Au collisions at \snn = 200 GeV}'',
  \href{http://dx.doi.org/10.1103/PhysRevC.69.034909}{{\em Phys. Rev. C}
  {\bfseries 69} (2004) 034909},
  \href{http://arxiv.org/abs/nucl-ex/0307022}{{\ttfamily
  arXiv:nucl-ex/0307022}}.

\bibitem{Tsallis:1987eu}
C.~Tsallis, ``{Possible Generalization of Boltzmann-Gibbs Statistics}'',
  \href{http://dx.doi.org/10.1007/BF01016429}{{\em J. Statist. Phys.}
  {\bfseries 52} (1988) 479--487}.

\bibitem{ALICE:2018cpu}
{\bfseries ALICE} Collaboration, S.~Acharya {\em et~al.}, ``{Centrality and
  pseudorapidity dependence of the charged-particle multiplicity density in
  Xe\textendash{}Xe collisions at $\sqrt{s_{\rm NN}}$~=~5.44~TeV}'',
  \href{http://dx.doi.org/10.1016/j.physletb.2018.12.048}{{\em Phys. Lett. B}
  {\bfseries 790} (2019) 35--48},
  \href{http://arxiv.org/abs/1805.04432}{{\ttfamily arXiv:1805.04432
  [nucl-ex]}}.

\bibitem{coalescenceSmallSystems}
K.-J. Sun, C.~M. Ko, and B.~D$\ddot{\mathrm{o}}$nigus, ``{Suppression of light
  nuclei production in collisions of small systems at the Large Hadron
  Collider}'', \href{http://dx.doi.org/10.1016/j.physletb.2019.03.033}{{\em
  Phys. Lett.} {\bfseries B792} (2019) 132--137},
\href{http://arxiv.org/abs/1812.05175}{{\ttfamily arXiv:1812.05175 [nucl-th]}}.

\bibitem{Vovchenko:2018fiy}
V.~Vovchenko, B.~D\"onigus, and H.~Stoecker, ``{Multiplicity dependence of
  light nuclei production at LHC energies in the canonical statistical
  model}'', \href{http://dx.doi.org/10.1016/j.physletb.2018.08.041}{{\em Phys.
  Lett. B} {\bfseries 785} (2018) 171--174},
  \href{http://arxiv.org/abs/1808.05245}{{\ttfamily arXiv:1808.05245
  [hep-ph]}}.

\bibitem{Vovchenko:2022xil}
V.~Vovchenko and V.~Koch, ``{Centrality dependence of proton and light nuclei
  yields as a consequence of baryon annihilation in the hadronic phase}'',
  \href{http://dx.doi.org/10.1016/j.physletb.2022.137577}{{\em Phys. Lett. B}
  {\bfseries 835} (2022) 137577},
  \href{http://arxiv.org/abs/2210.15641}{{\ttfamily arXiv:2210.15641
  [nucl-th]}}.

\bibitem{ALICE:2019nbs}
{\bfseries ALICE} Collaboration, S.~Acharya {\em et~al.}, ``{Global baryon
  number conservation encoded in net-proton fluctuations measured in Pb--Pb
  collisions at $\sqrt{s_{\rm NN}}$ = 2.76 TeV}'',
  \href{http://dx.doi.org/10.1016/j.physletb.2020.135564}{{\em Phys. Lett. B}
  {\bfseries 807} (2020) 135564},
  \href{http://arxiv.org/abs/1910.14396}{{\ttfamily arXiv:1910.14396
  [nucl-ex]}}.

\bibitem{Vovchenko:2020kwg}
V.~Vovchenko and V.~Koch, ``{Particlization of an interacting hadron resonance
  gas with global conservation laws for event-by-event fluctuations in
  heavy-ion collisions}'',
  \href{http://dx.doi.org/10.1103/PhysRevC.103.044903}{{\em Phys. Rev. C}
  {\bfseries 103} (2021) 044903},
  \href{http://arxiv.org/abs/2012.09954}{{\ttfamily arXiv:2012.09954
  [hep-ph]}}.

\bibitem{ALICE:2022xpf}
\textbf{ALICE} Collaboration, ``{Closing in on critical net-baryon fluctuations
  at LHC energies: Cumulants up to third order in Pb\textendash{}Pb
  collisions}'', \href{http://dx.doi.org/10.1016/j.physletb.2022.137545}{{\em
  Phys. Lett. B} {\bfseries 844} (2023) 137545},
  \href{http://arxiv.org/abs/2206.03343}{{\ttfamily arXiv:2206.03343
  [nucl-ex]}}.

\bibitem{vanillaCSM}
V.~Vovchenko, B.~D\"onigus, and H.~Stoecker, ``{Canonical statistical model
  analysis of p-p, p-Pb, and Pb--Pb collisions at energies available at the
  CERN Large Hadron Collider}'',
  \href{http://dx.doi.org/10.1103/PhysRevC.100.054906}{{\em Phys. Rev. C}
  {\bfseries 100} (2019) 054906},
  \href{http://arxiv.org/abs/1906.03145}{{\ttfamily arXiv:1906.03145
  [hep-ph]}}.

\bibitem{ALICE:2021lsv}
{\bfseries ALICE} Collaboration, S.~Acharya {\em et~al.}, ``{Production of
  pions, kaons, (anti-)protons and $\phi $ mesons in Xe\textendash{}Xe
  collisions at $\sqrt{s_{\mathrm{NN}}}$~=~5.44 TeV}'',
  \href{http://dx.doi.org/10.1140/epjc/s10052-021-09304-4}{{\em Eur. Phys. J.
  C} {\bfseries 81} (2021) 584},
  \href{http://arxiv.org/abs/2101.03100}{{\ttfamily arXiv:2101.03100
  [nucl-ex]}}.

\bibitem{STAR:2022hbp}
{\bfseries STAR} Collaboration, M.~Abdulhamid {\em et~al.}, ``{Beam Energy
  Dependence of Triton Production and Yield Ratio ($\mathrm{N}_t \times
  \mathrm{N}_p/\mathrm{N}_d^2$) in Au+Au Collisions at RHIC}'',
  \href{http://dx.doi.org/10.1103/PhysRevLett.130.202301}{{\em Phys. Rev.
  Lett.} {\bfseries 130} (2023) 202301},
  \href{http://arxiv.org/abs/2209.08058}{{\ttfamily arXiv:2209.08058
  [nucl-ex]}}.

\bibitem{ALICE:2020ibs}
{\bfseries ALICE} Collaboration, S.~Acharya {\em et~al.}, ``{Search for a
  common baryon source in high-multiplicity pp collisions at the LHC}'',
  \href{http://dx.doi.org/10.1016/j.physletb.2020.135849}{{\em Phys. Lett. B}
  {\bfseries 811} (2020) 135849},
  \href{http://arxiv.org/abs/2004.08018}{{\ttfamily arXiv:2004.08018
  [nucl-ex]}}.

\bibitem{relativisticHydro}
S.~Jeon and U.~Heinz, ``{Introduction to Hydrodynamics}'',
  \href{http://dx.doi.org/10.1142/S0218301315300106}{{\em Int. J. Mod. Phys.}
  {\bfseries E24} (2015) 1530010},
\href{http://arxiv.org/abs/1503.03931}{{\ttfamily arXiv:1503.03931 [hep-ph]}}.

\bibitem{Hirano:2007ei}
T.~Hirano, U.~W. Heinz, D.~Kharzeev, R.~Lacey, and Y.~Nara, ``{Mass ordering of
  differential elliptic flow and its violation for phi mesons}'',
  \href{http://dx.doi.org/10.1103/PhysRevC.77.044909}{{\em Phys. Rev. C}
  {\bfseries 77} (2008) 044909},
  \href{http://arxiv.org/abs/0710.5795}{{\ttfamily arXiv:0710.5795 [nucl-th]}}.

\bibitem{BlastWave2}
{\bfseries STAR} Collaboration, C.~Adler {\em et~al.}, ``{Identified particle
  elliptic flow in Au + Au collisions at $\sqrt{s_{\mathrm{NN}}}$ = 130 GeV}'',
  \href{http://dx.doi.org/10.1103/PhysRevLett.87.182301}{{\em Phys. Rev. Lett.}
  {\bfseries 87} (2001) 182301},
\href{http://arxiv.org/abs/nucl-ex/0107003}{{\ttfamily arXiv:nucl-ex/0107003
  [nucl-ex]}}.

\bibitem{Vovchenko:2019pjl}
V.~Vovchenko and H.~Stoecker, ``{Thermal-FIST: A package for heavy-ion
  collisions and hadronic equation of state}'',
  \href{http://dx.doi.org/10.1016/j.cpc.2019.06.024}{{\em Comput. Phys.
  Commun.} {\bfseries 244} (2019) 295--310},
  \href{http://arxiv.org/abs/1901.05249}{{\ttfamily arXiv:1901.05249
  [nucl-th]}}.

\bibitem{Vovchenko:2018fmh}
V.~Vovchenko, M.~I. Gorenstein, and H.~Stoecker, ``{Finite resonance widths
  influence the thermal-model description of hadron yields}'',
  \href{http://dx.doi.org/10.1103/PhysRevC.98.034906}{{\em Phys. Rev. C}
  {\bfseries 98} (2018) 034906},
  \href{http://arxiv.org/abs/1807.02079}{{\ttfamily arXiv:1807.02079
  [nucl-th]}}.

\bibitem{Andronic:2008gu}
A.~Andronic, P.~Braun-Munzinger, and J.~Stachel, ``{Thermal hadron production
  in relativistic nuclear collisions: The Hadron mass spectrum, the horn, and
  the QCD phase transition}'',
  \href{http://dx.doi.org/10.1016/j.physletb.2009.06.021}{{\em Phys. Lett. B}
  {\bfseries 673} (2009) 142--145},
  \href{http://arxiv.org/abs/0812.1186}{{\ttfamily arXiv:0812.1186 [nucl-th]}}.
  [Erratum: $Phys. Lett. B$ \textbf{678} (2009) 516].

\bibitem{Andronic:2018qqt}
A.~Andronic, P.~Braun-Munzinger, B.~Friman, P.~M. Lo, K.~Redlich, and
  J.~Stachel, ``{The thermal proton yield anomaly in Pb--Pb collisions at the
  LHC and its resolution}'',
  \href{http://dx.doi.org/10.1016/j.physletb.2019.03.052}{{\em Phys. Lett. B}
  {\bfseries 792} (2019) 304--309},
  \href{http://arxiv.org/abs/1808.03102}{{\ttfamily arXiv:1808.03102
  [hep-ph]}}.

\bibitem{Dashen:1969ep}
R.~Dashen, S.-K. Ma, and H.~J. Bernstein, ``{S Matrix formulation of
  statistical mechanics}'',
  \href{http://dx.doi.org/10.1103/PhysRev.187.345}{{\em Phys. Rev.} {\bfseries
  187} (1969) 345--370}.

\bibitem{Venugopalan:1992hy}
R.~Venugopalan and M.~Prakash, ``{Thermal properties of interacting hadrons}'',
  \href{http://dx.doi.org/10.1016/0375-9474(92)90005-5}{{\em Nucl. Phys. A}
  {\bfseries 546} (1992) 718--760}.

\bibitem{Weinhold:1997ig}
W.~Weinhold, B.~Friman, and W.~Norenberg, ``{Thermodynamics of Delta
  resonances}'', \href{http://dx.doi.org/10.1016/S0370-2693(98)00639-X}{{\em
  Phys. Lett. B} {\bfseries 433} (1998) 236--242},
  \href{http://arxiv.org/abs/nucl-th/9710014}{{\ttfamily
  arXiv:nucl-th/9710014}}.

\bibitem{Huovinen:2016xxq}
P.~Huovinen, P.~M. Lo, M.~Marczenko, K.~Morita, K.~Redlich, and C.~Sasaki,
  ``{Effects of \ensuremath{\rho}-meson width on pion distributions in
  heavy-ion collisions}'',
  \href{http://dx.doi.org/10.1016/j.physletb.2017.03.060}{{\em Phys. Lett. B}
  {\bfseries 769} (2017) 509--512},
  \href{http://arxiv.org/abs/1608.06817}{{\ttfamily arXiv:1608.06817
  [hep-ph]}}.

\bibitem{Lo:2017sde}
P.~M. Lo, ``{S-matrix formulation of thermodynamics with N-body scatterings}'',
  \href{http://dx.doi.org/10.1140/epjc/s10052-017-5106-0}{{\em Eur. Phys. J. C}
  {\bfseries 77} (2017) 533}, \href{http://arxiv.org/abs/1707.04490}{{\ttfamily
  arXiv:1707.04490 [hep-ph]}}.

\bibitem{Breit:1936zzb}
G.~Breit and E.~Wigner, ``{Capture of Slow Neutrons}'',
  \href{http://dx.doi.org/10.1103/PhysRev.49.519}{{\em Phys. Rev.} {\bfseries
  49} (1936) 519--531}.

\bibitem{HotQCD:2018pds}
{\bfseries HotQCD} Collaboration, A.~Bazavov {\em et~al.}, ``{Chiral crossover
  in QCD at zero and non-zero chemical potentials}'',
  \href{http://dx.doi.org/10.1016/j.physletb.2019.05.013}{{\em Phys. Lett. B}
  {\bfseries 795} (2019) 15--21},
  \href{http://arxiv.org/abs/1812.08235}{{\ttfamily arXiv:1812.08235
  [hep-lat]}}.

\end{thebibliography}\endgroup
\WarningFilter{bibliography}{empty author}

\newpage
\appendix

%
%

\section{The ALICE Collaboration}
\label{app:collab}
\begin{flushleft} 
\small

S.~Acharya\,\orcidlink{0000-0002-9213-5329}\,$^{\rm 127}$, 
D.~Adamov\'{a}\,\orcidlink{0000-0002-0504-7428}\,$^{\rm 86}$, 
A.~Agarwal$^{\rm 135}$, 
G.~Aglieri Rinella\,\orcidlink{0000-0002-9611-3696}\,$^{\rm 32}$, 
L.~Aglietta\,\orcidlink{0009-0003-0763-6802}\,$^{\rm 24}$, 
M.~Agnello\,\orcidlink{0000-0002-0760-5075}\,$^{\rm 29}$, 
N.~Agrawal\,\orcidlink{0000-0003-0348-9836}\,$^{\rm 25}$, 
Z.~Ahammed\,\orcidlink{0000-0001-5241-7412}\,$^{\rm 135}$, 
S.~Ahmad\,\orcidlink{0000-0003-0497-5705}\,$^{\rm 15}$, 
S.U.~Ahn\,\orcidlink{0000-0001-8847-489X}\,$^{\rm 71}$
% ...

\section*{Affiliation Notes}

$^{\rm I}$ Deceased\\
$^{\rm II}$ Also at: Max-Planck-Institut fur Physik, Munich, Germany\\
$^{\rm III}$ Also at: Italian National Agency for New Technologies, Energy and Sustainable Economic Development (ENEA), Bologna, Italy\\
$^{\rm IV}$ Also at: Dipartimento DET del Politecnico di Torino, Turin, Italy\\
$^{\rm V}$ Also at: Yildiz Technical University, Istanbul, T\"{u}rkiye\\
$^{\rm VI}$ Also at: Department of Applied Physics, Aligarh Muslim University, Aligarh, India\\
$^{\rm VII}$ Also at: Institute of Theoretical Physics, University of Wroclaw, Poland\\
$^{\rm VIII}$ Also at: Indian Institute of Science Education and Research (IISER) Berhampur, Odisha, India\\
$^{\rm IX}$ Also at: An institution covered by a cooperation agreement with CERN\\

\section*{Collaboration Institutes}

$^{1}$ A.I. Alikhanyan National Science Laboratory (Yerevan Physics Institute) Foundation, Yerevan, Armenia\\
$^{2}$ AGH University of Krakow, Cracow, Poland\\
$^{3}$ Bogolyubov Institute for Theoretical Physics, National Academy of Sciences of Ukraine, Kiev, Ukraine\\
$^{4}$ Bose Institute, Department of Physics  and Centre for Astroparticle Physics and Space Science (CAPSS), Kolkata, India\\
$^{5}$ California Polytechnic State University, San Luis Obispo, California, United States\\
$^{6}$ Central China Normal University, Wuhan, China\\
$^{7}$ Centro de Aplicaciones Tecnol\'{o}gicas y Desarrollo Nuclear (CEADEN), Havana, Cuba\\
$^{8}$ Centro de Investigaci\'{o}n y de Estudios Avanzados (CINVESTAV), Mexico City and M\'{e}rida, Mexico\\
$^{9}$ Chicago State University, Chicago, Illinois, United States\\
$^{10}$ China Institute of Atomic Energy, Beijing, China\\
$^{11}$ China University of Geosciences, Wuhan, China\\
$^{12}$ Chungbuk National University, Cheongju, Republic of Korea\\
$^{13}$ Comenius University Bratislava, Faculty of Mathematics, Physics and Informatics, Bratislava, Slovak Republic\\
$^{14}$ Creighton University, Omaha, Nebraska, United States\\
$^{15}$ Department of Physics, Aligarh Muslim University, Aligarh, India\\
$^{16}$ Department of Physics, Pusan National University, Pusan, Republic of Korea\\
$^{17}$ Department of Physics, Sejong University, Seoul, Republic of Korea\\
$^{18}$ Department of Physics, University of California, Berkeley, California, United States\\
$^{19}$ Department of Physics, University of Oslo, Oslo, Norway\\
$^{20}$ Department of Physics and Technology, University of Bergen, Bergen, Norway\\
$^{21}$ Dipartimento di Fisica, Universit\`{a} di Pavia, Pavia, Italy\\
$^{22}$ Dipartimento di Fisica dell'Universit\`{a} and Sezione INFN, Cagliari, Italy\\
$^{23}$ Dipartimento di Fisica dell'Universit\`{a} and Sezione INFN, Trieste, Italy\\
$^{24}$ Dipartimento di Fisica dell'Universit\`{a} and Sezione INFN, Turin, Italy\\
$^{25}$ Dipartimento di Fisica e Astronomia dell'Universit\`{a} and Sezione INFN, Bologna, Italy\\
$^{26}$ Dipartimento di Fisica e Astronomia dell'Universit\`{a} and Sezione INFN, Catania, Italy\\
$^{27}$ Dipartimento di Fisica e Astronomia dell'Universit\`{a} and Sezione INFN, Padova, Italy\\
$^{28}$ Dipartimento di Fisica `E.R.~Caianiello' dell'Universit\`{a} and Gruppo Collegato INFN, Salerno, Italy\\
$^{29}$ Dipartimento DISAT del Politecnico and Sezione INFN, Turin, Italy\\
$^{30}$ Dipartimento di Scienze MIFT, Universit\`{a} di Messina, Messina, Italy\\
$^{31}$ Dipartimento Interateneo di Fisica `M.~Merlin' and Sezione INFN, Bari, Italy\\
$^{32}$ European Organization for Nuclear Research (CERN), Geneva, Switzerland\\
$^{33}$ Faculty of Electrical Engineering, Mechanical Engineering and Naval Architecture, University of Split, Split, Croatia\\
$^{34}$ Faculty of Engineering and Science, Western Norway University of Applied Sciences, Bergen, Norway\\
$^{35}$ Faculty of Nuclear Sciences and Physical Engineering, Czech Technical University in Prague, Prague, Czech Republic\\
$^{36}$ Faculty of Physics, Sofia University, Sofia, Bulgaria\\
$^{37}$ Faculty of Science, P.J.~\v{S}af\'{a}rik University, Ko\v{s}ice, Slovak Republic\\
$^{38}$ Frankfurt Institute for Advanced Studies, Johann Wolfgang Goethe-Universit\"{a}t Frankfurt, Frankfurt, Germany\\
$^{39}$ Fudan University, Shanghai, China\\
$^{40}$ Gangneung-Wonju National University, Gangneung, Republic of Korea\\
$^{41}$ Gauhati University, Department of Physics, Guwahati, India\\
$^{42}$ Helmholtz-Institut f\"{u}r Strahlen- und Kernphysik, Rheinische Friedrich-Wilhelms-Universit\"{a}t Bonn, Bonn, Germany\\
$^{43}$ Helsinki Institute of Physics (HIP), Helsinki, Finland\\
$^{44}$ High Energy Physics Group,  Universidad Aut\'{o}noma de Puebla, Puebla, Mexico\\
$^{45}$ Horia Hulubei National Institute of Physics and Nuclear Engineering, Bucharest, Romania\\
$^{46}$ HUN-REN Wigner Research Centre for Physics, Budapest, Hungary\\
$^{47}$ Indian Institute of Technology Bombay (IIT), Mumbai, India\\
$^{48}$ Indian Institute of Technology Indore, Indore, India\\
$^{49}$ INFN, Laboratori Nazionali di Frascati, Frascati, Italy\\
$^{50}$ INFN, Sezione di Bari, Bari, Italy\\
$^{51}$ INFN, Sezione di Bologna, Bologna, Italy\\
$^{52}$ INFN, Sezione di Cagliari, Cagliari, Italy\\
$^{53}$ INFN, Sezione di Catania, Catania, Italy\\
$^{54}$ INFN, Sezione di Padova, Padova, Italy\\
$^{55}$ INFN, Sezione di Pavia, Pavia, Italy\\
$^{56}$ INFN, Sezione di Torino, Turin, Italy\\
$^{57}$ INFN, Sezione di Trieste, Trieste, Italy\\
$^{58}$ Inha University, Incheon, Republic of Korea\\
$^{59}$ Institute for Gravitational and Subatomic Physics (GRASP), Utrecht University/Nikhef, Utrecht, Netherlands\\
$^{60}$ Institute of Experimental Physics, Slovak Academy of Sciences, Ko\v{s}ice, Slovak Republic\\
$^{61}$ Institute of Physics, Homi Bhabha National Institute, Bhubaneswar, India\\
$^{62}$ Institute of Physics of the Czech Academy of Sciences, Prague, Czech Republic\\
$^{63}$ Institute of Space Science (ISS), Bucharest, Romania\\
$^{64}$ Institut f\"{u}r Kernphysik, Johann Wolfgang Goethe-Universit\"{a}t Frankfurt, Frankfurt, Germany\\
$^{65}$ Instituto de Ciencias Nucleares, Universidad Nacional Aut\'{o}noma de M\'{e}xico, Mexico City, Mexico\\
$^{66}$ Instituto de F\'{i}sica, Universidade Federal do Rio Grande do Sul (UFRGS), Porto Alegre, Brazil\\
$^{67}$ Instituto de F\'{\i}sica, Universidad Nacional Aut\'{o}noma de M\'{e}xico, Mexico City, Mexico\\
$^{68}$ iThemba LABS, National Research Foundation, Somerset West, South Africa\\
$^{69}$ Jeonbuk National University, Jeonju, Republic of Korea\\
$^{70}$ Johann-Wolfgang-Goethe Universit\"{a}t Frankfurt Institut f\"{u}r Informatik, Fachbereich Informatik und Mathematik, Frankfurt, Germany\\
$^{71}$ Korea Institute of Science and Technology Information, Daejeon, Republic of Korea\\
$^{72}$ KTO Karatay University, Konya, Turkey\\
$^{73}$ Laboratoire de Physique Subatomique et de Cosmologie, Universit\'{e} Grenoble-Alpes, CNRS-IN2P3, Grenoble, France\\
$^{74}$ Lawrence Berkeley National Laboratory, Berkeley, California, United States\\
$^{75}$ Lund University Department of Physics, Division of Particle Physics, Lund, Sweden\\
$^{76}$ Nagasaki Institute of Applied Science, Nagasaki, Japan\\
$^{77}$ Nara Women{'}s University (NWU), Nara, Japan\\
$^{78}$ National and Kapodistrian University of Athens, School of Science, Department of Physics , Athens, Greece\\
$^{79}$ National Centre for Nuclear Research, Warsaw, Poland\\
$^{80}$ National Institute of Science Education and Research, Homi Bhabha National Institute, Jatni, India\\
$^{81}$ National Nuclear Research Center, Baku, Azerbaijan\\
$^{82}$ National Research and Innovation Agency - BRIN, Jakarta, Indonesia\\
$^{83}$ Niels Bohr Institute, University of Copenhagen, Copenhagen, Denmark\\
$^{84}$ Nikhef, National institute for subatomic physics, Amsterdam, Netherlands\\
$^{85}$ Nuclear Physics Group, STFC Daresbury Laboratory, Daresbury, United Kingdom\\
$^{86}$ Nuclear Physics Institute of the Czech Academy of Sciences, Husinec-\v{R}e\v{z}, Czech Republic\\
$^{87}$ Oak Ridge National Laboratory, Oak Ridge, Tennessee, United States\\
$^{88}$ Ohio State University, Columbus, Ohio, United States\\
$^{89}$ Physics department, Faculty of science, University of Zagreb, Zagreb, Croatia\\
$^{90}$ Physics Department, Panjab University, Chandigarh, India\\
$^{91}$ Physics Department, University of Jammu, Jammu, India\\
$^{92}$ Physics Program and International Institute for Sustainability with Knotted Chiral Meta Matter (SKCM2), Hiroshima University, Hiroshima, Japan\\
$^{93}$ Physikalisches Institut, Eberhard-Karls-Universit\"{a}t T\"{u}bingen, T\"{u}bingen, Germany\\
$^{94}$ Physikalisches Institut, Ruprecht-Karls-Universit\"{a}t Heidelberg, Heidelberg, Germany\\
$^{95}$ Physik Department, Technische Universit\"{a}t M\"{u}nchen, Munich, Germany\\
$^{96}$ Politecnico di Bari and Sezione INFN, Bari, Italy\\
$^{97}$ Research Division and ExtreMe Matter Institute EMMI, GSI Helmholtzzentrum f\"ur Schwerionenforschung GmbH, Darmstadt, Germany\\
$^{98}$ Saga University, Saga, Japan\\
$^{99}$ Saha Institute of Nuclear Physics, Homi Bhabha National Institute, Kolkata, India\\
$^{100}$ School of Physics and Astronomy, University of Birmingham, Birmingham, United Kingdom\\
$^{101}$ Secci\'{o}n F\'{\i}sica, Departamento de Ciencias, Pontificia Universidad Cat\'{o}lica del Per\'{u}, Lima, Peru\\
$^{102}$ Stefan Meyer Institut f\"{u}r Subatomare Physik (SMI), Vienna, Austria\\
$^{103}$ SUBATECH, IMT Atlantique, Nantes Universit\'{e}, CNRS-IN2P3, Nantes, France\\
$^{104}$ Sungkyunkwan University, Suwon City, Republic of Korea\\
$^{105}$ Suranaree University of Technology, Nakhon Ratchasima, Thailand\\
$^{106}$ Technical University of Ko\v{s}ice, Ko\v{s}ice, Slovak Republic\\
$^{107}$ The Henryk Niewodniczanski Institute of Nuclear Physics, Polish Academy of Sciences, Cracow, Poland\\
$^{108}$ The University of Texas at Austin, Austin, Texas, United States\\
$^{109}$ Universidad Aut\'{o}noma de Sinaloa, Culiac\'{a}n, Mexico\\
$^{110}$ Universidade de S\~{a}o Paulo (USP), S\~{a}o Paulo, Brazil\\
$^{111}$ Universidade Estadual de Campinas (UNICAMP), Campinas, Brazil\\
$^{112}$ Universidade Federal do ABC, Santo Andre, Brazil\\
$^{113}$ Universitatea Nationala de Stiinta si Tehnologie Politehnica Bucuresti, Bucharest, Romania\\
$^{114}$ University of Cape Town, Cape Town, South Africa\\
$^{115}$ University of Derby, Derby, United Kingdom\\
$^{116}$ University of Houston, Houston, Texas, United States\\
$^{117}$ University of Jyv\"{a}skyl\"{a}, Jyv\"{a}skyl\"{a}, Finland\\
$^{118}$ University of Kansas, Lawrence, Kansas, United States\\
$^{119}$ University of Liverpool, Liverpool, United Kingdom\\
$^{120}$ University of Science and Technology of China, Hefei, China\\
$^{121}$ University of South-Eastern Norway, Kongsberg, Norway\\
$^{122}$ University of Tennessee, Knoxville, Tennessee, United States\\
$^{123}$ University of the Witwatersrand, Johannesburg, South Africa\\
$^{124}$ University of Tokyo, Tokyo, Japan\\
$^{125}$ University of Tsukuba, Tsukuba, Japan\\
$^{126}$ Universit\"{a}t M\"{u}nster, Institut f\"{u}r Kernphysik, M\"{u}nster, Germany\\
$^{127}$ Universit\'{e} Clermont Auvergne, CNRS/IN2P3, LPC, Clermont-Ferrand, France\\
$^{128}$ Universit\'{e} de Lyon, CNRS/IN2P3, Institut de Physique des 2 Infinis de Lyon, Lyon, France\\
$^{129}$ Universit\'{e} de Strasbourg, CNRS, IPHC UMR 7178, F-67000 Strasbourg, France, Strasbourg, France\\
$^{130}$ Universit\'{e} Paris-Saclay, Centre d'Etudes de Saclay (CEA), IRFU, D\'{e}partment de Physique Nucl\'{e}aire (DPhN), Saclay, France\\
$^{131}$ Universit\'{e}  Paris-Saclay, CNRS/IN2P3, IJCLab, Orsay, France\\
$^{132}$ Universit\`{a} degli Studi di Foggia, Foggia, Italy\\
$^{133}$ Universit\`{a} del Piemonte Orientale, Vercelli, Italy\\
$^{134}$ Universit\`{a} di Brescia, Brescia, Italy\\
$^{135}$ Variable Energy Cyclotron Centre, Homi Bhabha National Institute, Kolkata, India\\
$^{136}$ Warsaw University of Technology, Warsaw, Poland\\
$^{137}$ Wayne State University, Detroit, Michigan, United States\\
$^{138}$ Yale University, New Haven, Connecticut, United States\\
$^{139}$ Yonsei University, Seoul, Republic of Korea\\
$^{140}$  Zentrum  f\"{u}r Technologie und Transfer (ZTT), Worms, Germany\\
$^{141}$ Affiliated with an institute covered by a cooperation agreement with CERN\\
$^{142}$ Affiliated with an international laboratory covered by a cooperation agreement with CERN.\\

\end{flushleft} 

\end{document}